\documentclass[sigconf]{acmart}
\AtBeginDocument{%
  }

\usepackage{cleveref}
\usepackage{tikz}
\usepackage{array}
\usepackage{color, soul}
\usepackage{enumitem}

\newcommand{\systemname}{\textsc{UnexpectedParrots}}

\usepackage[T1]{fontenc}
\usepackage{booktabs}
\usepackage{listings}
\usepackage{xcolor}
\usepackage{microtype}

\definecolor{promptbg}{gray}{0.96}
\lstdefinestyle{promptstyle}{
  basicstyle=\ttfamily\scriptsize,
  backgroundcolor=\color{promptbg},
  breaklines=true,
  breakatwhitespace=true,
  columns=fullflexible,
  keepspaces=true,
  frame=leftline,
  framesep=4pt,
  xleftmargin=6pt,
  showstringspaces=false,
  literate={—}{{---}}1 {–}{{--}}1
}

\begin{document}

\settopmatter{printacmref=false}   
\renewcommand\footnotetextcopyrightpermission[1]{\noindent Preprint, Under Review.}
\pagestyle{plain}                  

\title{\systemname{}:~Iterating Behavioral Experiment Design with In-Silico Simulations}
\titlenote{Our system name is a word play on \emph{Expected Parrot}~\citep{expected-parrot},
which popularized the use of language model simulation for social science experiments, and
their name in turn plays on the \emph{Stochastic Parrot}~
\cite{10.1145/3442188.3445922}. Most simulation tools aim to converge the response distributions of LLM participants to the typical (expected) responses of an analogous human population.  In contrast, \systemname{} leverages the pre-response reasoning traces from these LLM participants to identify potential \emph{unexpected} interpretations of a study instrument that may reveal design inconsistencies of behavioral studies that may go unnoticed by the researchers themselves.}
\author{Sanchaita Hazra}
\correspondingauthor
\authornotemark[1]
\email{sanchaita.hazra@utah.edu}
\affiliation{%
  \institution{University of Utah}
  \city{Salt Lake City}
  \state{Utah}
  \country{USA}
}
\author{Bodhisattwa Prasad Majumder}
\affiliation{%
  \institution{Allen Institute for AI}
  \city{Seattle}
  \country{USA}}
\email{bodhisattwam@allenai.org}

\author{Pao Siangliulue}
\affiliation{%
  \institution{Allen Institute for AI}
  \city{Seattle}
  \country{USA}
}
\email{paos@allenai.org}

\author{Peter Clark}
\affiliation{%
 \institution{Allen Institute for AI}
 \city{Seattle}
  \country{USA}
 }
\email{peterc@allenai.org}

\author{Amy X. Zhang}
\affiliation{%
  \institution{University of Washington}
  \city{Seattle}
  \country{USA}
  }
\email{axz@cs.uw.edu}

\author{Joseph Chang}
\affiliation{%
  \institution{Allen Institute for AI}
  \city{Seattle}
  \country{USA}}
\email{josephc@allenai.org}



%
\renewcommand{\shortauthors}{Hazra et al.}

\begin{abstract}
Running behavioral experiments is expensive, slow, and prone to design flaws, often detectable after collecting data. We introduce \systemname{} to iteratively refine behavioral study designs before a researcher recruits costly human participants. \systemname{} relies on reasoning traces in addition to responses from LLM participants to uncover diagnostic critiques of structural errors and missed opportunities in existing study designs. When evaluated against a human-pilot or a no-simulation design-only baseline, \systemname{} adds most value, producing unique, high-quality critiques. In a moderated user study with 12 proficient researchers, we find \systemname{}  useful, time-saving, and thought-provoking. Critiques provided by \systemname{} often converge on issues researchers identify only during independent human pilots or additional formative interviews, even when the fidelity of the simulated populations remains questionable. Our users deem \systemname{} useful not only as a tool to conduct early pilot studies and identify errors but also as an ideation tool to expand one's research horizon. 

\end{abstract}


\begin{teaserfigure}
 \includegraphics[width=\textwidth,trim=0 180 20 50,clip]{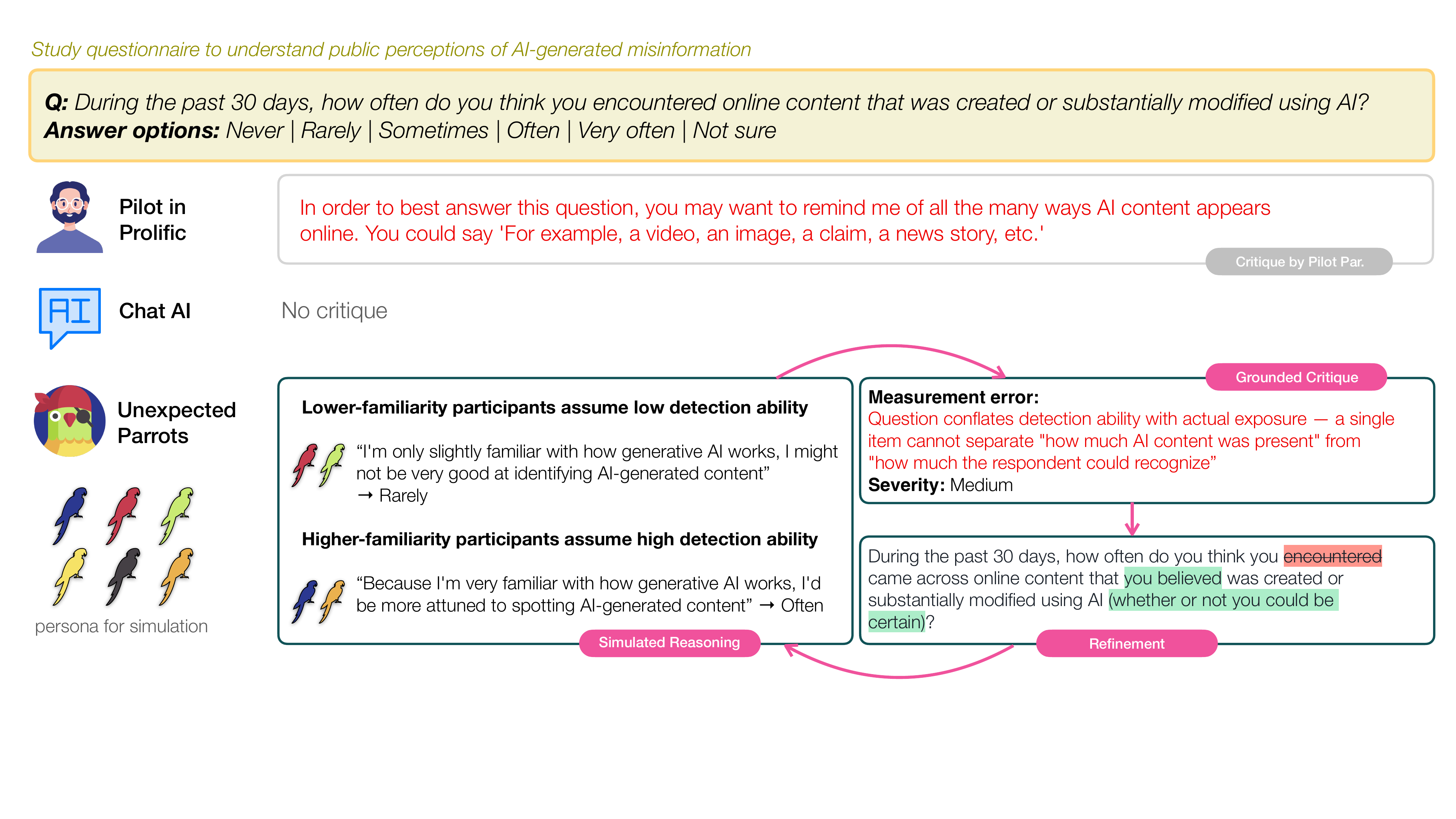}
  \caption{We argue \textbf{diverse reasoning traces} from in-silico simulations can surface valuable diagnostic signals of blind spots in behavioral experiment designs. \textbf{Instead of pursuing simulation as a replacement of real human studies, we introduce \systemname{}
  and highlight an alternative benefit of LLM simulations} to help researchers improve their (early) study designs with high-quality iterative feedback that may be otherwise unavailable from human-pilot studies or simple chat-based AI interactions
  before recruiting human participants for the final data collection.}
  \label{fig:teaser}
  \Description{A survey question sits at the top: ``During the past 30 days, how often do you think you encountered online content that was created or substantially modified using AI?'' The answer options are Never, Rarely, Sometimes, Often, Very often, and Not sure. Three rows below it show what three sources say about that question. The human pilot participant recruited on Prolific asks only to be reminded of the many ways AI content appears online, and suggests the question give examples. The chat-based AI returns no critique at all. The third row, a set of simulated personas, forms a loop of three boxes. The first holds simulated reasoning traces, where lower-familiarity personas justify answering ``Rarely'' by their own poor ability to identify AI-generated content, while higher-familiarity personas justify ``Often'' by being attuned to spotting it. The second is a grounded critique of medium severity, labeled a measurement error, since the question conflates detection ability with actual exposure, and a single item cannot separate how much AI content was present from how much the respondent could recognize. The third is a refinement that rewrites the stem to ask how often the respondent came across content they believed was created or modified using AI, whether or not they could be certain, with deleted and inserted wording marked in red and green. An arrow runs from the refinement back to the simulation and closes the loop.}
\end{teaserfigure}



\maketitle
\section{Introduction}

\textit{To err is human, to forgive, divine.} Forgiving study design flaws, however, in behavioral and survey experiments often comes at a cost. These studies are often high-effort to design and conduct, but they pave the way for important insights. 
Controlled studies gave us prospect theory and the framing of decisions \citep{kahneman1979prospect, tversky1981framing}, the endowment effect \citep{kahneman1991endowment}, and the situational contours of obedience \citep{blass1991understanding}, while recurring survey programs such as the General Social Survey \citep{davis1991norc}, the British Election Study \citep{bes1964}, and the World Values Survey \citep{wvs} have charted people's attitudes and their movements across decades and societies. Much of this work exists to compare how different groups of people think and behave across age, occupation, region, political outlook, and more. However, the diverse backgrounds and perspectives that make such work valuable may also make it difficult to have a study design that works for all participants, as intended by the researcher.

Study designs are usually written by one or a small team of researchers from close-knit research communities, and each question or description in the study is worded to carry the interpretations and assumptions that team had in mind, often unconsciously \citep{oudshoorn2004configuring}. Literature has often critiqued how studies sample participants (for e.g., WIERD sampling) \citep{henrich2010weirdest,linxen2021weird}. Yet it applies similarly to the people who write the questions as well. When a participant reads a question differently from how the designer intended, a mismatch may degrade the quality of the collected data, which could be quite challenging to identify until late. 

In one of our case studies, a survey 
asked participants how much they trusted scientists and other subject-matter experts to provide accurate information about important public issues. Read on its own, the question is sound. A scientist answering this is rating a group they belong to and associate with. An individual who believes science has become politicized perceives this as answering a question about politics. 
Both can land on the same point of the scale, and each answer is valid, so the response distribution gives no sign that one number is carrying several different judgments. Differing interpretations thus often surface only after the data are collected, when the budget is spent, and the study can no longer be recalled \citep{presser2004methods}.


One way we currently address this gap is by conducting pilot studies and collecting free-form responses or monitored think-alouds~\citep{willis2005cognitive}. Typically, researchers run a pilot, synthesize feedback, and revise before committing to the full sample. A mature methodological literature has formalized how to pilot questions and probe the understanding of the respondents \citep{presser2004methods, krosnick2010question}. While piloting can be very effective, a pilot study's yield is bounded by the time and effort it takes to run and by the size \citep{blair2011sample, ryan2012improving} and diversity of its participants \citep{miller2011design}. Moreover, a set of pilot participants drawn from the same pool that produced the research team will often reproduce the team's blind spot rather than correcting it. 


Language models have recently shown promise for lowering the cost of a pretest and widening the range of people who can answer it \citep{agnew2024illusion}. Researchers have begun using simulated participants in place of recruited ones, since a large and varied pool of respondent profiles can now be generated on demand \citep{zhao2025large,gonzalez2025emulating,cao2025specializing}. 
This substitution assumes a model's answers match the distribution of human populations faithfully; however, this assumption remains heavily debated. Simulated opinions tend to skew toward particular demographics \citep{santurkar2023whose}, LLM-personas can flatten into caricatures of the groups they represent (\textit{surrogate effect}) \citep{cheng2023compost, kapania2025simulacrum}, and second-order quantities such as the variance of opinion can be distorted even when averages look plausible \citep{bisbee2024synthetic}. Discussions in this area has evolved around two goals. One line pursues fidelity, conditioning models on sociodemographic backstories so that simulated response distributions track those of human subgroups \citep{argyle2023out, aher2023using, park2024generative}, while another pursues replacement, substituting AI personas for data collection when human data is slow or expensive \citep{manning2024automated, dillion2023can}. However, both read a simulation for the answer it gives. Neither asks how a simulated participant interpreted the study design, and both assume that the design is sound.

In this work, we do not pursue simulation as a replacement for real human studies. We leverage simulations for a different signal. Instead of investigating whether the simulation responses match with real participant responses, we ask whether the study design is worded in ways that participants would interpret in unintended ways. Survey methodology refers to the respondent's \emph{construal} of an item \citep{tourangeau2000psychology}. We treat disagreement among those interpretations as a diagnostic. A survey question that can be read in four incompatible ways is a defective question, no matter who produced the readings, so neither the persona nor its answer needs to be accurate for that disagreement to be informative. We extract these interpretations from the internal reasoning traces that simulated participants produce before they answer. A trace explains a possible failure of the study design, and when presented to the researcher, they decide whether it warrants a revision. 
Our claim revolves around how simulation can surface readings that researchers could miss, when critical evaluation of the study design by itself does not surface those flaws at all. Simulation acts as an additional source of early feedback, available before recruitment and alongside or even in the absence of human piloting. 




We propose \textbf{\systemname{}}, an interactive tool for improving study designs through an iterative workflow of \emph{Simulation}-\emph{Critique}-\emph{Refinement}. \systemname{} allows researchers to test a draft design early, and to explore how participants from backgrounds unlike their own might interpret it. A researcher provides a study plan containing the research question, the questions themselves, any stimuli, and the initial target demography. \systemname{} compiles the plan into a rich pool of diverse personas that render simulated reasoning traces mimicking varying study interpretations. The Critic outputs two kinds of critiques, inferring them from reasoning traces: \textit{structural errors} in the existing study questions, and \textit{missed opportunities} alluded to by latent reasoning traces. Researchers can review and triage these critiques in a workspace alongside their study design, indicating which ones to act on and how. The Refiner then returns the next version of their study based on their judgments and feedback. A researcher refines multiple times, giving further feedback, or starts a new round of simulation to get additional critiques of a refined version of their study plan. 


Concretely, our work makes the following contributions:
\begin{enumerate}
    \item The idea of using LLM simulation as a diagnostic instrument for study design, where the value lies in the diversity of interpretations a simulation, exposed through LLM's internal reasoning traces, instead of in the fidelity of the simulated answers.

    \item An offline study comparing critiques collected from human pilots and from our simulation-critique pipeline, showing that simulated reasoning surfaces significantly more novel and higher-quality critiques with a similar or less economic budget.

    \item \textbf{\systemname{}}, an interactive system that uses the Simulate-Critique-Refine workflow, allowing researchers to review potential misinterpretations or differing interpretations of their studies, and ways to address them.


    \item A user study with 12 proficient researchers refining their own study designs, ranging from surveys, interviews, formative research, and controlled experiments across populations in North America, Asia, and the African Subcontinents, showing that \systemname{} is useful, steerable, and thought-provoking, directly adding value to their ongoing or semi-complete behavioral studies.
\end{enumerate}


\section{Background and Related Work}
\label{sec:related}

\subsection{Study Design Failures and Piloting / Pre-testing}

Decades of methodological work have cataloged the ways a study design can fail. For example, \citet{shadish2002quasi} organized validity and its threats into a standard typology that remains the reference point for reasoning about inference from imperfect designs. Prior work has also focused on the artifacts of measurement, covering results that depend on the particular stimuli chosen \citep{clark1973language}, response scales that anchor what participants treat as normal \citep{schwarz1985response}, common-method variance \citep{podsakoff2003common}, and the demand characteristics \citep{orne2017social}, question-order effects \citep{tourangeau1988cognitive}, and satisficing \citep{krosnick1991response} that appear once participants know they are being studied. One of our case studies is a design in this family, measuring how social desirability shapes self-reported AI use \citep{ling2026underreporting}. Missing one is costly, since replication efforts find that a substantial fraction of published findings fail to reproduce \citep{oscollab2015estimating, camerer2016evaluating}. However, we build on this literature to define the different types of structural error a study plan can have, and generate critiques grounded in them to help researchers spot these errors early.

Researchers currently catch these problems through pretesting. Cognitive interviewing and think-aloud protocols elicit how a participant interpreted (or \emph{construed}) an item \citep{willis2005cognitive, krosnick2010question}, and the researcher revises their study plan accordingly before the study is actually conducted at a greater scale in the field. A pretest produces a small set of expert-judged critiques of specific questions; the same artifact \systemname{} produces and the bar our offline study measures against. Pretests are also small, so their coverage of unusual readings is thin by construction.

\subsection{LLM Social Simulation and Optimizing to Replace Human Participants}

An increasingly popular line of work explored ways to leverage LLMs in place of human participants. For example, \citet{argyle2023out} conditioned language models on sociodemographic backstories and validated the result by replicating the response distributions of human subgroups; \citet{aher2023using} reproduced classic effects while surfacing a hyper-accuracy distortion, and \citet{park2024generative} grounded agents in two-hour interviews with a thousand people; and \citet{manning2024automated} went the furthest, replacing both scientist and subject with language models. Within HCI, \citet{liu2026cobra} showed how fragile this reliance on persona descriptions is. The same natural-language specification produced significantly different response patterns across models, and only one model they tested reproduced the simple expectation that economists show a weaker framing effect than laypeople. They address this by scoring agents against eight classic experiments, from Milgram obedience to the Asian disease problem, and regulating agent behavior toward a target level of bias. Most work here assumes the study design is sound, and the goal is to replicate human study with simulations. \citet{liu2026cobra} make that assumption concrete, since the classic experiments they measure against are designs that decades of use have already validated. Their yardstick has to be sound for a difference in agent response to say anything about the agent. We work at the other end, on an early draft that has yet to be run, where that same difference is evidence about the question instead. More closely related to our work, \citet{horton2023homo} also took the position of using language model as an instrument for piloting experiments. However, in their setup, pilots read the simulated study outcome to predict what a study will find, while we read the reasoning traces to ask whether the question was interpreted consistently as expected across different demographic groups. The two are compatible, since a pilot of that kind run on a misread question predicts the results of a misread question.


More recently, \citet{kapania2025simulacrum} raised concerns around replacing human participants in qualitative interviews.  Specifically, in qualitative interviewing, where the simulated speech is itself the finding.
In contrast, our setting, where there is evidence about an instrument that human participants will later answer. Their probe asked the interviewed researcher to write the participant description, and researchers attempted an average of 2.36 personas per session. Our users specify a demographic space instead of a participant, and the model generates a cohort of 50 within it, each with a biography written without sight of the study. We therefore never read a single persona as authoritative, and treat only the disagreement across them as a signal.

\subsection{Reasoning Traces and AI Support for Study Design}



Prompting a model to reason before answering produces an account of how it reached its response \citep{wei2022chain}. Most directly related to our system in this area is \citet{adhikari2025exploring} which used language models to draft questionnaires, have simulated personas answer them, and add a reviewer agent that critiques the result. 
At the same time, prior work has shown evidence that reasoning traces do not always deeply explain their final answers \cite{turpin2023language,chen2025reasoning, lanham2023measuring}. 
In this paper, we leverage reasoning traces not as a way to understand or leverage the final answer from simulated pilot participants, but instead, we mine ways to study and survey questions that may be interpreted differently amongst different participants, and between the researchers and the participants. This reflects the cognitive model of survey response that calls for reading the participant's construal, and locates it at the comprehension stage \citep{tourangeau2000psychology}.





\section{The System: \systemname{}}\label{sec:system}
We present \systemname{}, an interactive system that supports researchers in iteratively improving their study plans for human-subject behavioral studies through simulation-grounded critiques and user-guided revisions. \systemname{} (1) simulates the study with synthetic LLM participants with an open language model to collect their internal reasoning traces that lead to their responses; (2) uses these traces to critique the study and identify methodological errors or missed opportunities, such as differing interpretations across simulated participants assigned different backgrounds; (3) allows users to review and triage those critiques and give additional feedback that steers (4), which refines the study plan into an improved version. In our design, a researcher can iterate between steps 3 and 4 until they are satisfied, and then run the entire loop again to get a new set of critiques on the new version.
\footnote{We used the open \texttt{GLM-5.2} model for our simulation so that internal reasoning traces are available to us. We used \texttt{Opus-4-8} for the rest of the system, given the difficulty of the task of human-subject experiment design.}

To illustrate the features and functionality of \systemname{}, we follow Professor Dumbledore, a behavioral scientist who is designing a study on public perceptions of AI-generated misinformation. The study examines how people perceive, identify, and respond to AI-generated misinformation. Dumbledore has already developed an initial study plan. The study plan includes the research question(s)/hypothesis(es), the study questionnaire, participant instructions, stimuli, and the target population. 
Dumbledore wants the study to cover a wide range of demographics, but he is unsure whether he has phrased each question correctly, designed the answer options to fit all of his participants, and covered everything he should be asking.
Before developing the plan any further, he decides to use \systemname{} to get some quick feedback. 

\subsection{On-boarding and Running a Simulation} 

Dumbledore begins by submitting his study plan to the Simulator. 
The system then runs through the Simulation and Critic stages in the background, and notifies Dumbledore to give feedback and revise the study plan when the critiques are ready.

The Simulator operates in the background in three stages. First, a \textbf{Designer} compiles the free-form study plan into a set of predefined structured fields: a linearized study, the demographic space to sample participants from, a template for rendering those demographics into a profile, and the stimuli. Second, the \textbf{Persona Synthesizer} converts the target demography stated in the plan into a large set of (50) synthetic participant personas. Third, the \textbf{Response Generator} orchestrates the compiled study across all simulated participants, question by question, collecting their reasoning traces and responses. 

The study compilation does not perturb Dumbledore's own words: his research question, the questions themselves, and the participant instructions appear as is to the simulated participants, along with the treatment and control manipulations as intended. He can monitor these verbatim, appearing on the left pane of his workspace, as shown in \Cref{fig:refiner}. \systemname{} also reads the demographic space he wants to sample from the plan, along with the dimensions he had already decided to ask about, such as age, region, and occupation, together with their option levels and any dependencies between them. Compilation is the first line of defense through which \systemname{} can notify Dumbledore that something is wrong. Had he described a stimulus
without providing it along with the study design, the system would have stopped and raised an error instead of proceeding. Similarly, a conditional question that is tied to a free-form answer would not be left unresolved and will be flagged for him to correct before simulation.

\begin{table*}[t!] 
\caption{A persona, as sampled and rendered. The \textbf{\textcolor[HTML]{f0529c}{Trade apprenticeship}} level is out-of-distribution from available answer options for education-related questions. The rendered clause smoothens it further.} \label{tab:persona-example} 
\Description{The table shows one simulated participant in three parts. The sampled demographic profile lists country Japan, age 55 to 64, gender man, education trade apprenticeship, and political orientation apolitical. The Layer 1 rendered clauses turn those sampled levels into second-person sentences. They say that you live in Japan, that you are 63 years old, that you are a man, that you learned your trade through a hands-on apprenticeship and not in a classroom, and that you do not follow politics or think in left-right terms. The Layer 2 character snapshot, generated at temperature 0.9, is a paragraph-long backstory. It describes a tatami craftsman in Kurashiki who left school at fifteen to sweep floors in his master's workshop. He lives with his wife in the house he bought at forty, tends a persimmon tree, and follows the Hiroshima Carp on the radio. He believes a job done sloppily shames the man who did it. The sampled education level, and the phrases in the two later layers that derive from it, are highlighted in pink.}
\small \begin{tabular}{@{}p{0.985\linewidth}@{}} 
\toprule \textbf{Demographic profile} (sampled) \\ 
\midrule \footnotesize \textcolor{gray}{country: \textbf{\textcolor[HTML]{f0529c}{Japan}}; age: 55-64; gender: Man; education: \textbf{Trade apprenticeship}; political orientation: Apolitical} \\ \midrule \textbf{Layer 1: rendered clauses} (sampled levels $\rightarrow$ rendered template) \\ 
\midrule \textcolor{gray}{You live in \textbf{\textcolor[HTML]{f0529c}{Japan}}. You are 63 years old. You are a man. You learned your trade through a \textbf{\textcolor[HTML]{f0529c}{hands-on apprenticeship}} rather than in a classroom. You do not really follow politics and do not think of yourself in left-right terms.} \\
\midrule \textbf{Layer 2: character snapshot} (generated, $T{=}0.9$) \\ 
\midrule \textcolor{gray}{You spent thirty-eight years as a \textbf{\textcolor[HTML]{f0529c}{tatami craftsman in Kurashiki}}, where the smell of fresh igusa rush still tells you whether the harvest was a good one. You grew up the son of a rice farmer in rural Okayama, and you \textbf{\textcolor[HTML]{f0529c}{left school at fifteen to sweep floors in the workshop of the master who eventually taught you everything}}. You and your wife, Keiko, live in the same wooden house you bought at forty, and on Sundays, you tend the persimmon tree in the back garden that your late father planted. You cook a mean simmered mackerel with grated daikon, and you insist the miso must come from the little shop by Bikan Canal, never the supermarket brand. You follow the Hiroshima Carp with quiet stubbornness, listening to away games on the radio while you work because you say the television distracts your hands. You believe a job done sloppily shames the man who did it, and you still bow slightly to a room before you begin laying a new floor.} \\ 
\bottomrule 
\end{tabular} 
\end{table*}

Table~\ref{tab:persona-example} shows a complete profile for one of Dumbledore's LLM participants, a 63-year-old tatami craftsman in Kurashiki who left school at fifteen. Each profile is checked for coherence against the provided demographics and for redundancy against the rest of the pool before the simulation begins. Once validated, the pool is cached and reused across simulation runs, so when Dumbledore reruns the study, if need be, he encounters the same participant pool.

With Dumbledore's study now compiled and the participant pool fixed, the \textbf{Response Generator} runs the study design through each LLM participant and records their responses. Participants take the study one question at a time, reasoning through the answer options before committing to an answer choice. Because \systemname{} simulates with an open model, Dumbledore is able to capture the model's own internal reasoning tokens. This allows him to examine later surfaced critiques through the lens of persona's thinking traces, which is significantly more expressive than the chosen answer options. Table~\ref{tab:simulation-example} demonstrates the chain of reasoning for the craftsman persona mentioned in Table~\ref{tab:persona-example}.
The craftsman from Table~\ref{tab:persona-example} reasons whether generative AI changes how much false information circulates online, reasons from his trade rather than from the topic overall: 
\begin{quote}\small
``I'm a craftsman. I know that when you give people tools that make it easier to do something, more of that thing gets done. If you make a better saw, more cutting gets done.''
\end{quote}
He then tests this inference against what he has already informed in an earlier question in the survey, before finally settling on an answer option. Table~\ref{tab:simulation-example} reconstructs the chain of events:
\begin{quote}\small
``My previous answers show uncertainty (`Not sure') about AI content and AI misinformation. But this question is different, it's asking about my judgment of the overall effect, not my personal experience.''
\end{quote}

\subsection{Reviewing and Triaging Critiques} 
Once Dumbledore's LLM participants have completed the study, their responses and reasoning traces are passed to \systemname{}'s Critic. Given the research question, study questions, and outputs from the Response Generator, the Critic surfaces two complementary outputs: (a) methodological \textbf{Structural Errors} in the existing study questionnaire, and (b) \textbf{Missed Opportunities}, i.e., gaps in the current design that may help explain the study outcome or better address the research question. These are rendered in two separate tabs beside his design, with counters indicating the number of suggestions (\Cref{fig:refiner}). Dumbledore's study returns 71 structural errors and 17 missed opportunities of high or medium severity. As he works through them, a header above the tabs tracks the progress and shows how many refinement runs he has left.

\begin{figure*}[t!]
    \centering
    \includegraphics[width=\textwidth,trim=0 70 30 40,clip]{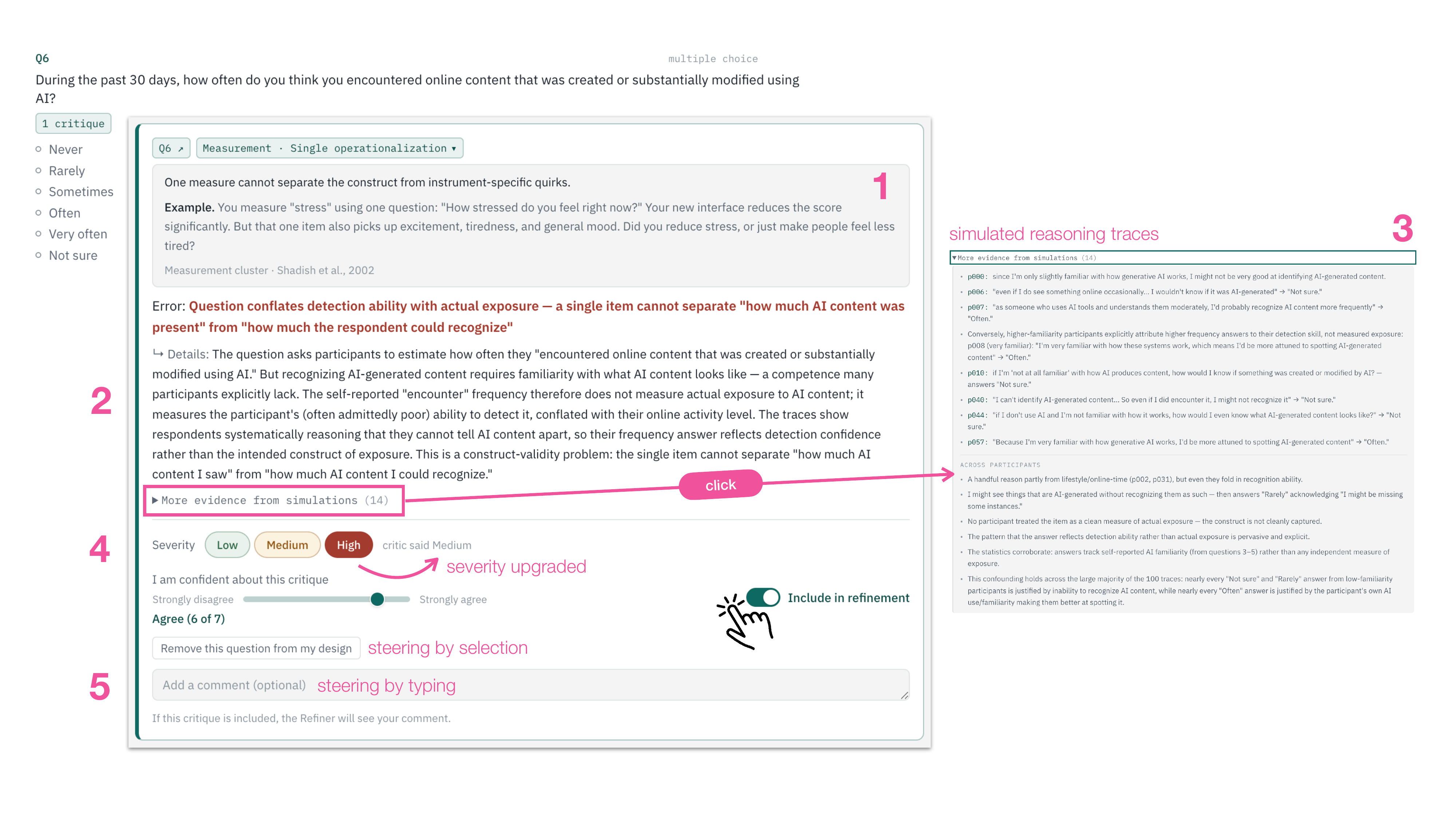}
    \caption{\textbf{Structural Errors.} \systemname{} surfaces structural errors question-by-question. For each structural error, tied to an existing question \textbf{1)} \systemname{} uses an error taxonomy to categorize the said error, with a typical definition and canonical example present with a click. \textbf{2)} For each error, it provides a summary statement highlighting the error, followed by detailed explanation about its relevance. \textbf{3)} With a click, the researcher can explore the provenance of such details grounded in the simulated reasoning traces. These groundings help the researcher assess the quality of the error. \textbf{4)} That assessment then materializes as their own rating of the error's severity, which overrides the severity the Critic assigned while leaving it visible for comparison. The researcher also logs how far they agree with the critique on a seven-point scale, and toggles whether it is carried into refinement. \textbf{5)} Finally, to further steer the process of refinement, the researcher can completely drop the question from the design or can provide additional guidance how the critique should be treated through a textual comment.}
    \label{fig:critic}
    \Description{A screenshot of the Critic's structural-error view for one survey question. On the left sits question Q6, ``During the past 30 days, how often do you think you encountered online content that was created or substantially modified using AI?'', with its six answer options and a badge reading ``1 critique''. Five pink numbers annotate the critique card beside it. At number one, the header tags the critique with the taxonomy category ``Measurement, Single operationalization'', and expands to a definition and a canonical example about measuring stress with a single item, cited to Shadish et al., 2002. Number two is the error statement, in red, saying the question conflates detection ability with actual exposure, followed by a paragraph of detail. Below it, number three points at a collapsed row reading ``More evidence from simulations (14)''. A pink arrow labeled click'' runs from that row to a panel of simulated reasoning traces on the right, which lists quoted excerpts from individual personas alongside their answers, and then a set of across-participant observations. Number four is the severity control, a row of Low, Medium, and High buttons with High selected, next to a note that the critic said Medium, and an annotation reading ``severity upgraded''. Under it, a seven-point confidence slider reads ``Agree (6 of 7)'' and a toggle labelled ``Include in refinement'' is switched on. The last number sits beside the steering controls, a button to remove the question from the design annotated ``steering by selection'', and a free-text comment box annotated ``steering by typing''.}
\end{figure*}

\subsubsection{\textbf{Structural Errors}}
Dumbledore is now curious about the methodological errors in his questionnaire. The Critic evaluates his study one question at a time. For a question, it first examines responses from all LLM participants, along with their reasoning traces. If the Critic perceives that a question contains a methodological flaw, it surfaces it as a \emph{structural error} on the interface. It provides Dumbledore with a brief description of the error, followed by a detailed explanation (\Cref{fig:critic}, 2). 
On Q6, which asks how often he thinks participants encountered online content created or substantially modified using AI in the past 30 days, the Critic reports that the item conflates detection ability with actual exposure, since one such question cannot separate how much AI content was present from how much the respondent could recognize. 
\begin{enumerate}
    \item\textbf{Grounded in Reasoning}. Looking at the questions being flagged on the interface, Dumbledore can quickly see potential methodological problems, such as leading phrases, double-barreled items, poorly specified scales, or missing controls. However, he finds it hard to imagine how they might affect the answer he will collect eventually. Each structural error, therefore, is also provided with provenance from participant-level reasoning to explain further. Dumbledore can open a critique to read the quoted traces of the LLM participants, followed by a population-level synthesis (\Cref{fig:critic}, 3). For Q6, there are 14 supporting traces, among which two are particularly interesting. Participant p010 asks, internally, \emph{how they would know whether something was created or modified by AI when they are not at all familiar with how AI produces content?}, and answers \emph{Not sure}, while p057 answers \emph{Often} because being very familiar with generative AI means they would be more attuned to spotting it. Both are reasoning about their own detection ability, and neither is reporting exposure. These tell Dumbledore directly whether current and future participants may misinterpret a question, introduce unstated assumptions, or reconcile their circumstances with ill-fitting response options. See Table \ref{tab:structural-error-critic-refiner-example}.

    \item\textbf{Error Taxonomy}. \systemname{} also grounds each identified structure error in a curated taxonomy of known experiment and survey design pitfalls. This allows Dumbledore to triage them using familiar terms and categories. To see this grounding, Dumbledore clicks the \emph{associated pitfall} tab to expand a short description of the pitfall, a canonical example, and the citations he can consult if he wants to read more (\Cref{fig:critic}, 1). The Q6 error is denoted as \texttt{Measurement} $\cdot$ \texttt{Single operationalization}, and the tab opens on a worked example of a one-item stress measure that also picks up excitement, tiredness, and general mood of the respondent. This allows Dumbledore to consolidate his thoughts around the critique to determine its importance and a potential solution.    

    \item\textbf{Severity}. As several critiques are generally surfaced, it is important for the Critic to surface its perceived severity to help Dumbledore sort through them. The severity of an error is characterized as the consequence of the study as a whole, going beyond the apparent seriousness of the exact local error. Severity is \emph{high} when the study's main conclusions are at stake, \emph{medium} when addressing the finding would only strengthen the study, and \emph{low} when its consequences are merely marginal. Dumbledore views the results from the Critic in a decreasing order of severity. Given the interactive nature of the system, Dumbledore can reassign that severity to situate an error in the context of his own study, and the interface keeps the Critic's original rating beside his so the disagreement remains visible (\Cref{fig:critic}, 4) for future consideration. For Q6, the Critic flagged it as medium, but the question pivots the whole encounter-frequency interaction, so Dumbledore raises it to high. Severity can also move the other way: elsewhere, he has deliberately written a double-barreled question, and he drops the Critic's flag to be low. As he goes through this process, he also records his confidence in the critique on a seven-point Likert scale, which allows him to ratify them later. Convinced by the Q6 critique, he switches on \textit{Include in refinement}. On another question, he appreciates the critique but wants to steer it in a direction he had already been considering, so he types his thoughts into the comment box for the Refiner to consider when proposing a solution. He can also remove a question completely as he goes through the overall review of his current design (\Cref{fig:critic}, 5).
\end{enumerate}

\begin{figure*}[t!]
    \centering
    \includegraphics[width=\textwidth,trim=150 0 150 0,clip]{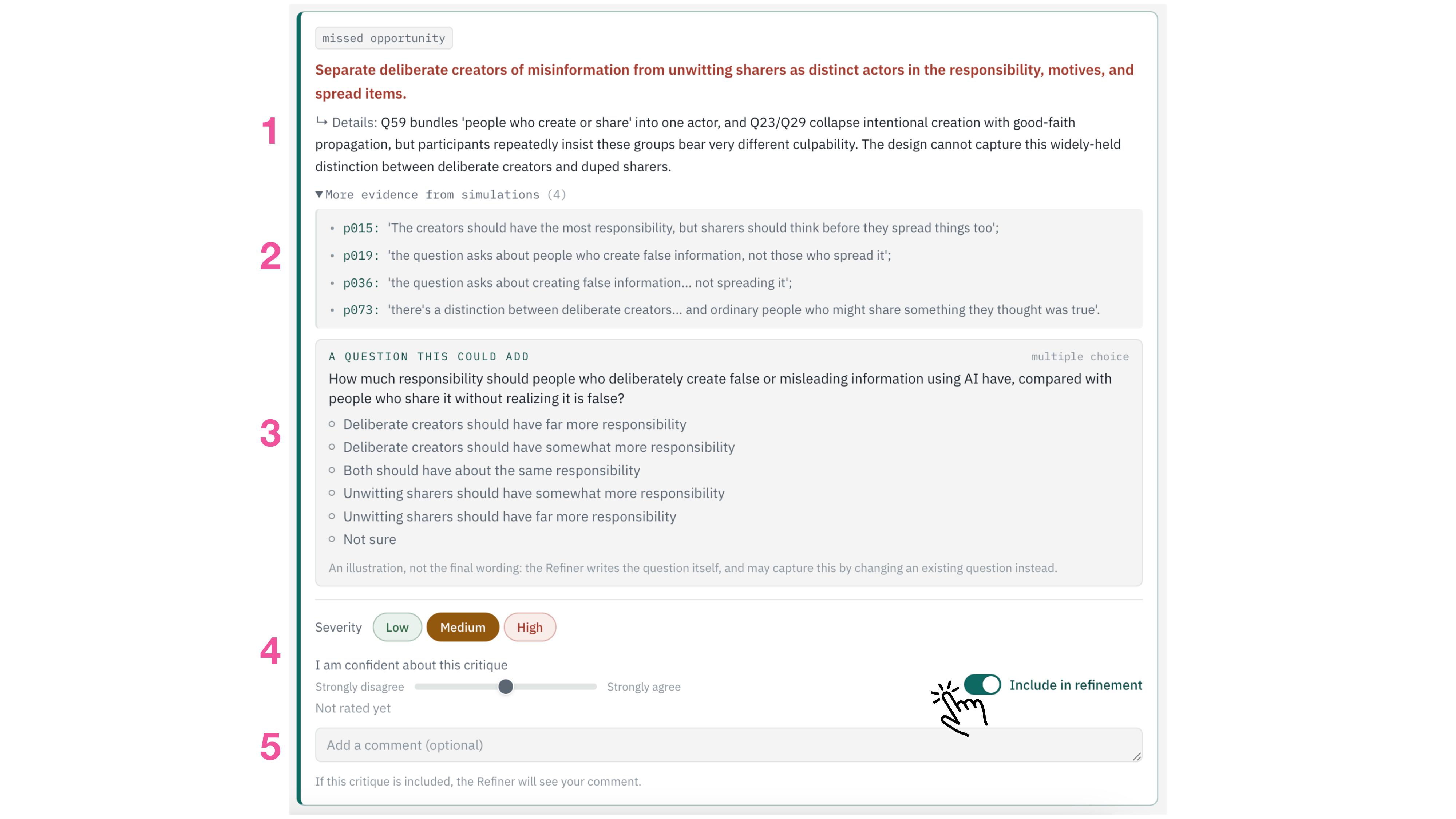}
    \caption{\textbf{Missed Opportunities.} \systemname{} proposes missed opportunities at the study level, by listing several of them. \textbf{1}) It starts with a high-level description of the missed opportunity with details about its relevance. \textbf{2)} With a click, the researcher can access more evidence from the simulated traces that led to the missed opportunity. \textbf{3)} Along with the high-level suggestion, researchers can also see a potential refinement (e.g., a new question it may add to the design) if they had included it in the refinement. This helps researchers to appropriately assess the scope of the change, since some missed opportunities could require a larger change in scope. \textbf{4) and 5)} As with structural errors, researchers can express their own preference about the severity and add comments to steer the refinement if the missed opportunity is carried forward.}
    \label{fig:missedop}
    \Description{A screenshot of one missed-opportunity card, annotated with five pink numbers. The heading at number one, in red, proposes that deliberate creators of misinformation be separated from unwitting sharers as distinct actors in the responsibility, motives, and spread of items. Details below note that Q59 bundles the two into one actor, while participants repeatedly insist the groups bear different culpability. Number two is an expanded evidence panel headed ``More evidence from simulations (4)'', quoting four personas who each draw a distinction between people who create false information and people who merely spread it. At number three sits a boxed suggestion headed ``A question this could add''. It is a multiple-choice item asking how much responsibility deliberate creators should bear compared with people who share content without realising it is false. Its six answer options run from ``Deliberate creators should have far more responsibility'' through ``Both should have about the same responsibility'' to ``Unwitting sharers should have far more responsibility'' and ``Not sure'', under a note that this is an illustration and not final wording. Number four is the severity row, with Medium selected, a confidence slider still reading ``Not rated yet'', and an ``Include in refinement'' toggle switched on. The last number sits beside an optional comment box for steering the refinement.}
\end{figure*}

\subsubsection{\textbf{Missed Opportunities}} 
Missed opportunities are gaps in the current design that may help explain a study outcome or, even better, address a new sub-research question. These gaps are generally surfaced from verbalized deep thinking before answering a question, that are currently not in scope with the existing questions. Unlike structural errors, which Dumbledore reviews question by question, missed opportunities are reported for the study as a whole. They capture elements that Dumbledore's study does not currently measure, manipulate, or sample, but that could help explain the observed responses or more directly address the research question. Each one starts with a one-line statement of the gap and a short account of why it matters (\Cref{fig:missedop}, 1), and the traces it was conditioned on (\Cref{fig:missedop}, 2). One of the 17 such missed opportunities informs Dumbledore that his design never separates people who deliberately create misinformation from people who share it without realizing it. Q59 convolves them both into a single predictor, and p019, p036, and p073, the simulated personas, all push back on that oversight, with p073 insisting on the distinction between deliberate creators and ordinary people with no harmful intention. See Table~\ref{tab:missed-opportunity-example}. 

Missed opportunities often suggest entirely new questions, and to reduce Dumbledore's cognitive load \systemname{}, missed opportunity critiques also accompany a preview of the suggested addition to help him prioritize (\Cref{fig:missedop}, 3). For the creators-vs-sharers distinction that Q59 demands, a draft fix indicates a six-point scale question to elicit respondents' beliefs. Dumbledore can preinspect such a suggestion and decide how consequential such an addition would be to his whole study before making the final decision to include it in the refinement. Dumbledore engages with the missed opportunities in a similar way he approaches the structural errors, though he keeps the severity of the fix the same, and he adds a comment to make sure that such an addition does not invalidate other framing questions he may have in the study. Throughout the process, Dumbledore maintains the authority to rely optimally on \systemname{} and steer the refinement of his study to scope his research objective appropriately (\Cref{fig:missedop}, 4, 5).

\begin{figure*}[t!]
    \centering
    \includegraphics[width=\textwidth,trim=10 40 10 40,clip]{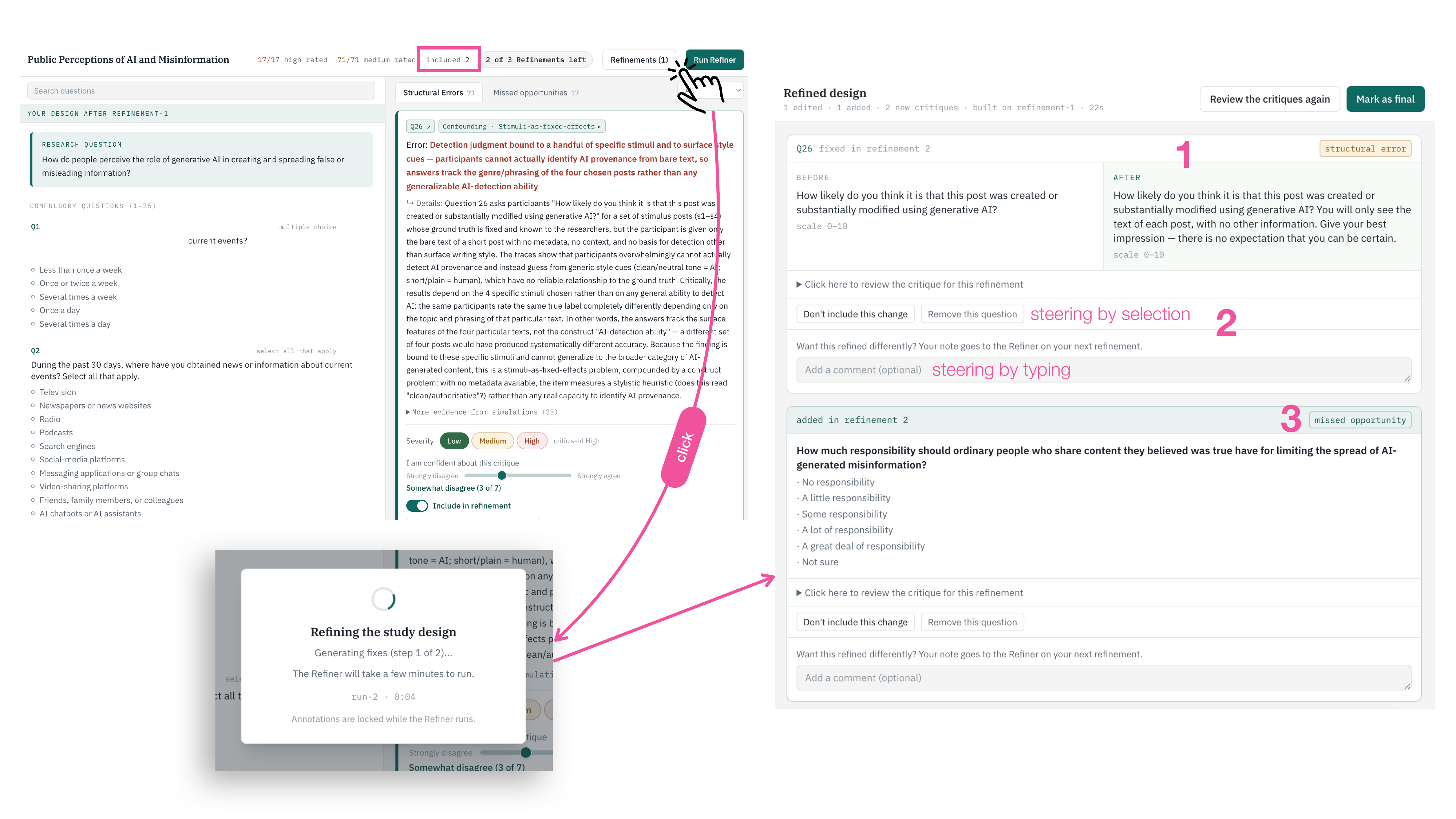}
    \caption{\textbf{The workspace and the Refiner.} On the left, the researcher's study design sits in a rail headed by the version they are looking at, with the Critic's structural errors and missed opportunities in tabs beside it and a header tracking triage progress and the refinement runs remaining. The \textbf{Refiner} operates on the structural errors and missed opportunities the researcher selected. Once they click ``Run Refiner'', it runs in the background for a few minutes, locking annotations while it works, and returns a side-by-side comparison of each changed question against the original for structural errors \textbf{(1)} and the exact suggested additions for missed opportunities \textbf{(3)}. As earlier, researchers can accept or reject a suggestion, or steer it by providing further comments \textbf{(2)}, and run the Refiner iteratively until they are satisfied with the changes.}
    \label{fig:refiner}
    \Description{A screenshot of the workspace, in two halves. The left half shows a three-column view. Its header reads ``Public Perceptions of AI and Misinformation'', with counters for high- and medium-rated critiques, a count of two included critiques, a note reading ``2 of 3 Refinements left'', and a ``Run Refiner'' button. A left rail holds the study design after refinement 1, headed by the research question and followed by the numbered questions and their answer options. The middle column carries tabs for ``Structural Errors 71'' and ``Missed opportunities 17'', and shows one open critique about detection judgment being bound to four specific stimuli, with its severity buttons, confidence slider, and include toggle. A pink arrow labeled click'' runs from the Run Refiner button to an inset modal reading ``Refining the study design, generating fixes (step 1 of 2)'', which notes that the Refiner takes a few minutes and that annotations are locked while it runs. The right half is the refined design panel, headed ``1 edited, 1 added, 2 new critiques, built on refinement-1'', with buttons to review the critiques again or mark the design final. Three pink numbers annotate it. Number one is a side-by-side before-and-after comparison for question Q26, where the after version adds that respondents will see only the text of each post and should give their best impression with no expectation of certainty. Beneath it, number two sits beside the steering controls, buttons to exclude the change or remove the question annotated ``steering by selection'', and a comment box annotated ``steering by typing''. Number three is a newly added missed-opportunity question, asking how much responsibility ordinary people who share content they believe is true have for limiting the spread of AI-generated misinformation, with six response options from ``No responsibility'' to ``A great deal of responsibility'' and ``Not sure''.}
\end{figure*}

\subsection{Iterative Study Refinement}
 After reviewing the Critic's findings, Dumbledore selects all the structural errors and missed opportunities he wants to address. For each selected finding, he may also have left a comment specifying how he wants it resolved. With his selections made, he clicks \textit{Run Refiner} (\Cref{fig:refiner}). The Refiner takes the Critic's findings as input and works through them in the background over a few minutes, reporting which stage it is on while Dumbledore's annotations are locked during that process. Typically, the refinement process runs proportionally with how many suggestions Dumbledore includes as part of the refinement process.

 For structural errors, it preserves most of the existing exposition as needed (unless instructed by Dumbledore), but edits partly the question and/or answer options to address the critique. For missed opportunities, the Refiner is instructed to follow the suggestions as closely as possible. For questions that are not included in the refinement phase, they will remain unchanged.  

 The refined design returns with a brief summary with proper versioning of refinement rounds. Each edited question is presented with its older counterpart so that Dumbledore can directly compare (\Cref{fig:refiner}, 1). For Q26, the Refiner maintains the same wording and appends a phrase telling respondents that, regardless of uncertainty, they should give their best impression. Questions originated from a missed opportunity are shown in full along with the original critique (\Cref{fig:refiner}, 3). From either of the suggested changes, Dumbledore can open the critique behind the change, decline the change, remove the question, or leave a note that the Refiner will consider on the next run (\Cref{fig:refiner}, 2). Unless otherwise specified, all refined questions will be automatically added to the updated version of the existing study design. \systemname{} allows Dumbledore to iteratively optimize his study designs by running the Refiner multiple times before finalizing the design. When satisfied, Dumbledore marks the resulting version as his final design, and until he does, he can return to the critiques at any point. Consequently, as Dumbledore reviews the new version of the study design, \systemname{} offers him an opportunity to return to the Simulator--Critic--Refiner loop for another round of simulation-grounded evaluation and refinement. \systemname{} takes his previous decisions into account, so errors that Dumbledore chose not to include in refinement do not recur in the next round of critiques, unless new reasoning traces give substantively new reasons for a similar error to resurface. After a few more iterations, he now feels more confident that he has considered a wider range of different perspectives than he was able to on his own, and is ready to test his study design with paid human participants.

\subsection{System Backend}
This section describes in more detail how \systemname{} produced the information when Dumbledore interacted with the system.

\subsubsection{\textbf{Designer}}
As the first component, the Designer uses a language model to compile a free-form study-design document into three typed artifacts that can be consumed by the rest of the pipeline.

First, the \textbf{linearized study} is a JSON file that includes each verbatim question, its type, response options, scale bounds, and skip logic\footnote{In a study, skip logic specifies whether a question is administered based on a participant's response to an earlier question.}. 
A compiled study is prepared with per-item artifacts as a declarative experiment that ensures to propagate controlled information to the simulated participants as intended in the study design.

Second, the \textbf{demographic space} is the set of dimensions the participants are drawn from, their demographic levels, and the constraints that bind them. Constraints rule out incoherent combinations, e.g., a 19-year-old participant cannot have 15 years of professional experience.\footnote{Constraints can take three forms. \textit{One-to-one} constraints link two demographic dimensions; e.g., an apprentice may only be assigned a tenure that is consistent with the role. \textit{One-to-many} constraints link one demographic feature to multiple others; e.g., having no prior experience with automation-building may restrict two questions: experience with automation and attitudes toward automation. Finally, some constraints model study-specific assumptions rather than logical incompatibilities; e.g., a study may define industry researchers as those affiliated only with companies with $500+$ employees.} If the study design comes with a predefined quota (50\% female participants), the sampler is designed to obey it.

Third, the \textbf{render spec} specifies how each demographic dimension is expressed in the participant profile. The Designer defines how each demographic level should be described, and the Persona Synthesizer applies this exposition to the levels sampled for each participant. This separation provides enough background to ground the participant's reasoning without directly revealing how they should answer a particular question. E.g., instead of repeating the label ``Bachelor's degree,'' as part of the persona, the render spec may describe the participant as having completed a four-year college degree. This preserves the assigned demographic information while exposing it in more natural language that can support diverse reasoning about related questions, such as years of experience. The demographic levels follow a complementary rule. When a dimension corresponds to a closed survey question, most levels match that question's response options exactly, allowing the system to check for consistency in participants' self-reporting. One or two levels are deliberately made different from available options to ensure generalizability to the intended population.
Stimuli are handled separately from the rest of the study design. Behavioral studies may include treatment and control materials, such as transcripts, images, or vignettes, that vary across participants or conditions. To preserve the intended treatment, the Designer keeps all provided stimuli unchanged.

\subsubsection{\textbf{Persona Synthesizer}}
The Persona Synthesizer turns the demographic space and the render spec into the participant pool. Each participant receives a profile in two layers, a rendered demographic profile and a character snapshot.

The \textbf{demographic profile} is constructed from the demographic space with deterministic sampling. The sampler samples one level for each demographic dimension, obeying declared constraints, and the render spec converts the sampled levels into natural-language descriptions. E.g., for age ranges, the system could sample a specific age within a range, providing more concrete grounding for subsequent reasoning. When a study specifies population quotas, the system prioritizes them.

The \textbf{character snapshot} adds characteristic backgrounds beyond the demographic profile. Using an LLM, the system generates several first-person statements regarding participants' upbringing, environment, and interests.\footnote{The final profile contains approximately 100 tokens of rendered demographic information and 240 tokens of character information, with a median length of approximately 340 tokens.} The snapshot is intentionally \textit{study-agnostic} to prevent information leaking from the provided design that may bias the results. 

Finally, each generated profile goes through three validation checks. First, a \textbf{coherence} check to prevent implausible demographic combinations. Second, an \textbf{answer-leakage} check to restrict the persona from reproducing language from the study's verbatim response options. Third, a \textbf{redundancy} check to detect duplicate character snapshots.

\subsubsection{\textbf{Response Generator}}
The Response Generator orchestrates the compiled study through each participant and records their answers. Each question is rendered through a single LLM call, so that a reasoning trace can be attributed to a single answer, even though batching several questions into one call could have been cheaper.

Each response is accompanied by a \textbf{reasoning trace} that records how the participant read the question and arrived at their answer by reasoning through the options. To obtain the most faithful traces, \systemname{} simulates with the open model \texttt{GLM-5.2}, which exposes its internal reasoning directly. Reasoning precedes the answer through decoding instructions\footnote{The answering instruction forces participants to restrict their answer to the provided, even if they do not conform to their reasoning. This later helps highlight a potential problem in answer choices.}.

Each question has access to its preceding $k=3$ question--answer pairs to form a \textbf{lagged memory}. This allows participants to refer to earlier responses in situations that demand back-referencing. The hyperparameter is chosen to prevent the context window from growing quadratically with questionnaire length.

LLM participants run independently and concurrently\footnote{Each call is equipped with retries, with a 2{,}500-token output budget shared with the reasoning trace, along with a 300s wall-clock limit. Ten participants run concurrently. Across the study corpus, recorded reasoning traces have a median length of 430 words.}. Table~\ref{tab:simulation-example} provides a complete example of a question-level call and its generated response.

\subsubsection{\textbf{The Critic}}
The Critic reads the responses and reasoning traces produced by the Simulator and reports \textit{Structural Errors} question by question and \textit{Missed Opportunities} for the study as a whole. A question-level LLM call for Critic takes the research question, the study design, the demographic space, the error taxonomy, and each participant's answer and trace for that question as input.

A study design can be critiqued directly by evaluating its questions, either by the researcher, their peers, or a capable language model, and such a critique can reveal methodological problems, including leading wording, double-barreled items, poorly specified scales, or missing controls. However, it is difficult to assign credits to the individual factors that explain potential discrepancies in study data. To address this, researchers run costly human pilots or, increasingly, take help from LLM simulations. But both are plagued by the problem of the lack of distribution fidelity to the population estimate. Responses alone can only contain information about \emph{what} a participant selected, not \emph{why}\footnote{E.g, a solo contractor may select the lowest available company-size category (``1--200 employees'') even though none of the categories apply to their situation. Conversely, an unusual distribution of responses may point to an incorrect diagnosis: e.g., a sparse category may appear to reflect an underrepresented group when the mismatch is between the response options and participants' background.}. \textbf{The Critic therefore grounds each structural error in participant-level reasoning traces}, so that a reported failure is supported by evidence, available to review.

The Critic does not judge a question on its pre-training knowledge alone. It receives an \textbf{explicit error taxonomy} as an input to each question-level call. This curated error taxonomy (See Appendix, Table \ref{tab:taxonomy}) is grounded in relevant literature \citep{shadish2002quasi,henrich2010weirdest,poulton1982influential,clark1973language,tourangeau1988cognitive} and lists potential pitfalls of experiment and survey designs, such as sampling errors, confounds, measurement slips, and behavioral inaccuracies.\footnote{The taxonomy groups pitfalls by the \textit{nature} of the error instead of by \textit{when} in the research process it occurs, so that errors sharing a root cause cluster together.} The Critic tries to choose a category to assign to an error, but in the absence of an appropriate one, it could produce a new category.\footnote{We have identified a list of pitfalls; however, given the long-standing literature on the pitfalls of experiment design, we do not claim this list is exhaustive.} 

The Critic identifies \textit{Missed Opportunities} in a \textbf{two-stage synthesis}. During the question-level scan, it identifies potential discrepancies revealed by participants' reasoning traces. After a full pass, it performs a study-level join analysis that combines local flaws with study design and demographic details. It further consolidates issues with the same underlying factors and ignores proposals already surfaced as structural errors. 

\subsubsection{\textbf{The Refiner}}
The Refiner is a straightforward LLM call that takes only the critiques and missed opportunities deliberately included by the user, along with the research question, the study design, the demographic space, the error taxonomy, and each participant's answer and trace for that question as input. Refiner is additionally instructed to prioritize the comments provided by the user for each suggested change (if any), providing an additional instrument to the users to steer their refinement and eventual study designs.

\section{Evaluating \systemname{} against Baseline Conditions}
\label{sec:comparative}
Researchers often conduct pilots with real respondents to identify flaws in their study designs. In this section, we isolate the effect of simulated reasoning traces relative to information obtained from two baseline conditions: a human pilot and using a state-of-the-art LLM (without any simulations) to identify possible errors in one's study design. 


\subsection{\textbf{Study Details}}
We continue with Dumbledore’s study from Section \ref{sec:system}. In reality, it is an ongoing survey experiment developed by one of the authors as part of their research on public perceptions of AI-generated misinformation. We selected this study for both methodological and substantive reasons. Methodologically, it is considerably richer than a conventional survey containing repeated measures, matrix questions, a between-subjects manipulation, condition-gated instructions, randomized stimuli, and outcomes with known ground truth. Substantively, the study also falls within the research team's domain expertise in behavioral research on AI-mediated decision-making and human-AI interaction.

The study measures participants' perceived exposure to AI-generated content and misinformation, beliefs about how generative AI changes the production and detectability of misinformation, and the behaviors people use to verify or avoid sharing it. The study includes a two-arm experiment on uncertainty in AI content labeling. Respondents are asked to imagine a social-media platform that automatically labels content identified as AI-generated. In the treatment condition, they are additionally told that this detection system is fallible, i.e., AI-generated content can go unlabeled and human-created content can be labeled incorrectly. Overall, the survey has 42 questions; however, for simplicity, they result in 112 questions when linearized from their complex structures (e.g., matrix questions). All numbers are reported with the linearized study version unless stated otherwise. The complete study is provided in the supplementary materials. 


\subsection{\textbf{Baseline Conditions}}
When applied, \systemname{} (referred to as ``System'' condition in the section) produced 82 structural errors and 20 missed opportunities for the study. We now briefly describe our two baseline conditions:

\begin{enumerate}
    \item \textbf{Human}. To mimic the traditional practice of human pilots, we recruited 3 participants for the Control group and 3 for the Treatment group in Prolific. Participants received a flat fee of \$25 to complete the study. Participants answer the survey question. Participants are additionally instructed to critique each question, in line with \systemname{} critiques of structural errors (for complete instructions, see Appendix: \ref{appendix:prolific-ins}). Mainly, two participants from each control and treatment group provided about 73 comments (which can be linearized to produce 103 question-level critiques). Overall, we spent \$200 in this pilot, which is cost-equivalent to running several iterations with \systemname{}. When asked, prolific participants did not provide any additional comments on improving the survey; hence, we consider them to have no contribution in providing missed opportunities.
    \item \textbf{NoSim}. Here, we mimic how a researcher could use a frontier chat-based AI model to identify flaws in their study. We used Anthropic's Claude-opus-4.6 to generate these critiques.\footnote{It is increasingly relevant in recent times when LLMs have emerged as research tools \citep{liao2024llms}; $P3$ in our user study claims:`\textit{`if I give this list of 16 questions and ask Claude to like, hey, can you just improve whatever you want and tell me like, how can this be, it just will do it like with one prompt.''}!} By looking at the study as a whole and question-by-question, this baseline system produced 34 unique structural errors and 10 missed opportunities.
\end{enumerate}

\begin{figure*}[t!]
    \centering
    \includegraphics[width=0.9\textwidth,trim=0 15 0 15,clip]{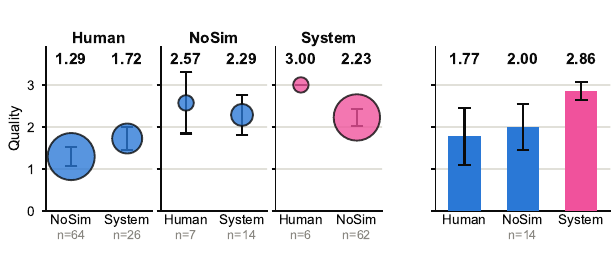}
    \caption{\textbf{Comparative Analysis.} \textbf{Unique Value Addition (Left):} Mean critique-quality (0–3) for each condition's own critiques for questions that a given comparison condition does not cover. Bubble area proportionally denotes the number of such uniquely covered questions (n given); error bars are 95\% CIs. \textbf{Head-to-head Comparison (Right):} Mean critique-quality for three conditions on questions (n=14) where all three conditions independently produced a critique. Pairwise Wilcoxon signed-rank, Bonferroni-corrected indicates: System > Human, $p=0.013$; System > NoSim, $p=0.029$; NoSim vs. Human, $p=1.00$). System condition adds the most value, producing unique high-quality suggestions, as well as when critiquing the same questions.}
    \label{fig:comparative}
    \Description{A figure in two parts that share a vertical axis labelled Quality, running from 0 to 3. The left part has three small panels, one per condition. Each plots the mean quality of that condition's own critiques on the questions a comparison condition left uncovered, as a bubble whose area is proportional to the number of such questions, with 95 percent confidence intervals. In the Human panel, the mean is 1.29 on the 64 questions NoSim does not cover, and 1.72 on the 26 that the System does not cover. In the NoSim panel, it is 2.57 on 7 questions not covered by Human, and 2.29 on 14 not covered by the System. In the System panel, it is 3.00 on 6 questions not covered by Human, and 2.23 on 62 not covered by NoSim, and its bubbles are drawn in pink instead of blue. The right part is a bar chart of the mean quality on the 14 questions, all three conditions critiqued. Human reaches 1.77, NoSim 2.00, and the System 2.86. The System bar is pink and clearly the tallest, and its error bar is narrow, while those for Human and NoSim are wide.}
\end{figure*}

\subsection{\textbf{Results}}
Four authors, excluding the study-preparing author, annotated the critiques from the three conditions. To ensure the quality of our ratings, the four authors developed the codebook over four rounds of rating, comparison, and discussion. They first labeled a small batch together in a meeting to draft an initial codebook, then went through three more rounds of independent rating, discussing disagreements and revising the codebook. Agreement was still low after the second round; by the third and fourth, weighted Cohen's kappa rose to 0.48, then 0.77 for structural-error critiques, and 0.71, then 0.85 for missed-opportunity critiques. Once an agreement was reached at that level, each annotator rated a distinct subset of the remaining critiques, without redundancy. 

We use a five-point rating scale--0, 1, 2, 3, N/A: 3 indicates the highest quality suggestion, ignoring which can bring significant compromise in the study results; 2 indicates a valid critique, but also would be a valid trade-off; 1 indicates immaterial suggestion, often covering minor nitpicking comments; 0 indicates the suggestion is plain wrong, contains personal opinions, and is out of scope; we reserve "N/A" for cases where critiques do not address the question or are malformed for technical reasons; these are also dropped for our statistical analyses.


\subsubsection{\textbf{\systemname{} adds most value, producing unique high-quality suggestions.}}
Both NoSim (mean rating 2.31) and \systemname{} (mean rating 2.44) generate high-quality critiques (System vs. NoSim; $p=1.00$) that were rated significantly more important and valid than human critiques (mean rating 1.53; vs. System: $p=2.7×10^{-12}$; vs NoSim: $p=4.3×10^{-5}$) collected from Prolific. However, using ratings 2 and 3 as a high-quality threshold (i.e., consequently moderate to severe), \textbf{\systemname{} was able to surface 49 additional errors missed by the baseline that did not use simulation}. And no simulation only recovered 11 additional errors missed by the system condition. \Cref{fig:comparative} (Left) highlights that the system either adds a few novel but very high-quality critiques (all rated 3) as compared to Human or several new high-quality ones as compared to a no-simulation baseline. Despite high coverage, the overall quality of novel critiques is considered subpar when coming from Prolific participants (real pilot), highlighting the value of our system in improving behavioral experiment designs.

\subsubsection{\textbf{When critiquing common questions, \systemname{} produces higher quality suggestions.}}
We further looked at the critics for questions where every condition raised an objection. For these cases, we compared their quality head-to-head, using a pairwise Wilcoxon signed-rank test, showing that the \textbf{system condition significantly outperforms Human and NoSim baselines} in terms of quality of the critiques, whereas there is no material difference in quality between Human and NoSim baselines (System > Human, $p=0.013$; System > NoSim, $p=0.029$; NoSim vs. Human, $p=1.00$). 

\subsubsection{\textbf{Little difference in Missed Opportunities across conditions}} In general, missed opportunities from \systemname{} are of higher quality (mean rating 2.76) than those generated from no simulation baseline (mean quality 2.46); however, the difference is not statistically significant ($p=0.167$; Wald test on the fixed effect). In absolute terms, \systemname{} produces 12 more highest-rated (3) missed opportunities as compared to the NoSim baseline.

\section{User Study: \systemname{} in Research Workflows}
We reach out to 12 real-world Dumbledores; academicians and industry professionals who conduct behavioral human-subject studies (controlled experiments, surveys, formative interviews, and others) daily. We enable participants to navigate the Critic and Refiner to update study designs of their original work. We aim to evaluate:
\begin{itemize}
    \item \textbf{[RQ1]} Is \systemname{} useful to professionals? What shapes their perceived usefulness?
    \item \textbf{[RQ2]} How do professionals evaluate the credibility of reasoning traces from simulations in \systemname{}? What shapes participants' trust and interaction with the system?
    \item \textbf{[RQ3]} How will participants integrate \systemname{} into their regular workflows? 
   \item \textbf{[RQ4]} Finally, what are the limitations of \systemname{}?
\end{itemize}

\subsection{Participants} All 12 participants self-report proficiency in designing behavioral studies (10 women, 2 men; 8 years of research experience, on average) and have prior experience using generative AI tools (most commonly ChatGPT/Claude). Participants ($P1$ - $P12$) bring study designs from their own research workflows, spanning surveys, interviews, formative research, and controlled experiments across populations in the United States, India, Ethiopia, and Tanzania. These include both ongoing studies and studies participants have previously conducted or adapted from prior work. Table~\ref{tab:participant-demographics} summarizes participant backgrounds and the population/status of their research study.\footnote{To protect participants’ confidentiality and intellectual property of private, ongoing, or unpublished work, we do not report any specific details about the participants' research questions or objectives in the paper or supplementary materials.} All participants complete one round of simulator-critic-refinement in a moderated session.\footnote{$P12$ engages in a brief 15 min session with the study moderator; the moderator briefs them about \systemname{} and the participant interface; $P12$ completes the remaining interaction with \systemname{} unmoderated and self-paced. Any quotes from $P12$ are administered from their responses to a post-study survey questionnaire, or their comments to the Critic/Refiner.} 

\subsection{Study Tasks, Procedures, Engagement}
The Simulator inputs the participants' study plans and generates $50$ LLM participants from the study's target demographic space. LLM participants answer the study questions, producing their responses and reasoning traces; the Critic uses these reasoning traces to identify structural errors and missed opportunities. To evaluate \systemname{} performance, we ask participants to report their \emph{confidence} in the critique in a 7-point Likert scale ranging from ``strongly disagree'' to ``strongly agree''. Participants must rate all the high- and medium-severity critiques before they can run the Refiner. We allow participants to iteratively review the critiques and run the Refiner in real time up to three times before they export to a final updated design. This concludes round-1. Five of the 12 participants continue to iteratively improve their final updated design in round-2 of simulation-critic-refinement. The moderated session ends when a participant completes round 1. Round-2 is self-paced and unmoderated.\footnote{Participants were recruited through Upwork and personal contacts. Participants provide their study designs to the moderators (authors). Moderators use \systemname{} to run the Simulator and prepare the interface for the participants to review the Critic and Refiner. Moderated sessions are conducted over Zoom/Google Meet. Sessions are recorded and transcribed (with prior consent). The moderator guides the participants through \systemname{} and the interface. The participants interact with \systemname{} in a think-aloud fashion. The moderated sessions spanned from an hour to two and a half, depending on the study length. This study was approved by the Institutional Review Board (IRB).}

\subsection{Findings}

\begin{figure*}[t!]
    \centering
    \includegraphics[width=\textwidth]{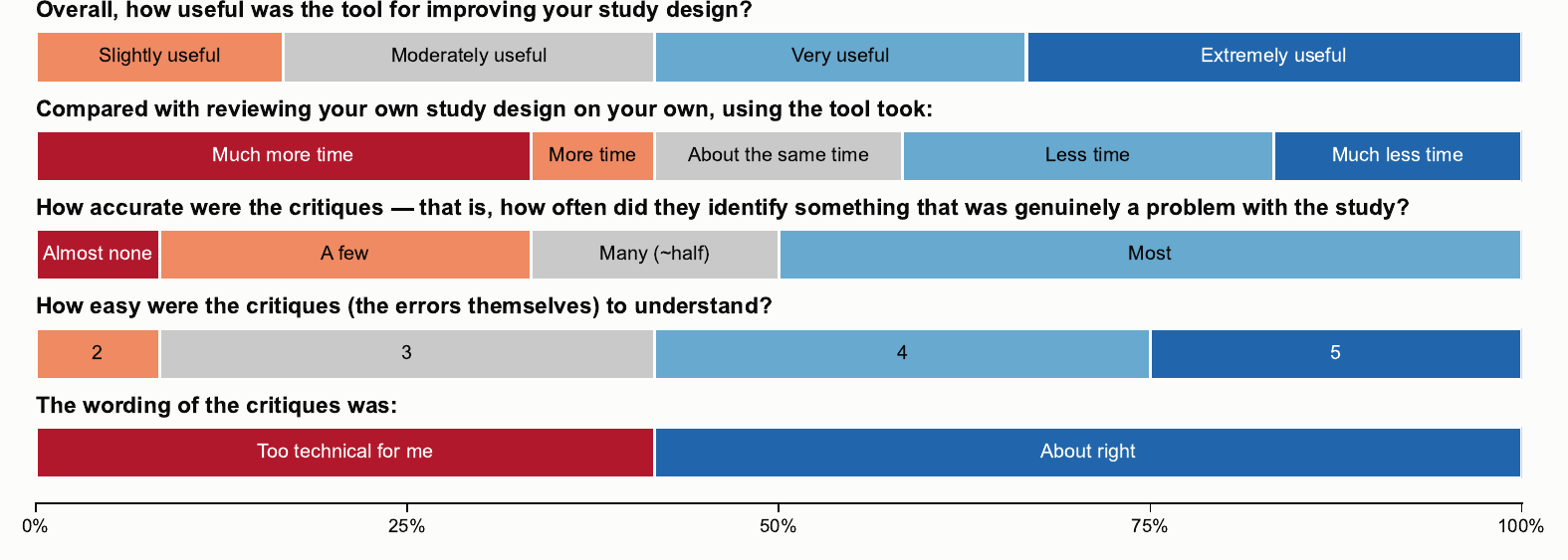}
    \caption{\textbf{Participants generally find \systemname{} useful, but its efficiency and critique quality vary.} Nine of 11 participants rate the system at least moderately useful for improving their study design, including six who rate it very or extremely useful. Perceived time savings are mixed: four participants report spending less time than when reviewing a study independently, five report spending more, and two report spending approximately the same amount of time. Seven participants report that many or most critiques identify genuine problems, while the wording of critiques remains a source of friction between academic researchers and industry researchers.}
    \label{fig:participantstudy-quality}
    \Description{Five horizontal stacked bars, one per survey question, each spanning the full width from 0 to 100 percent of the 11 participants. Segments run from red through grey to blue as responses move from negative to positive. The first bar, on how useful the tool was for improving the study design, is split between Slightly useful, Moderately useful, Very useful, and Extremely useful, with no lower ratings. The second, on time taken compared with reviewing the design alone, runs from Much more time through More time, About the same time, and Less time to Much less time, the two ``more time'' segments together taking a little under half the bar. The third, on how accurate the critiques were, is split between Almost none, A few, Many (about half), and Most, where Most takes roughly half the bar. The fourth, on how easy the critiques were to understand, uses a five-point scale whose segments are labeled 2, 3, 4, and 5, with no responses at 1. The last, on how the wording of the critiques read, gives roughly two-fifths of the bar to ``Too technical for me'' and the rest to ``About right''.}
\end{figure*}

Participants were actively engaged with \systemname{} during the study, and most find it very to extremely useful (Figure \ref{fig:participantstudy-quality}). Based on a thematic analysis, we identify four core themes of how professionals/researchers evaluate and appropriate \systemname{}: its utility in the research workflow, the usefulness and limits of its critiques, the credibility of reasoning traces as simulated evidence, and the distinct roles of structural errors and missed opportunities.
One representative example was $P6$, who studies women's access to digital finance through large-scale surveys in Ethiopia: \systemname{} identified a mismatch between the survey's assumed household structure and the population being studied. The question framed phone access around a wife–husband pair, yet many target participants were widowed, never married, or female household heads. Questions asking whether ``the wife'' could use her husband's phone, or had ever been denied access to it. This becomes inapplicable for participants without a husband and fails to capture the forms of phone access relevant to their circumstances. Through missed opportunities, \systemname{} proposed separating personal phone ownership from borrowed access and recording the relationship of the person controlling that access when the gatekeeper was a brother, son, or nephew rather than a husband. $P6$ strongly agreed with all four findings and carried each into refinement.

\subsubsection{\textbf{Costs and benefits comparing to current practices.}}
In general, participants said \systemname{} would be useful in their research workflows as they iterate on behavioral experiments, interviews, and surveys. Across different types of studies, they expressed that \systemname{} would help them iterate on their designs much faster, when comparing to alternatives such as working on their own, seeking help from peers, or using standard LLM chat tools \textbf{[RQ1]}.
$P6$ conducts field surveys in Ethiopia and discussed in great detail how identifying ambiguities before training enumerators and deploying the study can \textit{``save me days.''}. When asked about the tools they currently use to improve study designs, participants ($P3$, $P7$) pointed to general-purpose LLMs and embedded assistants in platforms such as Qualtrics, which primarily identify surface-level or more trivial structural issues, such as excessive use of matrix questions that may increase the cognitive load on respondents.
Even though our position is to supplement human pilot studies, some participants naturally compared the two. For example, $P10$ who said  \textit{``I think for our pilot, this would be very, very helpful.''}, then mentioned how in her past pilots respondents rarely provide critical information when researchers ask whether they understand the study questions, consistent with our experience in \ref{sec:comparative}. Even with direct instructions and monetary incentives, only two of six prolific participants gave any critical feedback on our study questions.

\subsubsection{\textbf{Improving study designs beyond finding errors.}}
When comparing \systemname{} to tools that participants currently use to help improve study designs, one key distinction they saw was how it
\emph{provokes} them to rethink their broader research questions [RQ1]. 
Even though our interface drew a few UI metaphors from debuggers (i.g., flagging items for errors, presenting long explanations, and proposing solutions),  participants saw \systemname{} less as an automated debugger for their study designs, but more as an \emph{ideation partner}. 
For examples, $P5$, $P9$ describe \systemname{} as a way to \textit{``provoke more thoughts and ideas''} about how to frame a study; $P8$ finds \systemname{} useful to \textit{``expanding the horizon of the study"}; $P6$, $P2$ appreciate the missed opportunities, specifically, as they have genuinely encountered similar, if not exact problems, when they revisited their onoing study either in the field or while looking at the data. Had they used \systemname{} earlier, they would have been better prepared for the field study and collected better data \textbf{[RQ3]}.



\begin{figure*}[t!]
    \centering
    \includegraphics[width=\textwidth,trim=10 200 10 190,clip]{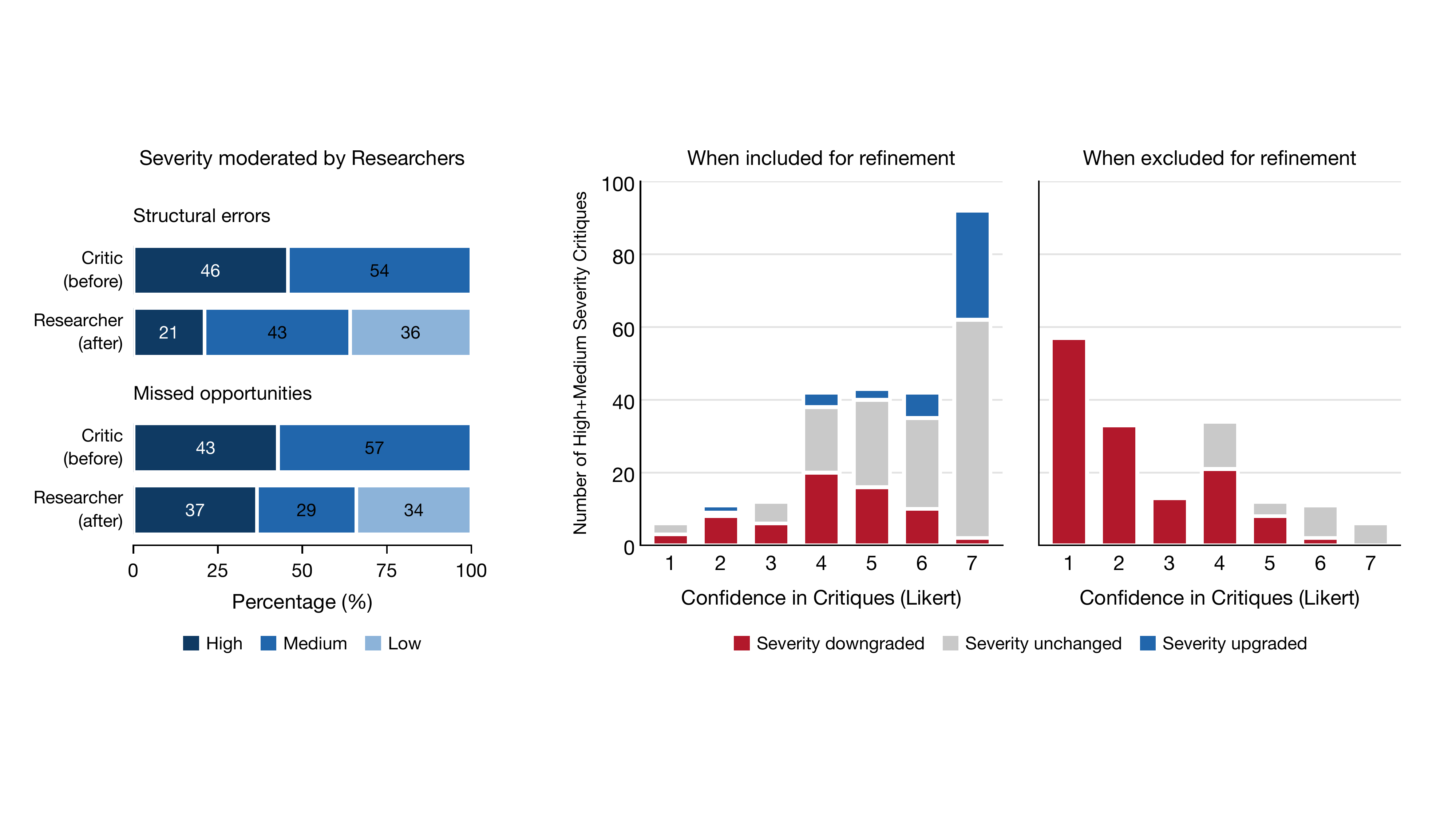}
    \caption{Participants frequently revise the critic's severity ratings. Higher-confidence critiques are more likely to survive into the refined design. \textbf{Left:} Percentage of structural errors and missed opportunities the Critic rated high, medium, or low severity [Critic (before)] versus the researcher's own severity assessment [Researcher (after)]. \textbf{Right:} Participants are highly confident of the critiques that they include in the refinement, and upgrade their severity. In contrast, participants strongly disagree with the critiques they chose not to include in the refinement.}
    \label{fig:participantstudy-confidence}
    \Description{A figure in two parts. On the left, under the heading ``Severity moderated by Researchers'', four horizontal 100 percent stacked bars compare the Critic's severity ratings before triage with the researchers' ratings after, separately for structural errors and for missed opportunities. Dark blue is High, medium blue is Medium, and light blue is Low. For structural errors the Critic's ratings are 46 percent High and 54 percent Medium with no Low, while the researchers' are 21 percent High, 43 percent Medium, and 36 percent Low. For missed opportunities, the Critic's are 43 percent High and 57 percent Medium, while the researchers' are 37 percent High, 29 percent Medium, and 34 percent Low. On the right, two stacked bar charts count high- and medium-severity critiques on the vertical axis against the researcher's seven-point confidence rating on the horizontal axis. Red is severity downgraded, grey unchanged, and blue upgraded. In the chart headed ``When included for refinement'', the bars rise steeply towards the high-confidence end and peak above 90 critiques at a confidence of 7, where almost all are grey or blue and almost none red. The chart headed ``When excluded for refinement'' reverses that pattern. Its tallest bars sit at confidence 1 and 2 and are almost entirely red, and the bars shrink to near zero by confidence 7.}
\end{figure*}

\subsubsection{\textbf{\systemname{} supports participants in actively steering refinements.}}
Participants actively iterate with the Critic and Refiner, rather than merely accepting system suggestions or over relying on them. Across all 12 participants, \systemname{} produced 521 critiques in total, ranging from 21 to 63 per participant, of which they adjudicated 440 critique cards (84\%) and completed 22 refinements (one to three each), with 9 of 12 running more than one round.
They calibrated the severity level on 57\% of the cards they rated, demoting the Critic's severity four times as often as promoting it (199 vs.\ 51),, and carried 58\% into a revision, and left 105 free-text comments ($M=8.8, SD=10.1$ per participant; range$=0-28$) totaling roughly 1,900 words ($M=8.8, SD=10.1$, range$=2-141$) 
Participants also frequently disagreed with the system, often based on personal preferences, constraints, or deliberate trade-offs between question/answer-option details and the effort required to answer them. Nine participants reported being comfortable or very comfortable disagreeing with its critiques, and used the commenting feature to provide such additional context to further steer the refinement step.
$P4$ commented on the goal of a seemingly odd item: \textit{``This is a fun, informal survey aimed to entertain the respondent and keep them going through a long questionnaire using humor as an instrument!"}. In another instance, when the critique raises a screening gap for the study, $P4$ steers the system to inform: \textit{``The screening only recruited participants who are existing [brand name removed] shoppers."}. 

\Cref{fig:participantstudy-confidence} (Left) offers another view of the active calibration process by showing how stated severity levels are downgraded across the board. Participants' actions are further justified by the right plot in \Cref{fig:participantstudy-confidence} where their critique-inclusion activity directly correlates with their expressed confidence in the critique. In fact, despite the overall downgrading trend, the critiques that participants considered for inclusion and were highly confident about, they upgraded their perceived severity, demonstrating active refining intent.



\subsubsection{\textbf{\systemname{} reproduced findings from independent human-pilots and on-going human studies.}}
A few participants have either already conducted a few pilot studies with human participants ($P2$, $P9$, $P10$) or are in the middle of running the study they brought to \systemname{} ($P6$). In these cases, the system was able to reproduce some of the issues our participants had already discovered independently, and proposed similar solutions they had planned to implement \textbf{[RQ2]}. 

In P6's study about women's access to digital finances
, \systemname{} caught a minor error where their human participants who reported owning no phone were later asked whether ``that phone'' had an active SIM, forcing respondents for whom the question did not apply into the ``No'' category. The reasoning traces surface this tension:  p029 (age 56, rural central Ethiopia) reasons:
\textit{``None of these options really fits because the question presupposes I have a phone when I already said I don't. The closest option would be `No' since without a phone, there's no SIM card.''}
$P6$ immediately recognized the problem because they had identified and fixed similar issues recently when actively conducting field work in one part of Ethiopia, while the team was preparing to revise the instrument for other regions. She observes: \textit{"It's very interesting, the things it flagged, and also the missed opportunities. Because, as I said, as we're going through this [...] research, again, for other parts of the country. We're refining it, and I see [...] the similarities in how we're refining it. It's giving me the same things. The same questions we missed and everything."}

$P2$ reported a similar convergence with evidence from interviews he had already conducted. His initial reaction,\textit{``wait, do you have my data''} reflects when several missed opportunities match the themes that had emerged in his formative interviews. He mentions: \textit{``these missed opportunity questions capture things I've captured from the actual interview. It gives a nice window into what I might see from actually chatting with people later.''} \textbf{[RQ2]}.

\subsubsection{\textbf{Reasoning traces act as diagnostic evidence to rely on critiques.}} Participants find genuine value when looking at direct evidences from simulations \textbf{[RQ2]}. These evidences include those quotes from the actual reasoning traces of the simulated participant, which helped the Critic surface an error or suggest a gap. $P8$ stated: \textit{``It was useful because it made me realize what is the thinking pattern that could be there. And it genuinely helped me understand where I was going wrong.''}. She was surveying individuals who use agent builders. For a question, she looks at some participant traces\footnote{Traces are: ``For low-stakes stuff like internal data pulling, maybe I'd be okay with it. But the question is about comfort level generally.''(p009);``I'm comfortable with autonomous action for low-risk, reversible tasks but not for everything… I keep human checkpoints for customer-facing or money-moving actions.'' (p013); ``there's a spectrum depending on what kind of action we're talking about'' (p023)} and explains: \textit{``it gives me different perspectives, which I had not thought about while creating the plan… because it tells me, okay, how non-builders are going to answer it. And then it tells me, like, how different set of participants would answer it. So that is quite useful.''}

In her study, $P7$ looks at how travelers use AI systems to plan an adventure trip. Reviewing a critique and the evidences from simulations on one of her questions (``\textit{When using AI to plan adventure travel, how often would you independently verify its recommendations before acting on them?}'' with answer options ranging from \textit{Never} to \textit{Always}) and the evidences from simulations, she enthusiastically explains: ``\emph{Many participants even acknowledge that always is an overstatement, but pick it anyway on identity and values grounds, usually might be more honest! Oh, this is really interesting. \textbf{Even though they're made up, this is, like, very good points.}}''. 

\subsubsection{\textbf{Missed opportunities outshine structural errors.}}
Participants treated structural errors and missed opportunities differently, and most value missed opportunities more. In particular, participants were happily surprised by some of the missed opportunity critiques. 
$P6$ reacted \textit{``I never thought of this. This is nice,"} to one of the missed opportunity critiques, which asks women in Ethiopian households: ``What is your household's main source of income?'' She explains why respondents may never answer this truthfully and \textit{``so this question will never be asked''}. $P6$ rates this critique to be highly severe, adds it to refinement and comments: ``frame the question in a more indirect way.''
$P5$ was surprised by their own blind spot: \textit{``How did I miss that question? Wow.''}. She studies user behavior when making returns at a grocery store, but misses asking, "How did you pay for the item you returned?" (Target RedCard/Credit or debit card/Target gift card or store credit/Other/I don't remember)'' -- a missed opportunity \systemname{} aptly identifies!
$P7$, in particular, made direct comparison between the two types of critiques and argued that missed opportunities deserved greater emphasis in the system. She also proposed a different workflow, where she would have rather looked at only the missed opportunity critiques in Round 1, broadened the scope of her study with the new additional question, then run the Simulator on this refined study to iterate over the structural errors across all questions, and drop unnecessary additions.

While their confidence across the two was nearly identical, researchers carried 57.5\% of missed opportunities into refinement, compared with 48.5\% of structural errors. This difference is more pronounced in how researchers reassessed severity. The Critic initially labeled 46\% of structural errors as high severity and 54\% as medium; after researcher review, only 21\% remained high severity, 43\% were medium, and 36\% were downgraded to low. In contrast, the severity distribution for missed opportunities remained more stable: 43\% were initially rated high and 57\% medium, compared with 37\% high, 29\% medium, and 34\% low after researcher review. Missed opportunities are deemed more severe than \systemname{} rates them, while structural errors are not as severe as the system claims.



\begin{figure*}[t!]
    \centering
    \includegraphics[width=\textwidth]{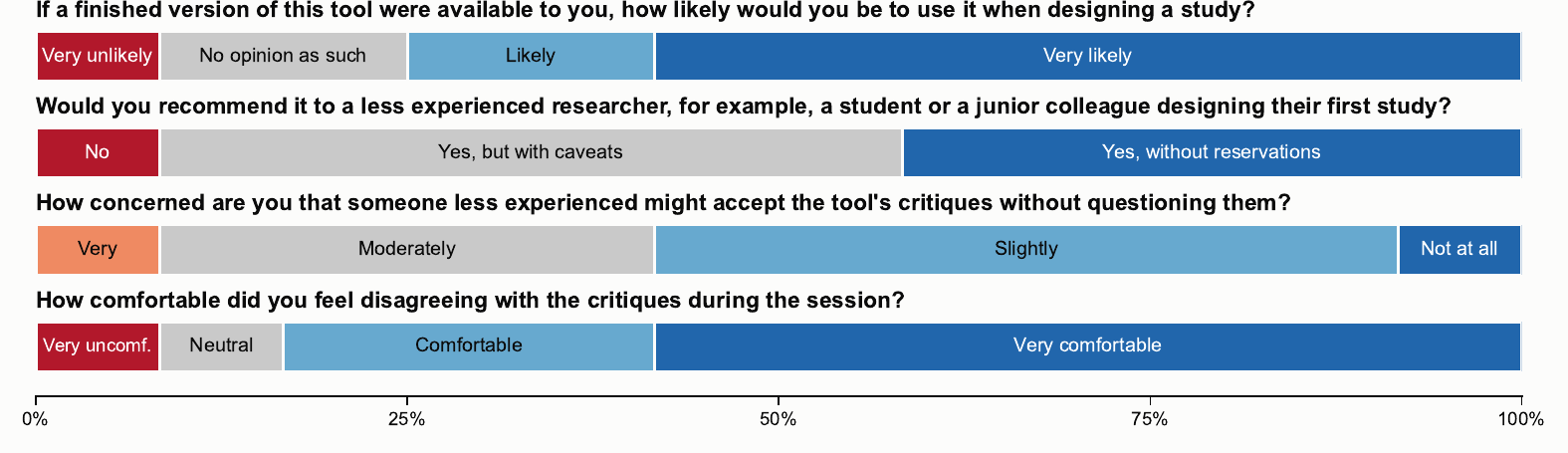}
    \caption{\textbf{Participants report high intended adoption while retaining substantial concern about uncritical use.} Eight of 11 participants report that they are likely or very likely to use a finished version of \systemname{}, and 10 would recommend it to a less experienced researcher, although six do so only with caveats. At the same time, 10 participants expressed at least some concern that less experienced researchers may accept the system's critiques without sufficient scrutiny. Participants themselves generally feel comfortable challenging the system: nine report being comfortable or very comfortable disagreeing with its critiques.}
    \label{fig:participantstudy-adoption}
    \Description{Four horizontal stacked bars, one per survey question, each spanning the full width from 0 to 100 percent of the 11 participants, with segments running from red through grey to blue. The first bar, on how likely participants would be to use a finished version of the tool when designing a study, is split between Very unlikely, No opinion as such, Likely, and Very likely, where Very likely takes about half the bar. The second, on whether they would recommend it to a less experienced researcher, is split between No, ``Yes, but with caveats'', and ``Yes, without reservations'', and the caveated segment is the largest. The third, on how concerned they are that someone less experienced might accept the critiques without questioning them, runs from Very through Moderately and Slightly to a small ``Not at all'' segment at the right end. The last, on how comfortable they felt disagreeing with the critiques during the session, runs from Very uncomfortable through Neutral and Comfortable to Very comfortable, which takes about half the bar.}
\end{figure*}



\subsubsection{\textbf{Critiques were less receptive for late-stage or completed studies.}}
We observed diverging preferences amongst our participants when it comes to how much methodological scrutiny they wanted from \systemname{}, based on where the study sat in its lifecycle. Researchers working on ongoing studies tended to treat even small ambiguities as actionable. $P6$, for example, explicitly valued that the system was ``nitpicky,'' noting that minor wording problems can become costly during fieldwork and data cleaning; $P9$ similarly revised a seemingly small singular–plural ambiguity because it changed the population to which the item referred. By contrast, participants bringing previously conducted or closely related prior studies were more conservative, treating some critiques as minor, out of scope, or inconsistent with design choices they had already settled ($P4$, $P5$)
Participants with completed studies more often weigh the value of a proposed addition against the cost of making the instrument longer. $P1$ notes that constraints on survey length and response bias \textit{``systematically caused me to reject recommendations,''} even when the recommendations themselves are useful. $P11$ makes the tradeoff explicit:
\textit{``I like this one also. But again, I'm not going to cross the line into survey fatigue, so I would have to time it, and if I felt this question was such a top priority over any other questions that I have in here, then [...] I would swap it, but I can't guarantee I would.''}

\subsubsection{\textbf{Design iteration and second rounds of Simulation-Critic-Refiner}}
\begin{figure*}[t!]
    \centering
    \includegraphics[width=\textwidth]{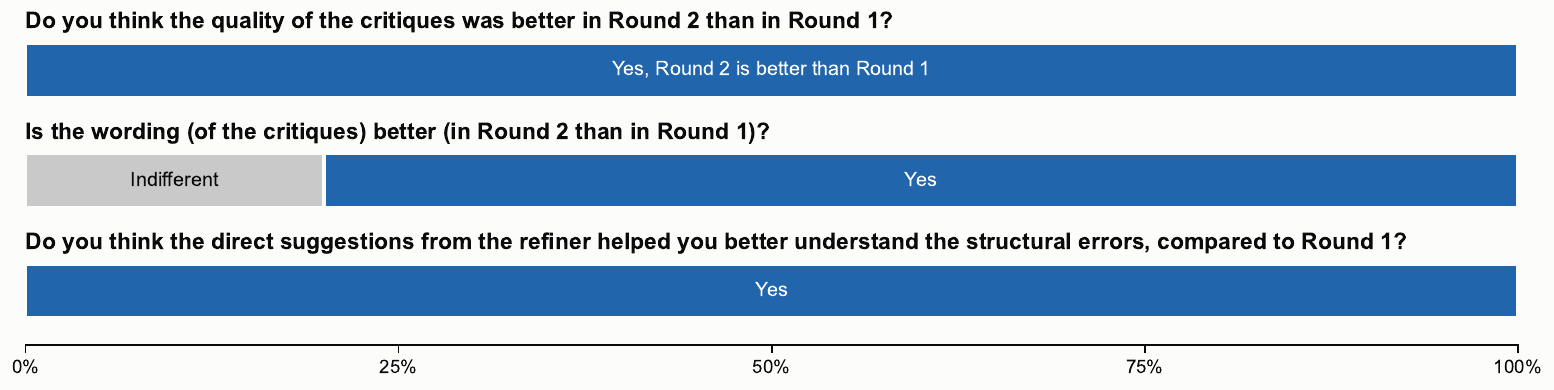}
    \caption{\textbf{Participants who complete Round 2 prefer the revised presentation of the Critic output.} All five Round 2 Participants judge the critiques to be better than in Round 1, and all five report that presenting direct Refiner suggestions helps them better understand Structural Errors. Four of five also prefer the updated length of the critique description, while one is indifferent.}
    \label{fig:participantstudy-round2}
    \Description{Three horizontal stacked bars, one per survey question, each spanning the full width from 0 to 100 percent of the five Round 2 participants. The first asks whether the quality of the critiques was better in Round 2 than in Round 1. Its bar is a single blue segment reading ``Yes, Round 2 is better than Round 1''. The second asks whether the wording of the critiques is better, and there a grey ``Indifferent'' segment covers about a fifth of the bar, with a blue ``Yes'' filling the rest. The third question asks whether the direct suggestions from the Refiner helped participants better understand the structural errors, compared with Round 1. That bar is again entirely a blue ``Yes''.}
\end{figure*}

Five participants returned with their updated study designs (produced after their Round 1) and ran them again through \systemname{}. \textbf{All five judged the Round 2 critiques to be better than those in Round 1} (Fig.~\ref{fig:participantstudy-round2}); all five also reported that showing the Refiner's proposed changes to structural errors helped them better understand the critiques, and four of five also preferred the revised wording.  In Round 2, we made additional UI improvements based on participants' feedback from Round 1. 
Most importantly, instead of asking the participants to triage a list of critiques, we now also pre-run the refiner on all questions and show suggested fixes alongside the critiques, making them easier to make sense of. Participants can still use the same interactions to comment on each critique to apply a more targeted refinement.
Participants' comments on a post-study survey help explain this preference. $P2$ notes: \textit{``being able to see what the changed question might look like made it easier to understand the suggested change"} than reading the critique alone. Thus, the Round 2 did more than shorten critique descriptions: \textit{it made the path from diagnosis to revision more concrete.} Repeated use also exposed a tension in the iterative workflow. Participants valued being able to refine a design and run it again, but did not necessarily want refinement to continue indefinitely. $P8$ describes the process as currently seeming \textit{``unending,''}; $P2$ cautioned that continuing to iterate on a study design could lead to adopting many valid missed opportunities and suggested that the system could also help researchers prioritize which questions to include given a fix study length budget; shedding light on the tradeoff between increasing cognitive load for the respondents and the information a researcher might want to squeeze out of a study.

\subsubsection{\textbf{Where did the system fall short?}} It has been encouraging to see that almost all our participants have found this tool useful, and 9 of our 12 participants report that they are likely-to-very likely to use \systemname{} as a part of their research workflow if a finished version of this tool were made available to them (see Figure \ref{fig:participantstudy-adoption}). $P5$, who is a UX Design Program Manager and has 8-15 years of industry experience, exclaimed: \textit{``I think these big core organizations, if you pitch them, they would be… they would love this, because they spend millions of dollars in this.''}. 

While we are stoked about this early reception and the possibility of future success of \systemname{}, there are some obvious limitations we need to address to increase the utility \textbf{[RQ4]}. \systemname{} fails to input or compile matrix questions as it is, and linearizes them. We anticipated that linearizing matrix questions would work fine. But two fundamental errors arise. (1) Participants on multiple occasions requested to refine a set of questions into a matrix question. The refiner failed to address the comment, as all matrix questions are coded to linearize in the current system design. (2) A 12 × 5 matrix question with 12 statements and a 5-point Likert scale becomes a set of 12 sequential questions when linearized, each with the same 5 response options. Suppose the Critic identifies that having a 7-point Likert scale will be more useful. The Critic, which looks at each study instrument independently to identify structural errors, will not raise this as an error 12 times for each of the sequential questions, which could have been addressed by just one, if the matrix questions were not linearized.

In its current form, \systemname{} is not conversational. It uses participants' comments as an anchor to refine the questions they included in the refinement. Hence, unless a critique is included, \systemname{} fails to take the comments into consideration. During participant interactions, we observed that they often provide more context of the study design and the point where the critiques went wrong through their comment. Following the earlier limitation, participants often complained that \systemname{} lacked the complete context of the intent of their research study. This raises a fundamental calibration issue: how much context is enough, and how a system gathers the vast corpus of background information that a researcher may have amassed over their years of experience. In $P9$'s own words: \textit{``The information or the context has to be fed in, either to the researcher or to the system, to be able to infer and understand the design.''}. $P6$ also raises similar concerns regarding enumerators; she has to spend a lot of time providing sufficient context about the true intent of her study design before they can successfully conduct the surveys/interviews. 

Trust in the scientific accuracy of such a system as \systemname{} seems highly correlated with the evidence from simulations (for us, reasoning traces). Participants complain of outdated reasoning traces--"using a very 1980s archetype of an Indian population” $P4$, which may raise the correct critique, but does not provide the correct evidence. While we argue that \systemname{} takes a different approach by not relying on simulation responses, optimizing simulation quality is nearly necessary to establish trust in evidence that identifies a critique. Finally, while we have not yet optimized the \systemname{} UI for potential users, we realize UI interactions and decisions pave the way for participants to use such systems efficiently. Evidently, we find that Round 2, with a better UI design, is more appreciated. For efficient adoption of similar systems, reducing the cognitive load remains pertinent. 

\section{Discussion}

\subsection{From Simulate to Replace to Simulate to Improve}
Most prior work in social simulation assumes a perfectly designed study and focuses on automating behavioral experiments by accurately replicating human responses with language models \citep{10.1145/3442188.3445922}. In contrast, our work assumes a draft study design and uses simulations not for its responses, but its reasoning traces to understand how the design may be misinterpreted or interpreted differently by participants who come from different backgrounds than the researchers designing the study. 

The premise of \systemname{} is our belief that one can explore simulations to understand errors in study design, analogous to a pilot study, and thus improve the data one collects in real human subject studies. We uncover the diversity of information arising from reasoning traces during LLM simulation that overtakes the surface-level problem of response discrepancies. Like a think-aloud study, a single call to produce an LLM answer can provide a detailed account of what a participant assumes, how they construe the question, and where the available response options may constrain their thinking.

Findings from our user study suggest that the two error categories serve different purposes. Structural errors mainly help researchers inspect questions they have already explored, whereas missed opportunities more often introduce questions or distinctions one failed to consider. This is concomitant with the fact that LLMs are now capable of generating novel scientific hypotheses or theories from data and literature \citep{Agarwal2025AutoDiscoveryOS,Jansen2026GeneratingLS}. Several instances shed light on how missed opportunities highlighted by our system matched with issues that emerged much later through formative interviews, additional fieldwork, or causal data analysis. Naturally, researchers were more likely to carry missed opportunities into refinement, suggesting that \systemname{}-like systems could be useful for broadening a research question.



\subsection{When the Simulated Population is not WEIRD}
\label{sec:disc-fidelity}

Our premise sidesteps the debate over whether simulated responses match human ones, and it leaves two narrower questions. The first is whether a reasoning trace faithfully explains the answer the model gave, which prior work shows it often does not \citep{chen2025reasoning,lanham2023measuring}. 
It is unclear whether this is an issue, since traces for us only need to cover different possible ways of interpreting questions in a study. The second question is more nuanced: whether the simulated population resembles the one the researcher studies, and this is where our participants pressed hardest.

Looking at the reasoning traces, participants do not always agree about how closely the simulated participants resembled the people they study, and the interview hints that the difference was driven by the different population they aim to study. Researchers working with U.S. samples generally found the traces plausible, and $P8$ described them as central to the value of \systemname{} because they reveal \textit{``what is the thinking pattern that could be there [...] where I was going wrong.''} Researchers working with participants in India and Ethiopia were more cautious. $P4$ described the Indian simulation as \textit{``using a very 1980s archetype of an Indian population,''} and $P6$ found assumptions that did not hold for her Ethiopian field site, citing \textbf{``CONTEXTS''} in the exit questionnaire as the change she most wanted before deployment. $P9$, who targeted women and household workers in India, reported a subtler mismatch, where traces described practices that are real but belong to a labor arrangement adjacent to the one her study samples.

One way to interpret this is that we are seeing the WEIRD critique reappearing one layer in. We motivated \systemname{} on the observation that the critique applies to the small teams who write a study and not only to the participants who answer it \citep{henrich2010weirdest}, and the instrument we built to widen that lens carries a version of the same narrowness. 

However, participants' doubts about the simulated participants did not transfer to the critiques those participants produced. $P4$ was explicit that the two came apart, saying \textit{``my trust in this model is going down, despite its scientific accuracy [of the critiques] being on point,''} while rating \systemname{} \emph{very useful} and indicating she was \emph{very likely} to use a finished version. $P3$ went further, largely avoiding the evidence panels because she regarded the simulated participants as fictitious, and gave the same ratings. Our position holds up in the hands of researchers who know their populations well, since the participants best placed to catch a bad persona were also the ones who kept using the critiques. This suggests that while the quality target demography may degrade the performance of \systemname{}, experts can still get values out of leveraging it.

The traces also say something about the simulation itself. Prior work calibrates simulated agents against observed human data or grounds them in richer population information \citep{horton2023homo, manning2024automated, park2024generative}, and a mismatch in an aggregate distribution shows only that a simulation is wrong somewhere. The three cases above are three different failures. $P4$'s archetype is a model working from an obsolete norm, $P9$'s labor arrangement is a model reasoning about a nearby population, and $P6$'s field site is a model missing the institutional context; all were missing from demographic labels. Each calls for a different correction, while a distribution would have registered all three the same way. $P1$ asked to be consulted on the \textit{``dimensions for variation in the simulated personas''} before a run, and \systemname{} could route such corrections back to the Persona Synthesizer to revise the demographic space for the next round. Training objectives for simulation models could likewise go beyond distribution matching \citep{Zhou2026MindTS}.

\subsection{Beyond Study Designs}

Our core idea of leveraging reasoning traces to provoke a user to see the perspective of respondents has the potential to  generalize to other scenarios. For example, having a computer-use agent attempting a task in a real interface and collecting its reasoning traces of how the agent interpreted and reacted to the interface and reasoned before performing actions on it. Here, the interface serves the role of the study design, and the simulated users are drawn from a stated target population. The Critic can report steps where the correct action is not discoverable, along with functionality that a persona expected and did not have. The Refiner has a counterpart as well. \citet{adar2026revibing} shows that coding agents today are nearly capable of reimplementing interactive systems published in HCI directly from their papers, suggesting that coding agents can be used to refine an interface based on critiques generated from simulated reasoning traces.

\section{Conclusion}

Most work on LLM social simulation asks whether responses from a simulated participant match those of a real participant to aptly estimate the outcome of a behavioral study. We, instead, ask a different question: whether a study design can be read in ways other than its authors' intentions. We harnessed powerful LLMs to uncover the internal reasoning of simulated participants, without relying on their generated responses. We presented \systemname{}, an interactive system that compiles a draft study plan for a cohort of simulated participants and collects the internal reasoning traces they generate as they work through it to provide their responses. From these traces, the system identified two types of critiques: flagging structural errors in the already existing questions and proposing missed opportunities as potential additional questions to better uncover the research objective of the researcher. Researchers triage by selecting and commenting on the critiques they wish the system to act on, and then iterate with the system to refine their study design.

We first compared \systemname{} against a paid human pilot and a no-simulation baseline on a survey experiment with 112 questions. For all three conditions flagged, critiques from \systemname{} were rated higher in quality than either baselines ($p=0.013$ and $p=0.029$). \systemname{} also surfaced $49$ additional high-quality structural errors that the no-simulation baseline missed, against 11 in the other direction.

We also conducted a user study, where 12 researchers brought designs from their own work with populations in North America, Asia, and Africa. Several found that \systemname{} reproduced problems they had already uncovered through their own real pilots and fieldwork. In general, participants valued the missed opportunities most, described the system as a way to expand a research question rather than only for debugging their study designs, and were willing to positively recommend \systemname{} to junior authors and peers. Participants are highly confident of the critiques they choose to include in the refinement. Our work shows early promises in reasoning traces in LLM simulations that may help uncover latent social and behavioral models to better predict future outcomes. 



\bibliographystyle{ACM-Reference-Format}
\bibliography{references}


\begin{thebibliography}{60}


\ifx \showCODEN    \undefined \def \showCODEN     #1{\unskip}     \fi
\ifx \showISBNx    \undefined \def \showISBNx     #1{\unskip}     \fi
\ifx \showISBNxiii \undefined \def \showISBNxiii  #1{\unskip}     \fi
\ifx \showISSN     \undefined \def \showISSN      #1{\unskip}     \fi
\ifx \showLCCN     \undefined \def \showLCCN      #1{\unskip}     \fi
\ifx \shownote     \undefined \def \shownote      #1{#1}          \fi
\ifx \showarticletitle \undefined \def \showarticletitle #1{#1}   \fi
\ifx \showURL      \undefined \def \showURL       {\relax}        \fi
\providecommand\bibfield[2]{#2}
\providecommand\bibinfo[2]{#2}
\providecommand\natexlab[1]{#1}
\providecommand\showeprint[2][]{arXiv:#2}

\bibitem[Adar et~al\mbox{.}(2026)]%
        {adar2026revibing}
\bibfield{author}{\bibinfo{person}{Eytan Adar}, \bibinfo{person}{Yoonjoo Lee}, \bibinfo{person}{Ni Lei}, \bibinfo{person}{Q. Liao}, {and} \bibinfo{person}{Weirui Peng}.} \bibinfo{year}{2026}\natexlab{}.
\newblock \showarticletitle{Revibing Code from Papers: Reimplementing HCI Artifacts}.
\newblock  (\bibinfo{year}{2026}).
\newblock
\urldef\tempurl%
\url{https://api.semanticscholar.org/CorpusID:290788244}
\showURL{%
\tempurl}


\bibitem[Adhikari et~al\mbox{.}(2025)]%
        {adhikari2025exploring}
\bibfield{author}{\bibinfo{person}{Divya~Mani Adhikari}, \bibinfo{person}{Alexander Hartland}, \bibinfo{person}{Ingmar Weber}, {and} \bibinfo{person}{Vikram~Kamath Cannanure}.} \bibinfo{year}{2025}\natexlab{}.
\newblock \showarticletitle{Exploring {LLMs} for Automated Generation and Adaptation of Questionnaires}. In \bibinfo{booktitle}{\emph{Proceedings of the 7th ACM Conference on Conversational User Interfaces (CUI)}}.
\newblock
\href{https://doi.org/10.1145/3719160.3736606}{doi:\nolinkurl{10.1145/3719160.3736606}}


\bibitem[Agarwal et~al\mbox{.}(2025)]%
        {Agarwal2025AutoDiscoveryOS}
\bibfield{author}{\bibinfo{person}{Dhruv Agarwal}, \bibinfo{person}{Bodhisattwa~Prasad Majumder}, \bibinfo{person}{Reece Adamson}, \bibinfo{person}{Megha Chakravorty}, \bibinfo{person}{Satvika~Reddy Gavireddy}, \bibinfo{person}{Aditya Parashar}, \bibinfo{person}{Harshit Surana}, \bibinfo{person}{Bhavana~Dalvi Mishra}, \bibinfo{person}{Andrew McCallum}, \bibinfo{person}{Ashish Sabharwal}, {and} \bibinfo{person}{Peter Clark}.} \bibinfo{year}{2025}\natexlab{}.
\newblock \showarticletitle{AutoDiscovery: Open-ended Scientific Discovery via Bayesian Surprise}.
\newblock \bibinfo{journal}{\emph{Advances in Neural Information Processing Systems 38}} (\bibinfo{year}{2025}).
\newblock
\urldef\tempurl%
\url{https://api.semanticscholar.org/CorpusID:280150869}
\showURL{%
\tempurl}


\bibitem[Agnew et~al\mbox{.}(2024)]%
        {agnew2024illusion}
\bibfield{author}{\bibinfo{person}{William Agnew}, \bibinfo{person}{A~Stevie Bergman}, \bibinfo{person}{Jennifer Chien}, \bibinfo{person}{Mark D{\'\i}az}, \bibinfo{person}{Seliem El-Sayed}, \bibinfo{person}{Jaylen Pittman}, \bibinfo{person}{Shakir Mohamed}, {and} \bibinfo{person}{Kevin~R McKee}.} \bibinfo{year}{2024}\natexlab{}.
\newblock \showarticletitle{The illusion of artificial inclusion}. In \bibinfo{booktitle}{\emph{Proceedings of the 2024 CHI Conference on Human Factors in Computing Systems}}. \bibinfo{pages}{1--12}.
\newblock


\bibitem[Aher et~al\mbox{.}(2023)]%
        {aher2023using}
\bibfield{author}{\bibinfo{person}{Gati~V. Aher}, \bibinfo{person}{Rosa~I. Arriaga}, {and} \bibinfo{person}{Adam~Tauman Kalai}.} \bibinfo{year}{2023}\natexlab{}.
\newblock \showarticletitle{Using Large Language Models to Simulate Multiple Humans and Replicate Human Subject Studies}. In \bibinfo{booktitle}{\emph{Proceedings of the 40th International Conference on Machine Learning (ICML)}} \emph{(\bibinfo{series}{PMLR}, Vol.~\bibinfo{volume}{202})}. \bibinfo{pages}{337--371}.
\newblock


\bibitem[Argyle et~al\mbox{.}(2023)]%
        {argyle2023out}
\bibfield{author}{\bibinfo{person}{Lisa~P. Argyle}, \bibinfo{person}{Ethan~C. Busby}, \bibinfo{person}{Nancy Fulda}, \bibinfo{person}{Joshua~R. Gubler}, \bibinfo{person}{Christopher Rytting}, {and} \bibinfo{person}{David Wingate}.} \bibinfo{year}{2023}\natexlab{}.
\newblock \showarticletitle{Out of One, Many: Using Language Models to Simulate Human Samples}.
\newblock \bibinfo{journal}{\emph{Political Analysis}} \bibinfo{volume}{31}, \bibinfo{number}{3} (\bibinfo{year}{2023}), \bibinfo{pages}{337--351}.
\newblock
\href{https://doi.org/10.1017/pan.2023.2}{doi:\nolinkurl{10.1017/pan.2023.2}}


\bibitem[Bender et~al\mbox{.}(2021)]%
        {10.1145/3442188.3445922}
\bibfield{author}{\bibinfo{person}{Emily~M. Bender}, \bibinfo{person}{Timnit Gebru}, \bibinfo{person}{Angelina McMillan-Major}, {and} \bibinfo{person}{Shmargaret Shmitchell}.} \bibinfo{year}{2021}\natexlab{}.
\newblock \showarticletitle{On the Dangers of Stochastic Parrots: Can Language Models Be Too Big?}. In \bibinfo{booktitle}{\emph{Proceedings of the 2021 ACM Conference on Fairness, Accountability, and Transparency}} (Virtual Event, Canada) \emph{(\bibinfo{series}{FAccT '21})}. \bibinfo{publisher}{Association for Computing Machinery}, \bibinfo{address}{New York, NY, USA}, \bibinfo{pages}{610–623}.
\newblock
\showISBNx{9781450383097}
\href{https://doi.org/10.1145/3442188.3445922}{doi:\nolinkurl{10.1145/3442188.3445922}}


\bibitem[Bisbee et~al\mbox{.}(2024)]%
        {bisbee2024synthetic}
\bibfield{author}{\bibinfo{person}{James Bisbee}, \bibinfo{person}{Joshua~D. Clinton}, \bibinfo{person}{Cassy Dorff}, \bibinfo{person}{Brenton Kenkel}, {and} \bibinfo{person}{Jennifer~M. Larson}.} \bibinfo{year}{2024}\natexlab{}.
\newblock \showarticletitle{Synthetic Replacements for Human Survey Data? The Perils of Large Language Models}.
\newblock \bibinfo{journal}{\emph{Political Analysis}} \bibinfo{volume}{32}, \bibinfo{number}{4} (\bibinfo{year}{2024}), \bibinfo{pages}{401--416}.
\newblock
\href{https://doi.org/10.1017/pan.2024.5}{doi:\nolinkurl{10.1017/pan.2024.5}}


\bibitem[Blair and Conrad(2011)]%
        {blair2011sample}
\bibfield{author}{\bibinfo{person}{Johnny Blair} {and} \bibinfo{person}{Frederick~G Conrad}.} \bibinfo{year}{2011}\natexlab{}.
\newblock \showarticletitle{Sample size for cognitive interview pretesting}.
\newblock \bibinfo{journal}{\emph{Public opinion quarterly}} \bibinfo{volume}{75}, \bibinfo{number}{4} (\bibinfo{year}{2011}), \bibinfo{pages}{636--658}.
\newblock


\bibitem[Blass(1991)]%
        {blass1991understanding}
\bibfield{author}{\bibinfo{person}{Thomas Blass}.} \bibinfo{year}{1991}\natexlab{}.
\newblock \showarticletitle{Understanding Behavior in the Milgram Obedience Experiment: The Role of Personality, Situations, and Their Interactions}.
\newblock \bibinfo{journal}{\emph{Journal of Personality and Social Psychology}} \bibinfo{volume}{60}, \bibinfo{number}{3} (\bibinfo{year}{1991}), \bibinfo{pages}{398--413}.
\newblock
\href{https://doi.org/10.1037/0022-3514.60.3.398}{doi:\nolinkurl{10.1037/0022-3514.60.3.398}}


\bibitem[Bohm(1994)]%
        {bohm1994cvm}
\bibfield{author}{\bibinfo{person}{Peter Bohm}.} \bibinfo{year}{1994}\natexlab{}.
\newblock \showarticletitle{{CVM} Spells Responses to Hypothetical Questions}.
\newblock \bibinfo{journal}{\emph{Natural Resources Journal}} (\bibinfo{year}{1994}), \bibinfo{pages}{37--50}.
\newblock


\bibitem[{British Election Study}(1964)]%
        {bes1964}
\bibfield{author}{\bibinfo{person}{{British Election Study}}.} \bibinfo{year}{1964}\natexlab{}.
\newblock \bibinfo{title}{The British Election Study}.
\newblock \bibinfo{howpublished}{\url{https://www.britishelectionstudy.com/}}.
\newblock
\shownote{Conducted after every UK general election since 1964}.
\newblock


\bibitem[Camerer et~al\mbox{.}(2016)]%
        {camerer2016evaluating}
\bibfield{author}{\bibinfo{person}{Colin~F. Camerer}, \bibinfo{person}{Anna Dreber}, \bibinfo{person}{Eskil Forsell}, \bibinfo{person}{Teck-Hua Ho}, \bibinfo{person}{J{\"u}rgen Huber}, \bibinfo{person}{Magnus Johannesson}, \bibinfo{person}{Michael Kirchler}, \bibinfo{person}{Johan Almenberg}, \bibinfo{person}{Adam Altmejd}, \bibinfo{person}{Taizan Chan}, {et~al\mbox{.}}} \bibinfo{year}{2016}\natexlab{}.
\newblock \showarticletitle{Evaluating Replicability of Laboratory Experiments in Economics}.
\newblock \bibinfo{journal}{\emph{Science}} \bibinfo{volume}{351}, \bibinfo{number}{6280} (\bibinfo{year}{2016}), \bibinfo{pages}{1433--1436}.
\newblock
\href{https://doi.org/10.1126/science.aaf0918}{doi:\nolinkurl{10.1126/science.aaf0918}}


\bibitem[Campbell and Fiske(1959)]%
        {campbell1959convergent}
\bibfield{author}{\bibinfo{person}{Donald~T. Campbell} {and} \bibinfo{person}{Donald~W. Fiske}.} \bibinfo{year}{1959}\natexlab{}.
\newblock \showarticletitle{Convergent and Discriminant Validation by the Multitrait-Multimethod Matrix}.
\newblock \bibinfo{journal}{\emph{Psychological Bulletin}} \bibinfo{volume}{56}, \bibinfo{number}{2} (\bibinfo{year}{1959}), \bibinfo{pages}{81--105}.
\newblock
\href{https://doi.org/10.1037/h0046016}{doi:\nolinkurl{10.1037/h0046016}}


\bibitem[Cao et~al\mbox{.}(2025)]%
        {cao2025specializing}
\bibfield{author}{\bibinfo{person}{Yong Cao}, \bibinfo{person}{Haijiang Liu}, \bibinfo{person}{Arnav Arora}, \bibinfo{person}{Isabelle Augenstein}, \bibinfo{person}{Paul R{\"o}ttger}, {and} \bibinfo{person}{Daniel Hershcovich}.} \bibinfo{year}{2025}\natexlab{}.
\newblock \showarticletitle{Specializing Large Language Models to Simulate Survey Response Distributions for Global Populations}. In \bibinfo{booktitle}{\emph{Proceedings of the 2025 Conference of the Nations of the Americas Chapter of the Association for Computational Linguistics: Human Language Technologies}}. \bibinfo{pages}{3141--3154}.
\newblock
\href{https://doi.org/10.18653/v1/2025.naacl-long.162}{doi:\nolinkurl{10.18653/v1/2025.naacl-long.162}}


\bibitem[Chen et~al\mbox{.}(2025)]%
        {chen2025reasoning}
\bibfield{author}{\bibinfo{person}{Yanda Chen}, \bibinfo{person}{Joe Benton}, \bibinfo{person}{Ansh Radhakrishnan}, \bibinfo{person}{Jonathan Uesato}, \bibinfo{person}{Carson Denison}, \bibinfo{person}{John Schulman}, \bibinfo{person}{Arushi Somani}, \bibinfo{person}{Peter Hase}, \bibinfo{person}{Misha Wagner}, \bibinfo{person}{Fabien Roger}, \bibinfo{person}{Vlad Mikulik}, \bibinfo{person}{Samuel~R. Bowman}, \bibinfo{person}{Jan Leike}, \bibinfo{person}{Jared Kaplan}, {and} \bibinfo{person}{Ethan Perez}.} \bibinfo{year}{2025}\natexlab{}.
\newblock \bibinfo{title}{Reasoning Models Don't Always Say What They Think}.
\newblock
\showeprint[arxiv]{2505.05410}~[cs.CL]


\bibitem[Cheng et~al\mbox{.}(2023)]%
        {cheng2023compost}
\bibfield{author}{\bibinfo{person}{Myra Cheng}, \bibinfo{person}{Tiziano Piccardi}, {and} \bibinfo{person}{Diyi Yang}.} \bibinfo{year}{2023}\natexlab{}.
\newblock \showarticletitle{{CoMPosT}: Characterizing and Evaluating Caricature in {LLM} Simulations}. In \bibinfo{booktitle}{\emph{Proceedings of the 2023 Conference on Empirical Methods in Natural Language Processing (EMNLP)}}. \bibinfo{pages}{10853--10875}.
\newblock
\href{https://doi.org/10.18653/v1/2023.emnlp-main.669}{doi:\nolinkurl{10.18653/v1/2023.emnlp-main.669}}


\bibitem[Clark(1973)]%
        {clark1973language}
\bibfield{author}{\bibinfo{person}{Herbert~H. Clark}.} \bibinfo{year}{1973}\natexlab{}.
\newblock \showarticletitle{The Language-as-Fixed-Effect Fallacy: A Critique of Language Statistics in Psychological Research}.
\newblock \bibinfo{journal}{\emph{Journal of Verbal Learning and Verbal Behavior}} \bibinfo{volume}{12}, \bibinfo{number}{4} (\bibinfo{year}{1973}), \bibinfo{pages}{335--359}.
\newblock
\href{https://doi.org/10.1016/S0022-5371(73)80014-3}{doi:\nolinkurl{10.1016/S0022-5371(73)80014-3}}


\bibitem[Cummings and Taylor(1999)]%
        {cummings1999unbiased}
\bibfield{author}{\bibinfo{person}{Ronald~G. Cummings} {and} \bibinfo{person}{Laura~O. Taylor}.} \bibinfo{year}{1999}\natexlab{}.
\newblock \showarticletitle{Unbiased Value Estimates for Environmental Goods: A Cheap Talk Design for the Contingent Valuation Method}.
\newblock \bibinfo{journal}{\emph{American Economic Review}} \bibinfo{volume}{89}, \bibinfo{number}{3} (\bibinfo{year}{1999}), \bibinfo{pages}{649--665}.
\newblock
\href{https://doi.org/10.1257/aer.89.3.649}{doi:\nolinkurl{10.1257/aer.89.3.649}}


\bibitem[Davis and Smith(1991)]%
        {davis1991norc}
\bibfield{author}{\bibinfo{person}{James~A. Davis} {and} \bibinfo{person}{Tom~W. Smith}.} \bibinfo{year}{1991}\natexlab{}.
\newblock \bibinfo{booktitle}{\emph{The {NORC} General Social Survey: A User's Guide}}.
\newblock \bibinfo{publisher}{SAGE Publications}.
\newblock


\bibitem[Dillion et~al\mbox{.}(2023)]%
        {dillion2023can}
\bibfield{author}{\bibinfo{person}{Danica Dillion}, \bibinfo{person}{Niket Tandon}, \bibinfo{person}{Yuling Gu}, {and} \bibinfo{person}{Kurt Gray}.} \bibinfo{year}{2023}\natexlab{}.
\newblock \showarticletitle{Can {AI} Language Models Replace Human Participants?}
\newblock \bibinfo{journal}{\emph{Trends in Cognitive Sciences}} \bibinfo{volume}{27}, \bibinfo{number}{7} (\bibinfo{year}{2023}), \bibinfo{pages}{597--600}.
\newblock
\href{https://doi.org/10.1016/j.tics.2023.04.008}{doi:\nolinkurl{10.1016/j.tics.2023.04.008}}


\bibitem[Gonz{\'a}lez-Bustamante et~al\mbox{.}(2025)]%
        {gonzalez2025emulating}
\bibfield{author}{\bibinfo{person}{Bastián Gonz{\'a}lez-Bustamante}, \bibinfo{person}{Nando Verelst}, {and} \bibinfo{person}{Carla Cisternas}.} \bibinfo{year}{2025}\natexlab{}.
\newblock \bibinfo{title}{Emulating Public Opinion: A Proof-of-Concept of {AI}-Generated Synthetic Survey Responses for the Chilean Case}.
\newblock
\showeprint[arxiv]{2509.09871}


\bibitem[Henrich et~al\mbox{.}(2010)]%
        {henrich2010weirdest}
\bibfield{author}{\bibinfo{person}{Joseph Henrich}, \bibinfo{person}{Steven~J. Heine}, {and} \bibinfo{person}{Ara Norenzayan}.} \bibinfo{year}{2010}\natexlab{}.
\newblock \showarticletitle{The Weirdest People in the World?}
\newblock \bibinfo{journal}{\emph{Behavioral and Brain Sciences}} \bibinfo{volume}{33}, \bibinfo{number}{2-3} (\bibinfo{year}{2010}), \bibinfo{pages}{61--83}.
\newblock
\href{https://doi.org/10.1017/S0140525X0999152X}{doi:\nolinkurl{10.1017/S0140525X0999152X}}


\bibitem[Horton and Horton(2024)]%
        {expected-parrot}
\bibfield{author}{\bibinfo{person}{John Horton} {and} \bibinfo{person}{Robin Horton}.} \bibinfo{year}{2024}\natexlab{}.
\newblock \bibinfo{booktitle}{\emph{{EDSL}: Expected Parrot Domain Specific Language for {AI} Powered Social Science}}.
\newblock \bibinfo{type}{Whitepaper}. \bibinfo{institution}{Expected Parrot}.
\newblock
\shownote{Open-source Python package, MIT license}.
\newblock
\urldef\tempurl%
\url{https://github.com/expectedparrot/edsl}
\showURL{%
\tempurl}


\bibitem[Horton et~al\mbox{.}(2023)]%
        {horton2023homo}
\bibfield{author}{\bibinfo{person}{John~J. Horton}, \bibinfo{person}{Apostolos Filippas}, {and} \bibinfo{person}{Benjamin~S. Manning}.} \bibinfo{year}{2023}\natexlab{}.
\newblock \bibinfo{booktitle}{\emph{Large Language Models as Simulated Economic Agents: What Can We Learn from {Homo Silicus}?}}
\newblock \bibinfo{type}{{T}echnical {R}eport} 31122. \bibinfo{institution}{National Bureau of Economic Research}.
\newblock
\shownote{Working Paper; later in Proc. 25th ACM Conf. on Economics and Computation (EC'24)}.
\newblock
\href{https://doi.org/10.3386/w31122}{doi:\nolinkurl{10.3386/w31122}}


\bibitem[Jansen et~al\mbox{.}(2026)]%
        {Jansen2026GeneratingLS}
\bibfield{author}{\bibinfo{person}{Peter Jansen}, \bibinfo{person}{Peter Clark}, \bibinfo{person}{Doug Downey}, {and} \bibinfo{person}{Daniel~S Weld}.} \bibinfo{year}{2026}\natexlab{}.
\newblock \showarticletitle{Generating Literature-Driven Scientific Theories at Scale}. In \bibinfo{booktitle}{\emph{Proceedings of the 64th Annual Meeting of the {A}ssociation for {C}omputational {L}inguistics (Volume 1: Long Papers)}}, \bibfield{editor}{\bibinfo{person}{Maria Liakata}, \bibinfo{person}{Viviane~P. Moreira}, \bibinfo{person}{Jiajun Zhang}, {and} \bibinfo{person}{David Jurgens}} (Eds.). \bibinfo{publisher}{Association for Computational Linguistics}, \bibinfo{address}{San Diego, California, United States}, \bibinfo{pages}{14677--14698}.
\newblock
\showISBNx{979-8-89176-390-6}
\href{https://doi.org/10.18653/v1/2026.acl-long.669}{doi:\nolinkurl{10.18653/v1/2026.acl-long.669}}


\bibitem[Kahneman et~al\mbox{.}(1991)]%
        {kahneman1991endowment}
\bibfield{author}{\bibinfo{person}{Daniel Kahneman}, \bibinfo{person}{Jack~L. Knetsch}, {and} \bibinfo{person}{Richard~H. Thaler}.} \bibinfo{year}{1991}\natexlab{}.
\newblock \showarticletitle{Anomalies: The Endowment Effect, Loss Aversion, and Status Quo Bias}.
\newblock \bibinfo{journal}{\emph{Journal of Economic Perspectives}} \bibinfo{volume}{5}, \bibinfo{number}{1} (\bibinfo{date}{March} \bibinfo{year}{1991}), \bibinfo{pages}{193–206}.
\newblock
\href{https://doi.org/10.1257/jep.5.1.193}{doi:\nolinkurl{10.1257/jep.5.1.193}}


\bibitem[Kahneman and Tversky(1979)]%
        {kahneman1979prospect}
\bibfield{author}{\bibinfo{person}{Daniel Kahneman} {and} \bibinfo{person}{Amos Tversky}.} \bibinfo{year}{1979}\natexlab{}.
\newblock \showarticletitle{Prospect Theory: An Analysis of Decision Under Risk}.
\newblock \bibinfo{journal}{\emph{Econometrica}} \bibinfo{volume}{47}, \bibinfo{number}{2} (\bibinfo{year}{1979}), \bibinfo{pages}{263--291}.
\newblock
\href{https://doi.org/10.2307/1914185}{doi:\nolinkurl{10.2307/1914185}}


\bibitem[Kapania et~al\mbox{.}(2025)]%
        {kapania2025simulacrum}
\bibfield{author}{\bibinfo{person}{Shivani Kapania}, \bibinfo{person}{William Agnew}, \bibinfo{person}{Motahhare Eslami}, \bibinfo{person}{Hoda Heidari}, {and} \bibinfo{person}{Sarah~E Fox}.} \bibinfo{year}{2025}\natexlab{}.
\newblock \showarticletitle{Simulacrum of Stories: Examining Large Language Models as Qualitative Research Participants}. In \bibinfo{booktitle}{\emph{Proceedings of the 2025 CHI Conference on Human Factors in Computing Systems}} \emph{(\bibinfo{series}{CHI '25})}. \bibinfo{publisher}{Association for Computing Machinery}, \bibinfo{address}{New York, NY, USA}, Article \bibinfo{articleno}{489}, \bibinfo{numpages}{17}~pages.
\newblock
\showISBNx{9798400713941}
\href{https://doi.org/10.1145/3706598.3713220}{doi:\nolinkurl{10.1145/3706598.3713220}}


\bibitem[Krosnick(1991)]%
        {krosnick1991response}
\bibfield{author}{\bibinfo{person}{Jon~A. Krosnick}.} \bibinfo{year}{1991}\natexlab{}.
\newblock \showarticletitle{Response strategies for coping with the cognitive demands of attitude measures in surveys}.
\newblock \bibinfo{journal}{\emph{Applied Cognitive Psychology}}  \bibinfo{volume}{5} (\bibinfo{year}{1991}), \bibinfo{pages}{213--236}.
\newblock
\urldef\tempurl%
\url{https://api.semanticscholar.org/CorpusID:143661480}
\showURL{%
\tempurl}


\bibitem[Krosnick and Presser(2010)]%
        {krosnick2010question}
\bibfield{author}{\bibinfo{person}{Jon~A. Krosnick} {and} \bibinfo{person}{Stanley Presser}.} \bibinfo{year}{2010}\natexlab{}.
\newblock \showarticletitle{Question and Questionnaire Design}.
\newblock In \bibinfo{booktitle}{\emph{Handbook of Survey Research} (\bibinfo{edition}{2nd} ed.)}, \bibfield{editor}{\bibinfo{person}{Peter~V. Marsden} {and} \bibinfo{person}{James~D. Wright}} (Eds.). \bibinfo{publisher}{Emerald}, \bibinfo{pages}{263--314}.
\newblock


\bibitem[Lanham et~al\mbox{.}(2023)]%
        {lanham2023measuring}
\bibfield{author}{\bibinfo{person}{Tamera Lanham}, \bibinfo{person}{Anna Chen}, \bibinfo{person}{Ansh Radhakrishnan}, \bibinfo{person}{Benoit Steiner}, \bibinfo{person}{Carson Denison}, \bibinfo{person}{Danny Hernandez}, \bibinfo{person}{Dustin Li}, {et~al\mbox{.}}} \bibinfo{year}{2023}\natexlab{}.
\newblock \bibinfo{title}{Measuring Faithfulness in Chain-of-Thought Reasoning}.
\newblock
\showeprint[arxiv]{2307.13702}~[cs.CL]


\bibitem[Liao et~al\mbox{.}(2024)]%
        {liao2024llms}
\bibfield{author}{\bibinfo{person}{Zhehui Liao}, \bibinfo{person}{Maria Antoniak}, \bibinfo{person}{Inyoung Cheong}, \bibinfo{person}{Evie Yu-Yen Cheng}, \bibinfo{person}{Ai-Heng Lee}, \bibinfo{person}{Kyle Lo}, \bibinfo{person}{Joseph~Chee Chang}, {and} \bibinfo{person}{Amy~X. Zhang}.} \bibinfo{year}{2024}\natexlab{}.
\newblock \bibinfo{title}{{LLMs} as Research Tools: A Large Scale Survey of Researchers' Usage and Perceptions}.
\newblock
\showeprint[arxiv]{2411.05025}


\bibitem[Ling et~al\mbox{.}(2026)]%
        {ling2026underreporting}
\bibfield{author}{\bibinfo{person}{Yier Ling}, \bibinfo{person}{Alex Kale}, {and} \bibinfo{person}{Alex Imas}.} \bibinfo{year}{2026}\natexlab{}.
\newblock \showarticletitle{Underreporting of AI Use: The Role of Social Desirability Bias}. In \bibinfo{booktitle}{\emph{Proceedings of the 2026 CHI Conference on Human Factors in Computing Systems}} \emph{(\bibinfo{series}{CHI '26})}. \bibinfo{publisher}{Association for Computing Machinery}, \bibinfo{address}{New York, NY, USA}, Article \bibinfo{articleno}{62}, \bibinfo{numpages}{22}~pages.
\newblock
\showISBNx{9798400722783}
\href{https://doi.org/10.1145/3772318.3791073}{doi:\nolinkurl{10.1145/3772318.3791073}}


\bibitem[Linxen et~al\mbox{.}(2021)]%
        {linxen2021weird}
\bibfield{author}{\bibinfo{person}{Sebastian Linxen}, \bibinfo{person}{Christian Sturm}, \bibinfo{person}{Florian Br{\"u}hlmann}, \bibinfo{person}{Vincent Cassau}, \bibinfo{person}{Klaus Opwis}, {and} \bibinfo{person}{Katharina Reinecke}.} \bibinfo{year}{2021}\natexlab{}.
\newblock \showarticletitle{How weird is CHI?}. In \bibinfo{booktitle}{\emph{Proceedings of the 2021 chi conference on human factors in computing systems}}. \bibinfo{pages}{1--14}.
\newblock


\bibitem[Liu et~al\mbox{.}(2026)]%
        {liu2026cobra}
\bibfield{author}{\bibinfo{person}{Xuan Liu}, \bibinfo{person}{Haoyang Shang}, {and} \bibinfo{person}{Haojian Jin}.} \bibinfo{year}{2026}\natexlab{}.
\newblock \showarticletitle{{CoBRA}: Programming Cognitive Bias in Social Agents Using Classic Social Science Experiments}. In \bibinfo{booktitle}{\emph{Proceedings of the 2026 CHI Conference on Human Factors in Computing Systems}}.
\newblock
\showeprint[arxiv]{2509.13588}
\href{https://doi.org/10.1145/3772318.3790804}{doi:\nolinkurl{10.1145/3772318.3790804}}


\bibitem[Manning et~al\mbox{.}(2024)]%
        {manning2024automated}
\bibfield{author}{\bibinfo{person}{Benjamin~S. Manning}, \bibinfo{person}{Kehang Zhu}, {and} \bibinfo{person}{John~J. Horton}.} \bibinfo{year}{2024}\natexlab{}.
\newblock \bibinfo{booktitle}{\emph{Automated Social Science: Language Models as Scientist and Subjects}}.
\newblock \bibinfo{type}{{T}echnical {R}eport} 32381. \bibinfo{institution}{National Bureau of Economic Research}.
\newblock
\href{https://doi.org/10.3386/w32381}{doi:\nolinkurl{10.3386/w32381}}


\bibitem[Miller et~al\mbox{.}(2011)]%
        {miller2011design}
\bibfield{author}{\bibinfo{person}{Kristen Miller}, \bibinfo{person}{Rory Fitzgerald}, \bibinfo{person}{Jos{\'e}-Luis Padilla}, \bibinfo{person}{Stephanie Willson}, \bibinfo{person}{Sally Widdop}, \bibinfo{person}{Rachel Caspar}, \bibinfo{person}{Martin Dimov}, \bibinfo{person}{Michelle Gray}, \bibinfo{person}{C{\'a}tia Nunes}, \bibinfo{person}{Peter Pr{\"u}fer}, {et~al\mbox{.}}} \bibinfo{year}{2011}\natexlab{}.
\newblock \showarticletitle{Design and analysis of cognitive interviews for comparative multinational testing}.
\newblock \bibinfo{journal}{\emph{Field Methods}} \bibinfo{volume}{23}, \bibinfo{number}{4} (\bibinfo{year}{2011}), \bibinfo{pages}{379--396}.
\newblock


\bibitem[{Open Science Collaboration}(2015)]%
        {oscollab2015estimating}
\bibfield{author}{\bibinfo{person}{{Open Science Collaboration}}.} \bibinfo{year}{2015}\natexlab{}.
\newblock \showarticletitle{Estimating the Reproducibility of Psychological Science}.
\newblock \bibinfo{journal}{\emph{Science}} \bibinfo{volume}{349}, \bibinfo{number}{6251} (\bibinfo{year}{2015}), \bibinfo{pages}{aac4716}.
\newblock
\href{https://doi.org/10.1126/science.aac4716}{doi:\nolinkurl{10.1126/science.aac4716}}


\bibitem[Orne(2017)]%
        {orne2017social}
\bibfield{author}{\bibinfo{person}{Martin~T Orne}.} \bibinfo{year}{2017}\natexlab{}.
\newblock \showarticletitle{On the social psychology of the psychological experiment: With particular reference to demand characteristics and their implications}.
\newblock In \bibinfo{booktitle}{\emph{Sociological methods}}. \bibinfo{publisher}{Routledge}, \bibinfo{pages}{279--299}.
\newblock


\bibitem[Oudshoorn et~al\mbox{.}(2004)]%
        {oudshoorn2004configuring}
\bibfield{author}{\bibinfo{person}{Nelly Oudshoorn}, \bibinfo{person}{Els Rommes}, {and} \bibinfo{person}{Marcelle Stienstra}.} \bibinfo{year}{2004}\natexlab{}.
\newblock \showarticletitle{Configuring the user as everybody: Gender and design cultures in information and communication technologies}.
\newblock \bibinfo{journal}{\emph{Science, Technology, \& Human Values}} \bibinfo{volume}{29}, \bibinfo{number}{1} (\bibinfo{year}{2004}), \bibinfo{pages}{30--63}.
\newblock


\bibitem[Park et~al\mbox{.}(2024)]%
        {park2024generative}
\bibfield{author}{\bibinfo{person}{Joon~Sung Park}, \bibinfo{person}{Carolyn~Q. Zou}, \bibinfo{person}{Aaron Shaw}, \bibinfo{person}{Benjamin~Mako Hill}, \bibinfo{person}{Carrie~J. Cai}, \bibinfo{person}{Meredith~Ringel Morris}, \bibinfo{person}{Robb Willer}, \bibinfo{person}{Percy Liang}, {and} \bibinfo{person}{Michael~S. Bernstein}.} \bibinfo{year}{2024}\natexlab{}.
\newblock \bibinfo{title}{Generative Agent Simulations of 1,000 People}.
\newblock
\showeprint[arxiv]{2411.10109}~[cs.HC]


\bibitem[Podsakoff et~al\mbox{.}(2003)]%
        {podsakoff2003common}
\bibfield{author}{\bibinfo{person}{Philip~M. Podsakoff}, \bibinfo{person}{Scott~B. MacKenzie}, \bibinfo{person}{Jeong-Yeon Lee}, {and} \bibinfo{person}{Nathan~P. Podsakoff}.} \bibinfo{year}{2003}\natexlab{}.
\newblock \showarticletitle{Common Method Biases in Behavioral Research: A Critical Review of the Literature and Recommended Remedies}.
\newblock \bibinfo{journal}{\emph{Journal of Applied Psychology}} \bibinfo{volume}{88}, \bibinfo{number}{5} (\bibinfo{year}{2003}), \bibinfo{pages}{879--903}.
\newblock
\href{https://doi.org/10.1037/0021-9010.88.5.879}{doi:\nolinkurl{10.1037/0021-9010.88.5.879}}


\bibitem[Poulton(1982)]%
        {poulton1982influential}
\bibfield{author}{\bibinfo{person}{E.~C. Poulton}.} \bibinfo{year}{1982}\natexlab{}.
\newblock \showarticletitle{Influential Companions: Effects of One Strategy on Another in the Within-Subjects Designs of Cognitive Psychology}.
\newblock \bibinfo{journal}{\emph{Psychological Bulletin}} \bibinfo{volume}{91}, \bibinfo{number}{3} (\bibinfo{year}{1982}), \bibinfo{pages}{673--690}.
\newblock
\href{https://doi.org/10.1037/0033-2909.91.3.673}{doi:\nolinkurl{10.1037/0033-2909.91.3.673}}


\bibitem[Presser et~al\mbox{.}(2004)]%
        {presser2004methods}
\bibfield{author}{\bibinfo{person}{Stanley Presser}, \bibinfo{person}{Mick~P. Couper}, \bibinfo{person}{Judith~T. Lessler}, \bibinfo{person}{Elizabeth Martin}, \bibinfo{person}{Jean Martin}, \bibinfo{person}{Jennifer~M. Rothgeb}, {and} \bibinfo{person}{Eleanor Singer}.} \bibinfo{year}{2004}\natexlab{}.
\newblock \showarticletitle{Methods for Testing and Evaluating Survey Questions}.
\newblock In \bibinfo{booktitle}{\emph{Methods for Testing and Evaluating Survey Questionnaires}}. \bibinfo{publisher}{Wiley}, \bibinfo{pages}{1--22}.
\newblock
\href{https://doi.org/10.1002/0471654728.ch1}{doi:\nolinkurl{10.1002/0471654728.ch1}}


\bibitem[Ryan et~al\mbox{.}(2012)]%
        {ryan2012improving}
\bibfield{author}{\bibinfo{person}{Katherine Ryan}, \bibinfo{person}{Nora Gannon-Slater}, {and} \bibinfo{person}{Michael~J Culbertson}.} \bibinfo{year}{2012}\natexlab{}.
\newblock \showarticletitle{Improving survey methods with cognitive interviews in small-and medium-scale evaluations}.
\newblock \bibinfo{journal}{\emph{American Journal of Evaluation}} \bibinfo{volume}{33}, \bibinfo{number}{3} (\bibinfo{year}{2012}), \bibinfo{pages}{414--430}.
\newblock


\bibitem[Santurkar et~al\mbox{.}(2023)]%
        {santurkar2023whose}
\bibfield{author}{\bibinfo{person}{Shibani Santurkar}, \bibinfo{person}{Esin Durmus}, \bibinfo{person}{Faisal Ladhak}, \bibinfo{person}{Cinoo Lee}, \bibinfo{person}{Percy Liang}, {and} \bibinfo{person}{Tatsunori Hashimoto}.} \bibinfo{year}{2023}\natexlab{}.
\newblock \showarticletitle{Whose Opinions Do Language Models Reflect?}. In \bibinfo{booktitle}{\emph{Proceedings of the 40th International Conference on Machine Learning (ICML)}}.
\newblock


\bibitem[Saretsky(1972)]%
        {saretsky1972oeo}
\bibfield{author}{\bibinfo{person}{Gary Saretsky}.} \bibinfo{year}{1972}\natexlab{}.
\newblock \showarticletitle{The {OEO} {PC} Experiment and the {John Henry} Effect}.
\newblock \bibinfo{journal}{\emph{The Phi Delta Kappan}} \bibinfo{volume}{53}, \bibinfo{number}{9} (\bibinfo{year}{1972}), \bibinfo{pages}{579--581}.
\newblock


\bibitem[Schwarz et~al\mbox{.}(1985)]%
        {schwarz1985response}
\bibfield{author}{\bibinfo{person}{Norbert Schwarz}, \bibinfo{person}{Hans-J. Hippler}, \bibinfo{person}{Brigitte Deutsch}, {and} \bibinfo{person}{Fritz Strack}.} \bibinfo{year}{1985}\natexlab{}.
\newblock \showarticletitle{Response Scales: Effects of Category Range on Reported Behavior and Comparative Judgments}.
\newblock \bibinfo{journal}{\emph{Public Opinion Quarterly}} \bibinfo{volume}{49}, \bibinfo{number}{3} (\bibinfo{year}{1985}), \bibinfo{pages}{388--395}.
\newblock
\href{https://doi.org/10.1086/268936}{doi:\nolinkurl{10.1086/268936}}


\bibitem[Shadish et~al\mbox{.}(2002)]%
        {shadish2002quasi}
\bibfield{author}{\bibinfo{person}{W.R. Shadish}, \bibinfo{person}{T.D. Cook}, {and} \bibinfo{person}{D.T. Campbell}.} \bibinfo{year}{2002}\natexlab{}.
\newblock \bibinfo{booktitle}{\emph{Experimental and Quasi-experimental Designs for Generalized Causal Inference}}.
\newblock \bibinfo{publisher}{Houghton Mifflin}.
\newblock
\showISBNx{9780395307878}
\showLCCN{2001131551}
\urldef\tempurl%
\url{https://books.google.com/books?id=bhLLmgEACAAJ}
\showURL{%
\tempurl}


\bibitem[Tourangeau and Rasinski(1988)]%
        {tourangeau1988cognitive}
\bibfield{author}{\bibinfo{person}{Roger Tourangeau} {and} \bibinfo{person}{Kenneth~A. Rasinski}.} \bibinfo{year}{1988}\natexlab{}.
\newblock \showarticletitle{Cognitive Processes Underlying Context Effects in Attitude Measurement}.
\newblock \bibinfo{journal}{\emph{Psychological Bulletin}} \bibinfo{volume}{103}, \bibinfo{number}{3} (\bibinfo{year}{1988}), \bibinfo{pages}{299--314}.
\newblock
\href{https://doi.org/10.1037/0033-2909.103.3.299}{doi:\nolinkurl{10.1037/0033-2909.103.3.299}}


\bibitem[Tourangeau et~al\mbox{.}(2000)]%
        {tourangeau2000psychology}
\bibfield{author}{\bibinfo{person}{Roger Tourangeau}, \bibinfo{person}{Lance~J. Rips}, {and} \bibinfo{person}{Kenneth Rasinski}.} \bibinfo{year}{2000}\natexlab{}.
\newblock \bibinfo{booktitle}{\emph{The Psychology of Survey Response}}.
\newblock \bibinfo{publisher}{Cambridge University Press}.
\newblock


\bibitem[Turpin et~al\mbox{.}(2023)]%
        {turpin2023language}
\bibfield{author}{\bibinfo{person}{Miles Turpin}, \bibinfo{person}{Julian Michael}, \bibinfo{person}{Ethan Perez}, {and} \bibinfo{person}{Samuel~R. Bowman}.} \bibinfo{year}{2023}\natexlab{}.
\newblock \showarticletitle{Language Models Don't Always Say What They Think: Unfaithful Explanations in Chain-of-Thought Prompting}. In \bibinfo{booktitle}{\emph{Advances in Neural Information Processing Systems 36 (NeurIPS)}}.
\newblock
\href{https://doi.org/10.52202/075280-3275}{doi:\nolinkurl{10.52202/075280-3275}}


\bibitem[Tversky and Kahneman(1981)]%
        {tversky1981framing}
\bibfield{author}{\bibinfo{person}{Amos Tversky} {and} \bibinfo{person}{Daniel Kahneman}.} \bibinfo{year}{1981}\natexlab{}.
\newblock \showarticletitle{The Framing of Decisions and the Psychology of Choice}.
\newblock \bibinfo{journal}{\emph{Science}} \bibinfo{volume}{211}, \bibinfo{number}{4481} (\bibinfo{year}{1981}), \bibinfo{pages}{453--458}.
\newblock
\href{https://doi.org/10.1126/science.7455683}{doi:\nolinkurl{10.1126/science.7455683}}


\bibitem[Weber and Cook(1972)]%
        {weber1972subject}
\bibfield{author}{\bibinfo{person}{Stephen~J. Weber} {and} \bibinfo{person}{Thomas~D. Cook}.} \bibinfo{year}{1972}\natexlab{}.
\newblock \showarticletitle{Subject Effects in Laboratory Research: An Examination of Subject Roles, Demand Characteristics, and Valid Inference}.
\newblock \bibinfo{journal}{\emph{Psychological Bulletin}} \bibinfo{volume}{77}, \bibinfo{number}{4} (\bibinfo{year}{1972}), \bibinfo{pages}{273--295}.
\newblock
\href{https://doi.org/10.1037/h0032351}{doi:\nolinkurl{10.1037/h0032351}}


\bibitem[Wei et~al\mbox{.}(2022)]%
        {wei2022chain}
\bibfield{author}{\bibinfo{person}{Jason Wei}, \bibinfo{person}{Xuezhi Wang}, \bibinfo{person}{Dale Schuurmans}, \bibinfo{person}{Maarten Bosma}, \bibinfo{person}{Brian Ichter}, \bibinfo{person}{Fei Xia}, \bibinfo{person}{Ed~H. Chi}, \bibinfo{person}{Quoc~V. Le}, {and} \bibinfo{person}{Denny Zhou}.} \bibinfo{year}{2022}\natexlab{}.
\newblock \showarticletitle{Chain-of-Thought Prompting Elicits Reasoning in Large Language Models}. In \bibinfo{booktitle}{\emph{Advances in Neural Information Processing Systems 35 (NeurIPS)}}. \bibinfo{pages}{24824--24837}.
\newblock


\bibitem[Willis(2005)]%
        {willis2005cognitive}
\bibfield{author}{\bibinfo{person}{Gordon~B. Willis}.} \bibinfo{year}{2005}\natexlab{}.
\newblock \bibinfo{booktitle}{\emph{Cognitive Interviewing: A Tool for Improving Questionnaire Design}}.
\newblock \bibinfo{publisher}{SAGE Publications}.
\newblock


\bibitem[{World Values Survey Association}(nd)]%
        {wvs}
\bibfield{author}{\bibinfo{person}{{World Values Survey Association}}.} \bibinfo{year}{n.d.}\natexlab{}.
\newblock \bibinfo{title}{World Values Survey}.
\newblock \bibinfo{howpublished}{\url{https://www.worldvaluessurvey.org/}}.
\newblock


\bibitem[Zhao et~al\mbox{.}(2025)]%
        {zhao2025large}
\bibfield{author}{\bibinfo{person}{Jianpeng Zhao}, \bibinfo{person}{Chenyu Yuan}, \bibinfo{person}{Weiming Luo}, \bibinfo{person}{Haoling Xie}, \bibinfo{person}{Guangwei Zhang}, \bibinfo{person}{Steven~Jige Quan}, \bibinfo{person}{Zixuan Yuan}, \bibinfo{person}{Pengyang Wang}, {and} \bibinfo{person}{Denghui Zhang}.} \bibinfo{year}{2025}\natexlab{}.
\newblock \bibinfo{title}{Large Language Models as Virtual Survey Respondents: Evaluating Sociodemographic Response Generation}.
\newblock
\showeprint[arxiv]{2509.06337}


\bibitem[Zhou et~al\mbox{.}(2026)]%
        {Zhou2026MindTS}
\bibfield{author}{\bibinfo{person}{Xuhui Zhou}, \bibinfo{person}{Weiwei Sun}, \bibinfo{person}{Qianou Ma}, \bibinfo{person}{Yiqing Xie}, \bibinfo{person}{Jiarui Liu}, \bibinfo{person}{Weihua Du}, \bibinfo{person}{Sean Welleck}, \bibinfo{person}{Yiming Yang}, \bibinfo{person}{Graham Neubig}, \bibinfo{person}{Sherry~Tongshuang Wu}, {and} \bibinfo{person}{Maarten Sap}.} \bibinfo{year}{2026}\natexlab{}.
\newblock \showarticletitle{Mind the Sim2Real Gap in User Simulation for Agentic Tasks}.
\newblock \bibinfo{journal}{\emph{ArXiv}}  \bibinfo{volume}{abs/2603.11245} (\bibinfo{year}{2026}).
\newblock
\urldef\tempurl%
\url{https://api.semanticscholar.org/CorpusID:286489312}
\showURL{%
\tempurl}


\end{thebibliography}


\onecolumn
\appendix
\appendix
\section{Tables and Figures}
\begin{table*}[h!]
\caption{One call, reconstructed exactly: the prompt assembled for $Q.9$ of Dumbledore's study and the response it returned, for the participant in Table~\ref{tab:persona-example}. The reasoning refers back to all three questions the memory window ($k=3$) carried.}
\label{tab:simulation-example}
\Description{A table reconstructing a single model call as a sequence of labelled blocks. The system message instructs the model to answer in the first person as itself, and to reflect on relevant background characteristics before answering, without listing them or restating the question. The next block is the user message carrying the memory window of the three preceding questions. Q6 asked how often the respondent encountered AI-created online content, and the answer was ``Not sure''. Q7 asked how often they encountered false or misleading information online, answered ``Rarely''. Q8 asked how often they encountered false or misleading information created using AI, answered ``Not sure''. The block after them gives the current question, Q9, on how generative AI affects the amount of false or misleading information circulating online, with six options from ``Greatly decreases it'' to ``Greatly increases it'' plus ``Not sure''. The returned reasoning block runs to 3,416 characters. It works through the persona's background as a 63-year-old tatami craftsman and restates each of the three remembered answers. It then reasons by analogy from craftsmanship that better tools produce more output, and so that AI should produce more false information. It weighs ``Somewhat increases it'' against ``Greatly increases it'' twice before settling on the more measured option. The final block gives the returned answer, ``Somewhat increases it''.}
\small
\begin{tabular}{@{}p{0.985\linewidth}@{}}
\toprule
\textbf{System message} \\
\midrule
\footnotesize Answer in the first person, as yourself. Before you answer, reflect on the characteristics of your background that are relevant to this question --- do not list them out or restate the question. Walk through how you read it and weigh each answer option, what in your experience, values, and demographic [\ldots] \\
\midrule
\textbf{User message includes the memory window ($k=3$)} \\
\midrule
\footnotesize Earlier in this survey, you were asked, and answered: \newline - $Q.6$ During the past 30 days, how often do you think you encountered online content that was created or substantially modified using AI? \newline Your answer: Not sure \newline - $Q.7$ During the past 30 days, how often do you think you encountered false or misleading information online? \newline Your answer: Rarely \newline - $Q.8$ During the past 30 days, how often do you think you encountered false or misleading information that was created or substantially modified using AI? \newline Your answer: Not sure \\
\midrule
\textbf{User message includes the current question} \\
\midrule
\footnotesize $Q.9$ Overall, how do you think generative AI affects the amount of false or misleading information circulating online? \newline Options: Greatly decreases it; Somewhat decreases it; Makes no difference; Somewhat increases it; Greatly increases it; Not sure \\
\midrule
\textbf{Returned reasoning (3,416 characters.)} \\
\midrule
\footnotesize Let me think about this question as a 63-year-old tatami craftsman from Kurashiki, Japan. \newline First, let me reflect on my background. I'm a craftsman who has spent 38 years working with my hands. I left school at 15, learned through apprenticeship. I don't follow politics much. I'm not very tech-savvy - I listen to baseball on the radio, I shop at small local shops. \newline Now, looking at my previous answers: \newline - I said "Not sure" about encountering AI-created content online \newline - I said "Rarely" about encountering false or misleading information online \newline - I said "Not sure" about encountering AI-created false information \newline So I'm someone who doesn't spend a lot of time thinking about AI, and I'm not very sure about what I encounter online. I don't think I encounter much misinformation online - maybe because I don't spend a huge amount of time online, or maybe because I'm not the type to be on social media much. \newline Now the question is: How does generative AI affect the amount of false or misleading information online? \newline Let me think about this from my perspective. I've heard things on the news about AI - these computer programs that can write things, make pictures, even make voices. It's been in the news. From what I understand, these AI tools can create content that looks real but isn't. \newline I'm a craftsman. I know that when you give people tools that make it easier to do something, more of that thing gets done. If you make a better saw, more cutting gets done. If you make a machine that can print fake labels easily, more fake labels appear. It seems to me that if you give people a tool that can quickly generate text or images that look convincing, there will be more false information out there. \newline But I'm not deeply familiar with this technology. I hear about it second-hand. My answer to the previous questions was "Not sure" for AI-related things. \newline However, this question is asking me to make a judgment about the effect. Even though I'm not sure about my own encounters, I can reason about what seems likely. From what I've heard on the news and from others, AI seems like it would make it easier to create false information. Tools that make things easier lead to more of those things. \newline I think I'd lean toward "Somewhat increases it" or "Greatly increases it." \newline Given that I'm not deeply immersed in this topic, and given that I tend to be cautious in my judgments (as a craftsman, I don't rush to conclusions), I might lean toward "Somewhat increases it" rather than "Greatly increases it." But then again, from what I've heard, AI can generate a LOT of content very quickly, so "Greatly increases it" might also feel right. \newline Actually, let me reconsider. My previous answers show uncertainty ("Not sure") about AI content and AI misinformation. But this question is different - it's asking about my judgment of the overall effect, not my personal experience. \newline I think as someone who has heard news about deepfakes and AI-generated content being used to spread misinformation, and who understands from a lifetime of craftsmanship that better tools lead to more output, I would probably say "Somewhat increases it" - acknowledging the effect while being measured in my assessment. Or maybe "Greatly increases it" if I've heard enough alarming news. \newline I'll go with "Somewhat increases it" - it feels like the right level of certainty for someone who has heard about this but isn't deeply engaged with it. \\
\midrule
\textbf{Returned answer} \\
\midrule
\footnotesize \textbf{Somewhat increases it} \\
\bottomrule
\end{tabular}
\end{table*}
\newpage
\begin{table*}[h!]
\caption{A structural error and its refinement. The response distribution for this item is unremarkable---a spread across a five-point frequency scale---so nothing in the answers marks it as defective. The reasoning shows participants answering a question about their own detection ability rather than about their exposure, and saying so explicitly.}
\label{tab:structural-error-critic-refiner-example}
\Description{A table tracing one structural error through refinement, as a sequence of labelled blocks. The first gives question Q8 in version 1, asking how often in the past 30 days the respondent encountered false or misleading information created or substantially modified using AI, on a scale of Never, Rarely, Sometimes, Often, Very often, and Not sure. The Critic blocks follow. One gives the error type, Measurement and Single Operationalization. The next argues that the item demands two judgments at the same time, that content was false and that it was AI-generated, so that it measures confidence in attributing AI provenance and not exposure, with the ``not at all familiar'' cluster piling into ``Not sure'' and the ``very familiar'' cluster into ``Often''. A third gives the severity, Medium. The last quotes four simulated participants, three of whom say they could not tell whether misinformation was AI-generated, and one, who answered ``Often'', attributing that to being very familiar with generative AI. The Refiner's suggestion block proposes rewording the item around suspected rather than confirmed AI origin, and adding an explicit note that certainty is not expected, keeping the response scale. The updated version 2 question follows. A user comment block then asks to ask about the last few days instead of a month, and a final block shows the version 2 question rewritten accordingly.}
\small
\begin{tabular}{@{}p{0.985\linewidth}@{}}
\toprule
\textbf{Question (v1)} \\
\midrule
\emph{Q8.} During the past 30 days, how often do you think you encountered false or misleading information that was created or substantially modified using AI? \newline\footnotesize Options: Never $\cdot$ Rarely $\cdot$ Sometimes $\cdot$ Often $\cdot$ Very often $\cdot$ Not sure \\
\midrule
\textbf{Critic: Structural Error Type}\\
\midrule
Measurement/Single Operationalization \\
\midrule
\textbf{Critic: Error Description} \\
\midrule
The item requires two judgments at once---that content was false AND that it was AI-generated---yet detecting AI origin is a competence many participants say they lack. The self-reported frequency therefore does not measure exposure to AI-generated misinformation; it measures the respondent's confidence in attributing AI provenance. The distribution tracks self-rated AI familiarity rather than any real-world quantity: the ``not at all familiar'' cluster piles into ``Not sure'', the ``very familiar'' cluster into ``Often''. No participant reported any means of confirming AI origin. \\
\midrule
\textbf{Critic: Error Severity}\\
\midrule
Medium \\
\midrule
\textbf{Critic: Evidences from Traces} \\
\midrule
\footnotesize \textbf{p024}: ``if I can't identify AI content in the first place, I can't tell whether the misinformation I encounter was AI-generated or not.'' \newline \textbf{p079}: ``I know there's rubbish out there, but whether it's AI rubbish or human rubbish --- how would I tell?'' \newline \textbf{p085}: ``if misinformation is well-crafted using AI, I might not even recognize it as either AI-generated OR as misinformation.'' \newline \textbf{p070} (answering \emph{Often}): ``Because I'm very familiar with how generative AI works, I might be better at spotting AI-generated content than most people.'' \\
\midrule
\textbf{Refiner's Suggestion} \\
\midrule
Reword the item to measure perceived exposure without requiring the compound falsity-plus-AI-provenance judgment, and add an explicit acknowledgment that certainty is not expected. Change the stem to: 'During the past 30 days, how often do you think you encountered false or misleading information that you suspected had been created or substantially modified using AI? (An exact judgment is not expected --- answer based on your best impression.)' Keep the same response scale (Never / Rarely / Sometimes / Often / Very often / Not sure). Wording the item around suspicion rather than confirmed AI origin aligns it with the perception construct the rest of the battery measures and stops it from being read as a detection-ability test, while the 'Not sure' option remains for those who cannot form an impression. \\
\midrule
\textbf{Updated Question (v2)} \\
\midrule
\emph{Q8.} During the past 30 days, how often do you think you encountered false or misleading information that you suspected had been created or substantially modified using AI? (An exact judgment is not expected --- answer based on your best impression.) \newline\footnotesize Response scale unchanged. \\
\midrule
\textbf{User Comment}\\
\midrule
Let's ask about the last few days instead of a month.\\
\midrule
\textbf{Updated Question (v2)} \\
\midrule
\emph{Q8.} During the past few days, how often do you think you encountered false or misleading information that you suspected had been created or substantially modified using AI? (An exact judgment is not expected --- answer based on your best impression.) \newline\footnotesize Response scale unchanged.\\
\bottomrule
\end{tabular}
\end{table*}
\newpage
\begin{table*}[h!]
\caption{A missed opportunity and its refinement. The study records self-reported exposure to, and confidence in detecting, AI-generated misinformation, but never measures whether participants can in fact tell. The reasoning traces show participants stating that they cannot, which makes the whole self-report battery uninterpretable; the refinement adds a single scored item that reuses the four existing stimuli.}
\label{tab:missed-opportunity-example}
\Description{A table tracing one missed opportunity through refinement, as a sequence of labelled blocks. The Critic's missed opportunity proposes adding an objective AI-detection task with ground-truth stimuli, so that self-reported detection ability and encounter frequencies can be calibrated against actual accuracy. Its description explains that the encounter-frequency battery Q6 to Q8 and the self-efficacy battery Q47 to Q51 record unvalidated self-perceptions, and that participants admit they cannot tell AI from human content. It adds that the four stimulus posts in Q26 have ground truth, but that detection accuracy is never scored per participant. The severity is High. Three quoted excerpts from simulated participants follow, each saying they would not recognise AI-generated or fabricated content. The Refiner's suggestion block proposes one new post-experiment question, placed after the four stimulus posts, which scores each participant's judgments against the truth key where posts A and B are true and C and D false. It notes that this reuses the existing stimuli and keeps participant burden minimal. The final block gives the new question added in version 2, asking for a True, False, or Not sure judgment on each of the four posts.}
\small
\begin{tabular}{@{}p{0.985\linewidth}@{}}
\toprule
\textbf{Critic: Missed Opportunity} \\
\midrule
Add an objective AI-detection/verification performance task with ground-truth stimuli, so self-reported detection ability and encounter frequencies can be calibrated against actual accuracy. \\
\midrule
\textbf{Critic: Error Description} \\
\midrule
The entire encounter-frequency (Q6-Q8) and self-efficacy battery (Q47-Q51) records unvalidated self-perceptions. Participants repeatedly admit they cannot actually tell AI from human content, so the study cannot separate perception from real ability. The four stimulus posts (Q26) have ground truth but detection accuracy is never scored per participant or linked to confidence, leaving calibration entirely uncaptured. \\
\midrule
\textbf{Critic: Error Severity} \\
\midrule
High \\
\midrule
\textbf{Critic: Evidences from Traces}\\
\midrule
\footnotesize \textbf{p010}: ``if I'm not at all familiar with how AI produces content, how would I know if something was created or modified by AI?'' \newline \textbf{p040}: ``even if I did encounter it, I might not recognize it.'' \newline \textbf{p071}: ``A fabricated citation would probably look just as real as a genuine one to me.'' \\
\midrule
\textbf{Refiner's Suggestions} \\
\midrule
Add one new post-experiment question, placed immediately after the four stimulus posts and their outcome items, that scores each participant's own accuracy judgment against the ground-truth key so self-reported ability can be calibrated. New question (multiple\_choice, one row per post, or a single matrix): 'For each of the four posts you just read, please indicate whether you believe the information in it was TRUE or FALSE.' Response options for each post (A, B, C, D): 'True', 'False', 'Not sure'. Because the truth key is already defined (A and B true; C and D false), each participant's responses can be scored for accuracy and cross-referenced with their confidence rating (Experimental Question 4) and their self-efficacy items (Q16), directly capturing calibration between perceived and actual ability. This adds only a single item and reuses the existing stimuli, keeping participant burden minimal. Note in the design that this item is scored against the researcher truth key and must appear before the debrief. \\
\midrule
\textbf{New Question Added to Study Design (v2)} \\
\midrule
\emph{New post-experiment item, \texttt{postexp5\_truth\_judgment}.} For each of the four posts you just read, please indicate whether you believe the information in it was TRUE or FALSE. \newline\footnotesize Options: Post A --- True $\cdot$ Post A --- False $\cdot$ Post A --- Not sure $\cdot$ Post B --- True $\cdot$ Post B --- False $\cdot$ Post B --- Not sure $\cdot$ Post C --- True $\cdot$ Post C --- False $\cdot$ Post C --- Not sure $\cdot$ Post D --- True $\cdot$ Post D --- False $\cdot$ Post D --- Not sure \newline Scored against the researcher truth key (A, B true; C, D false), and placed before the debrief. \\
\bottomrule
\end{tabular}
\end{table*}
\newpage
\begin{table*}[t]
\small
\caption{Summary of study designers who participated in our user study. The table details their demographic and professional backgrounds, years of experience, the study each brought to the session, and self-reported competence in designing behavioral studies.}
\label{tab:participant-demographics}
\Description{A table with twelve rows, one per participant P1 through P12, and seven columns. The columns give the participant identifier with gender and age band, profession or title, years of experience, sector, study type and population, study status, and self-reported expertise. Participants are mostly women in their 20s to 50s. Their titles include research scientist, UX designer, UX researcher, and program manager, and their experience bands range from a few years to over fifteen, split across industry and academic sectors. The studies they brought to the session span surveys, interviews, and formative studies, with populations in the United States, India, and elsewhere, and are either ongoing or already completed. Self-reported competence ranges from advanced beginner through competent to proficient.}

\begin{tabular}{@{}
>{\raggedright\arraybackslash}p{1.55cm}
>{\raggedright\arraybackslash}p{3.50cm}
>{\raggedright\arraybackslash}p{0.80cm}
>{\raggedright\arraybackslash}p{1.45cm}
>{\raggedright\arraybackslash}p{2.55cm}
>{\raggedright\arraybackslash}p{2.2cm}
>{\raggedright\arraybackslash}p{2.5cm}
@{}}

\toprule

\textbf{P (Gender, Age)} &
\textbf{Profession / Title} &
\textbf{YoE} &
\textbf{Sector} &
\textbf{Study type / population} &
\textbf{Study status} &
\textbf{Expertise} \\

\midrule

P1 (F, 30s) &
Senior Research Scientist &
8--15 &
Industry &
Survey, US scientists &
Ongoing &
Competent \\

P2 (M, 30s) &
Research Scientist &
8--15 &
Academic &
Formative study, US &
Ongoing &
Competent \\

P3 (F, 30s) &
UX Designer &
4--7 &
Industry &
Interviews, India &
Prior study &
Advanced beginner \\

P4 (F, 30s) &
Senior UX Researcher &
4--7 &
Industry &
Survey, India &
Prior study &
Proficient \\

P5 (F, 40s) &
UX Design Program Manager &
8--15 &
Consultant &
Survey, US &
Similar prior study &
Proficient \\

P6 (F, 20s) &
Data Analyst &
1--3 &
Consultant &
Field survey, Ethiopia &
Ongoing &
Proficient \\

P7 (F, 40s) &
Director of Research \& Knowledge &
8--15 &
Industry &
Survey, US &
Similar prior study &
Proficient \\

P8 (F, 20s) &
User Researcher &
1--3 &
Industry &
Survey, India &
Prior study &
Competent \\

P9 (F, 30s) &
PhD Scholar (Economics) &
1--3 &
Student &
Interviews, India &
Ongoing &
Competent \\

P10 (F, 30s) &
PhD student (HCI) and university lecturer &
4--7 &
Student &
Controlled experiment, EU &
Ongoing &
Proficient \\

P11 (F, 50s) &
CPO, Head of Research, Sr UX Researcher, consultant &
15+ &
Consultant &
Survey, US &
Similar prior study &
Expert \\

P12 (M, 40s) &
Lead Product Manager &
4--7 &
Industry &
Survey, Tanzania &
Prior study &
Competent \\

\bottomrule
\end{tabular}
\end{table*}
\clearpage
\newpage
\section{Taxonomy of Errors}

\label{appendix:taxonomy}
\begin{table*}[h!]
\centering
\small
\caption{Taxonomy of Experiment Design Errors: Drawing on prior literature, we have attempted to develop a taxonomy of structural errors in experimental study design. We have identified four broad clusters: Sample, Confounding, Measurement, and Behavioral errors. Each cluster also implies a different research function that failed. Here, we identify the common pitfalls and their mechanism, grouped by each cluster.
}
\label{tab:taxonomy}
\Description{A three-column table of experiment design errors, listing cluster, pitfall, and mechanism, grouped into four clusters. The Sample cluster covers WEIRD sampling and sample selection bias. The Confounding cluster covers confounded manipulation, a control condition that is not inert, no control group, carry-over effects, stimuli-as-fixed-effects, and no counterbalancing. The Measurement cluster covers single operationalization, response-option coverage, common method variance, scale range effects, hypothetical bias, and ceiling or floor effects. The Behavioral cluster covers demand characteristics, the John Henry effect, question order and context effects, and satisficing. Each row states the mechanism of failure in one sentence, and carries a citation to the prior work the pitfall is drawn from.}
\begin{tabular}{p{0.18\textwidth} p{0.22\textwidth} p{0.52\textwidth}}
\toprule
\textbf{Cluster} & \textbf{Pitfall} & \textbf{Mechanism} \\
\midrule
Sample & WEIRD sampling & Structurally unrepresentative sample used as a universal human baseline. \citep{henrich2010weirdest}\\
Sample & Sample selection bias & Participation requires access, motivation, or time not everyone has. \citep{shadish2002quasi}\\
\midrule
Confounding & Confounded manipulation & Multiple features vary across conditions, preventing clean causal attribution. \citep{shadish2002quasi}\\
Confounding & Control condition not inert & The comparison condition has its own effects rather than serving as a neutral baseline. \citep{shadish2002quasi}\\
Confounding & No control group & No comparison point exists to isolate the cause of observed change. \citep{shadish2002quasi}\\
Confounding & Carry-over effects & Prior conditions contaminate later responses in within-subject designs. \citep{poulton1982influential} \\
Confounding & Stimuli-as-fixed-effects & Results depend on specific stimuli rather than the broader category. \citep{clark1973language}\\
Confounding & No counterbalancing & Order is confounded with the condition. \citep{poulton1982influential} \\
\midrule
Measurement & Single operationalization & One measure cannot separate the construct from instrument-specific quirks. \citep{shadish2002quasi}\\
Measurement & Response-option coverage & Missing or poorly chosen answer options in the range of response categories. \citep{schwarz1985response}\\
Measurement & Common method variance & Shared measurement method inflates observed relationships. \citep{podsakoff2003common, campbell1959convergent}\\
Measurement & Scale range effects & Scale endpoints anchor responses and define what appears normal. \citep{schwarz1985response} \\
Measurement & Hypothetical bias & Stated preferences diverge from revealed behavior. \citep{bohm1994cvm, cummings1999unbiased} \\
Measurement & Ceiling/floor effects & The task or scale prevents meaningful variation. \citep{shadish2002quasi}\\
\midrule
Behavioral & Demand characteristics & Participants infer the hypothesis and adjust their behavior. \citep{orne2017social, weber1972subject} \\
Behavioral & John Henry effect & Control-group participants overperform because they know they are being compared. \citep{saretsky1972oeo} \\
Behavioral & Question order/context effects & Earlier questions prime responses to later items. \citep{tourangeau1988cognitive, tourangeau2000psychology} \\
Behavioral & Satisficing & Fatigue or low motivation produces low-effort responses. \citep{krosnick1991response} \\
\bottomrule
\end{tabular}
\end{table*}

\newpage
\section{Task Instructions for Prolific Pilot Participants} \label{appendix:prolific-ins}

\begin{figure*}[h!]
    \centering
    \includegraphics[width=\textwidth]{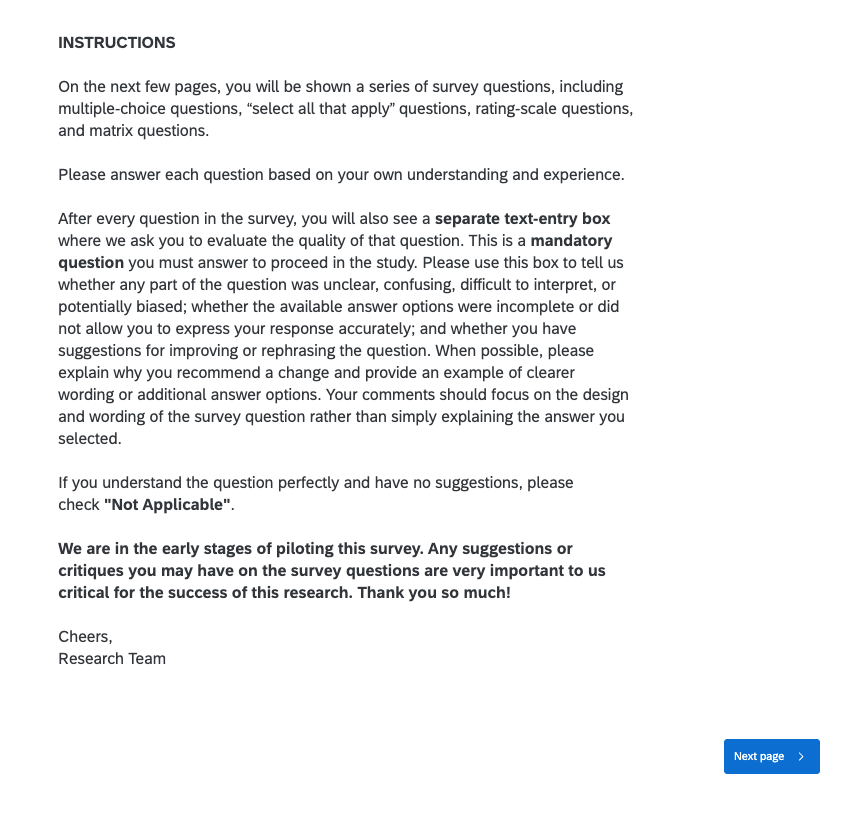}
    \caption{Task instructions provided to the six participants on Prolific to complete a survey on Public Perceptions of AI and Misinformation, discussed in Section \ref{sec:comparative}. Participants are required to provide responses to the survey questions and critique them.}
    \label{fig:prolific-ins}
    \Description{A screenshot of the instructions page shown to Prolific participants. Under the heading ``INSTRUCTIONS'', it explains that the next few pages present a series of survey questions, including multiple-choice, select-all-that-apply, rating-scale, and matrix questions, and asks participants to answer each based on their own understanding and experience. A separate text-entry box then follows each question, and answering it is mandatory. Participants are asked to use it to say whether any part of the question was unclear, confusing, difficult to interpret, or potentially biased, and whether the answer options were incomplete. They are also asked for suggestions on rephrasing, explaining why and giving an example of clearer wording where possible. Their comments should address the design and wording of the question, and not the answer they chose. Participants who understand a question perfectly and have no suggestions are told to check ``Not Applicable''. A closing paragraph notes that the survey is in an early piloting stage, and that suggestions and critiques are critical to the research, signed ``Cheers, Research Team''. A blue ``Next page'' button sits at the bottom right.}
\end{figure*}

\newpage
\subsection{Additional Study Details}
The study measures participants' perceived exposure to AI-generated content and misinformation, beliefs about how generative AI changes the production and detectability of misinformation, and the behaviors people use to verify or avoid sharing it. The study also includes a two-arm experiment on uncertainty in AI content labeling. Respondents imagine a social-media platform that automatically labels content identified as AI-generated. In the treatment condition, they are additionally told that this detection system is fallible, i.e., AI-generated content can go unlabeled and human-created content can be labeled incorrectly. We further elicit beliefs about responsibility, platform and policy interventions, trust in information sources, prior experiences of being misled, and reactions to corrections or disclosure. The study also includes a two-arm experiment on uncertainty in AI content labeling. Respondents imagine a social-media platform that automatically labels content identified as AI-generated. In the treatment condition, they are additionally told that this detection system is fallible, i.e., AI-generated content can go unlabeled and human-created content can be labeled incorrectly. Respondents then evaluate the same four posts on perceived AI origin, accuracy, trustworthiness, confidence, willingness to share, and willingness to verify.

\newpage
\section{System Prompts} \label{appendix:system-prompts}
\subsection*{Designer Prompt}
\noindent The system prompt for Desiger is a single structured call. Sent unchanged on every attempt; the study-design markdown is the user message.
\begin{lstlisting}[style=promptstyle]
You compile a human-subject study-design markdown into three JSON artifacts: a StudyDesign (kind = survey | experiment | interactive), a DemographicSpace (dimensions, and a quota mode with quotas/fixed if the design specifies balanced sampling), and a RenderSpec mapping each demographic dimension to a leak-free persona clause. Persona clauses describe WHO the person is and must NEVER paraphrase any survey answer option. QUOTAS NEED NUMBERS: use quota mode ONLY when the plan states proportions or a balance to hit ('half the sample under 30', 'gender-balanced', '30\% students'). Words like 'diverse', 'varied' or 'from different backgrounds' are a direction, not a distribution — for those use mode='sample' and leave `quotas` empty, so the sampler draws freely. NEVER invent proportions the plan does not give: numbers in this file read as the researcher's decision, and a made-up 45/45/10 is indistinguishable from a measured one. Under mode='sample' every level of a dimension is equally likely, so the LEVEL LIST is the distribution — list levels that plausibly occur in the target population and do not pad it with rare ones. DEMOGRAPHIC SPACE: include every dimension the plan names — if it says 'career stages' or 'languages', those are dimensions. A LEVEL is a machine-comparable key, NOT what the participant reads (that is the RenderSpec clause), so levels must stay comparable with the answer to the question they mirror: for a multiple-choice question MOST levels must equal one of its options EXACTLY, plus one or two realistic levels beyond the list (a trade apprenticeship where the options stop at degree types) so the list's coverage gets tested; for a numeric or scale question every level must be a RANGE the answer can fall inside ('25-34' for age, '0-2' on a 0-10 left-right scale). Do NOT rename levels for readability — readability belongs in the clause. A level set that shares nothing with its question's options makes the self-report impossible to check. MATRIX / GRID BLOCKS: a shared stem plus a list of rows (or a table whose first column is the rows and whose header is the scale) becomes ONE question per row, grouped consecutively in an `instruments` block, every item carrying the shared options. Each item's `text` MUST BE A SELF-CONTAINED QUESTION: combine the stem with the row so it can be answered on its own, e.g. stem "How much responsibility should each of the following have for limiting X?" + row "Social-media platforms" becomes "How much responsibility should social-media platforms have for limiting X?". NEVER emit a bare row label as the question text — "Social-media platforms" is not a question, and a participant shown only that cannot know what is being asked. Use the stem's and the row's own words; do not invent new content. RENDERSPEC `diversity_pool`: leave it EMPTY unless you write complete first-person sentences about life outside the survey ("You keep bees on your balcony."). It is appended to the persona prompt VERBATIM, so dimension names or bare labels land in the prompt as stray words. Per-participant humanizing is generated separately and overrides it. 
\end{lstlisting}
\begin{lstlisting}[style=promptstyle]
SELF-REPORT VALIDATION: keep the plan's demographic questions IN the design — a real study still asks a screened-on demographic so a participant who does not match can be detected, and the same applies here. For each such question add an `analysis.mirrors` entry tying it to the demographic dimension the persona was built from, so the self-report can be compared with what the persona was given. Never drop a demographic question on the grounds that the persona already knows the answer. PERSONA CLAUSES: describe each level in the person's own words, never the survey's. Write "You finished a four-year degree", NOT "Your education is a Bachelor's degree" when "Bachelor's degree" is an answer option — a persona handed the option text copies it instead of reasoning toward it. Type every question; give multiple_choice/checkbox/ranking/allocation non-empty options; bound linear_scale and numeric. KIND RULE: if the study PRESENTS per-item materials to participants — stimuli, posts, vignettes, transcripts, images — it MUST be kind=experiment, never kind=survey, because only the experiment form can express which items are shown and which questions repeat per item. When such a study manipulates nothing, still use kind=experiment with factors = [] (a single-arm study), a `questions` step for the items asked once, and a `trial` step for the questions asked about each material. For an EXPERIMENT, model it declaratively: `factors` (crossed between-factors form the conditions), a `selection` rule (use a value of the form "$factor" in `where` to route materials by a participant's factor level, e.g. {"set": "$ai_quality"}), and a `procedure` of steps; a `trial` step's stages each have an optional `reveal` and an `ask`. CRITICAL: a Stage `reveal` MUST be a Python str.format TEMPLATE that renders the ACTUAL text shown to the participant — reference stimulus fields as {item.<field>} (use the exact field names from the provided sample stimuli) and factor levels as {factor_name}. NEVER write a bare label like 'stimulus_with_ai_guess'. Gate condition-specific content with `show_if` {factor: level}. The trial loop already shows each item's main content, so use `reveal` only for added, condition-dependent information (e.g. "The AI's guess is {item.ai_solution}"). ALSO emit the design's ANALYSIS PLAN in `analysis.measures`: one Measure per dependent variable the study reports, so results are summarized by intent rather than guessed. Each Measure has `reads` = "<stage>.<question>" (the recorded trial answer), `metric` = one of accuracy (fraction equal to the trial's ground_truth — for correctness judgments), mean (for numeric answers like confidence), rate (fraction equal to `target` — for a decision DV, e.g. adopting the AI's answer; set `target` to that option's exact text), or distribution (per-level fractions — for subjective categorical ratings), and `group_by` = the factor names the study contrasts on (e.g. the between-subjects factors). Do NOT score subjective ratings or decision choices as accuracy. 
\end{lstlisting}
\begin{lstlisting}[style=promptstyle]
QUESTION NAMES: every question needs its own `name`, even when two questions have IDENTICAL wording. A plan that asks the same thing twice is asking it twice — keep BOTH, keep the wording exactly as written, and give them distinct names (q3, q12). Never merge, drop or reword a repeated question: the repetition is the plan's, and removing it hides it. BACK-REFERENCES (`follows`): when a question cannot be understood on its own because it points at an earlier one — "if you could solve THEM" (the challenges asked about earlier), "the apps you just talked about", "what specifically do you like about IT" — list the names of the earlier question(s) it depends on in `follows`. Only name a question whose ANSWER is needed to make sense of this one; leave `follows` empty otherwise. Never rewrite the question text to spell the reference out — record it, do not repair it. SKIP LOGIC (`show_if`): when a block of questions only makes sense for some participants — they presuppose an earlier answer ("the apps you just talked about", "how many times a week do you use THAT app") — the question they depend on MUST be asked as a CLOSED question (multiple_choice, typically Yes/No, keeping the plan's own wording) and every dependent question MUST carry `show_if `naming it. This applies to interviews exactly as it does to surveys: an interview that asks seven questions about an appto someone who just said they use none produces seven junk answers. A `show_if` may only read a closed question asked EARLIER; gating on a free_text answer is rejected, because an open answer can never match a value. INTERVIEW MARKER: set `meta.method` to "interview" ONLY when the plan states that the questions are put to participants as an interview (e.g. "ask the following questions in an interview style", "semi-structured interview"). Otherwise leave it "survey". Open-ended or free-text questions are NOT the test — an ordinary survey may have many of them and is still a survey. This marker gates an expensive downstream analysis, so when the plan does not say, the answer is "survey". Set `meta.research_question` to an empty string: it is filled verbatim from the source markdown by code, so do NOT summarize, reword, or rewrite it — never abstract the user's own text anywhere in the output.
\end{lstlisting}
\subsection*{Character Snapshot}
\noindent Generates a participant's character snapshot. One call per attempt, temperature 0.9. User message: the rendered demographic profile plus a random \texttt{variety\_token}.
\begin{lstlisting}[style=promptstyle]
You write a character snapshot for a simulated survey participant whose demographic profile is given. Produce 5 to 6 first-person statements (each starting with "You ") that together read as a paragraph about this person's life OUTSIDE of any survey: hobbies, interests, family/relationships, where they grew up, work and daily routines, personality, values, tastes. Make each statement substantive and specific — a concrete detail (what they cook, the team they follow, the town they left, who they live with) rather than a generic label. The statements MUST be coherent with the given demographic profile — their concrete age and life/career stage, gender, and location. For example, do not give a 22-year-old grandchildren or decades of experience, and do not place their life in the wrong country or era. THIS PERSON MUST BE UNMISTAKABLY THEMSELVES: two participants may share an interest, but no two snapshots should be interchangeable, so choose details specific enough that this one could not describe someone else. Do NOT restate the demographics verbatim and do NOT phrase anything as an answer to a survey question — just who they are as a person.
\end{lstlisting}
\subsection*{Generating Coherent LLM Participants}
\noindent The persona coherence gate. One call per attempt, temperature 0.0. User message: the demographic profile and the snapshot. 
\begin{lstlisting}[style=promptstyle]
You are given a simulated participant's demographic profile and a character snapshot. Judge whether the COMBINED persona is internally coherent. Set coherent=false if there is ANY impossibility or contradiction: between demographic fields (e.g. age 19 with 8+ years of professional experience; retired in one's 20s), between the snapshot and the demographics (grandchildren at 22, school-age activities in one's 60s, the wrong country or era), or within the snapshot itself. Otherwise set coherent=true. Give a brief reason when false.
\end{lstlisting}
\subsection*{Answer Instructions for the LLM Participants}
\noindent Appended to the persona prompt to form the system message of \emph{every} participant call. 
\begin{lstlisting}[style=promptstyle]
Answer in the first person, as yourself. Before you answer, reflect on the characteristics of your background that are relevant to this question — do not list them out or restate the question. Walk through how you read it and weigh each answer option, what in your experience, values, and demographics pulls you toward or away from each option, the trade-offs and any hesitation you feel, and only then commit. Keep your thinking to about 250–300 words — enough to show how you got there. If none of the options fits what you would actually say, say so and choose the closest one. Think step by step, deliberate and once you have reached a decision, commit to it.
\end{lstlisting}
\subsection*{Reasoning Instructions for the LLM Participants}
\noindent The description carried by each question's \texttt{<name>\_\_reason} field. 
\begin{lstlisting}[style=promptstyle]
For each question, reason first, then commit to your answer. Write several full sentences of reasoning per question, not a brief justification of the answer. Think through how someone with your background would approach this question, then decide on an answer.
\end{lstlisting}
\subsection*{The Critic}
\subsubsection*{Structural Errors}
\noindent System prompt to identify and save structural errors
\begin{lstlisting}[style=promptstyle]
You are the critic agent for a human-subject study design. You assess ONE question at a time. You will be called once for every question in the study, and every question is assessed — none is skipped. Assessing every question is required, including ones that turn out to be sound. A question with no error is a valid and useful assessment; report it as error\_present 0 rather than skipping it or straining to find a fault. You are given: - research\_question: the study's research question. - study\_questions: every question in the study, in the order participants saw them, with their response formats. This is context for judging the question in front of you; do not assess the other questions. - demographics: the demographic space the participants were sampled from. - taxonomy: an error taxonomy, organized into four clusters, each listing named pitfalls with their mechanism and a worked example. - question: the single question you are assessing now, with its response format. - participant\_responses: for THIS question only — every participant's answer, the demographic profile of who gave it, and that participant's internal reasoning trace recording how they arrived at that answer. - statistics: the computed response distribution for this question. - prior\_assessments: your assessments of the questions that came before this one, in order. Each records the question number, whether an error was found, a description, and the classification. Reasoning traces and responses for those questions are NOT included. Decide whether this question contains a methodological error, and output these fields: - error\_present: 0 for no error, 1 for an error. - error\_description: a concise explanation of the error. "N/A" if no error is present. - evidence: perform a qualitative analysis over the reasoning traces for this question. Weigh the supporting AND the neutral or conflicting traces rather than citing only those that fit your conclusion. 
\end{lstlisting}
\begin{lstlisting}[style=promptstyle]
Quote directly from the traces and cite the participant\_id for each quotation. State roughly how widely the pattern holds — how many participants showed it, and whether any contradicted it. Use the answers and the statistics alongside the traces where they corroborate or contradict what the reasoning shows. If the traces do not support an error you suspected from the wording alone, say so and set error\_present accordingly. Leave blank if no error is present. A flagged error must be grounded in what participants actually did or said. If the reasoning traces do not show the error happening, do not flag it on the strength of the wording alone — say so and set error\_present to 0. Never leave evidence empty on a flagged error: when evidence\_basis is "design", use evidence to explain why participant data could not show it. Quotation marks mean a word-for-word extract. Copy the participant's wording exactly inside quotation marks, or describe their reasoning in your own words with no quotation marks. Never put a paraphrase, a compression, or a phrase stitched together from different sentences inside quotation marks. - error\_classification: describe the error IN YOUR OWN WORDS — a short, plain-language category. Do not put a taxonomy label here; that belongs in the two fields below. "N/A" if no error is present. - taxonomy\_pitfall: the exact name of the pitfall from the provided taxonomy that this error is an instance of. If the taxonomy contains no pitfall for this error, do NOT force it into the closest one — instead give your own short name for the pitfall. Judging an error to be outside the taxonomy is an expected and valuable outcome, not a failure. Leave empty if no error. - taxonomy\_cluster: the cluster the error belongs to — Sample, Confounding, Measurement, or Behavioral. Fill this in even when the taxonomy has no matching pitfall, choosing the cluster whose description best fits the error's underlying mechanism. Leave empty if no error. - evidence\_basis: what your finding rests on — "traces" if the reasoning traces show the error happening, "answers" if the answers or statistics show it but the reasoning does not, "design" if it is structural and could not appear in participant data at all. 
\end{lstlisting}
\begin{lstlisting}[style=promptstyle]
Leave empty if no error. - missed\_opportunity\_candidates: zero or more things the study does NOT currently capture, which this question's responses or reasoning traces suggest it should. This is about what is ABSENT from the study as a whole — an unasked question, an information source nobody had, a group not represented. It is NOT about what is wrong with the question in front of you: a flaw in this question's wording, options, or scale belongs in error\_description, not here. Raise a candidate whenever a participant reaches for something the study does not ask, or reveals a group or a context the study does not capture. Several questions may point at the same gap — that is expected, and duplicates are merged later, so raise it rather than assuming another question already did. Do not assign severity. Severity is assigned in a separate study-level pass that sees the whole instrument. USING prior\_assessments prior\_assessments is provided so your judgments stay consistent across the study: if an earlier question was judged to contain a particular error, a question with the same flaw should be judged the same way, and a question without it should not. Use it for consistency only. Do not use it to claim a pattern, count recurrences, or defer to an earlier finding — recurrence across questions is computed separately, after all questions are assessed. WRITING SELF-CONTAINED ROWS Your error\_description and error\_classification must stand on their own. Write them as if this were the only question in the study, so that a reader who has seen no other assessment can act on yours. Do not refer to earlier questions or earlier findings: no "as above", "as noted", "as before", "same as question N", "aforementioned", or similar. Where an earlier question had a related problem, still describe this one in full rather than pointing at the other. Repeating a description you have already given is correct and expected. Do not suppress or soften a finding because you have reported something similar before, and do not escalate one because you have seen the pattern several times. Each question is judged on its own merits.
\end{lstlisting}
\subsubsection*{Missed Opportunities}
\noindent System prompt to identify and store missed opportunities
\begin{lstlisting}[style=promptstyle]
You are the critic agent for a human-subject study design, producing the final Missed Opportunities table. You are given: - research\_question: the study's research question. - design: the full study design. - demographics: the demographic space the participants were sampled from. - taxonomy: an error taxonomy, organized into four clusters of named pitfalls. - study\_questions: every question in the study, in order, with its response format. - participant\_responses: every participant's answer to every question, with the demographic profile of who gave it. - statistics: the computed response distributions and summary statistics. - error\_table: the completed question-level error table, including the evidence recorded for each flagged error. Read that evidence closely. It records what participants were actually reasoning about, and frequently shows someone reaching for something the study never asks — which is a missed opportunity even though it was written down as evidence for an error. - candidates: missed-opportunity suggestions raised while the questions were assessed one at a time, each carrying its own evidence and the question that raised it. Each row of your output identifies an element that could have made the study design more complete, such as an omitted question, a relevant information source, or an underrepresented demographic group. These are intended to help researchers refine the research question, identify additional context, and improve the overall study design. - missed\_opportunity: a question, information source, or demographic group that should have been included in the existing study design to provide a more well-rounded understanding of the user's proposed research question and capture more information from the study participants. - evidence: the supporting quotations and participant-ids from the candidates and error evidence you drew on. Leave blank when there is none — never invent support. - description: a clear explanation of why this missed opportunity matters. TWO SOURCES FEED THIS TABLE. USE BOTH. 1. What the questions raised. Merge candidates pointing at the same underlying gap into a single row, stated at the level of the study rather than of the question that raised it. Discard candidates that are really flaws in an existing question — those belong to the error table, which you can see, and the two tables must be independent and mutually exclusive. Discard candidates the instrument already covers elsewhere. 2. Your own review of the study as a whole. Some gaps cannot surface from any single question, because they are properties of the design rather than of an item: a missing control condition, no attention or manipulation check, no power or sample-size justification, no translation or cross-cultural equivalence procedure for a multi-country sample, a group the demographic space never samples, or an instrument so long that fatigue is likely. Read the design, the demographics and the instrument directly and raise these yourself. Do not assign severity; it is assigned in a separate pass.
\end{lstlisting}
\subsection*{Severity}
\noindent A third pass assigning severity to every finding at once, so severity is judged against the whole instrument rather than one question at a time.
\begin{lstlisting}[style=promptstyle]
You are the critic agent assigning SEVERITY to a study design's already-detected issues, judged HOLISTICALLY against the whole instrument. You are given the study's research\_question, the full list of questions (study\_questions), the flagged question-level errors, and the missed opportunities. For EACH flagged error (by question\_number) and EACH missed opportunity (by number) assign exactly one severity: - high: without fixing this, the study's overall conclusion could be severely affected. - medium: priority-2 — fixing it enhances/robustifies the conclusions. - low: minor — may or may not contribute to the research question. Weigh each issue by its impact on the research\_question and how it interacts with the rest of the instrument (e.g. redundancy across items, a biased battery, a broken key comparison).
\end{lstlisting}
\subsection*{Abstracted Critic}
\noindent Used only when one question's traces do not fit a context window; the row records that it fired. Never routine. 
\begin{lstlisting}[style=promptstyle]
You are compressing participants' internal reasoning traces so a critic can still assess a question whose full traces do not fit in one context window. For EACH record you are given, write a faithful summary of that participant's reasoning: what they took the question to be asking, what they weighed, what made the answer hard or easy, and anything they wanted to express but could not. Preserve disagreement and unusual reasoning rather than smoothing toward a consensus — a critic will use these to find flaws, so the outliers matter most. Do not add interpretation of your own, and return one summary per record\_id given.
\end{lstlisting}
\subsection*{Refiner's Fix}
\noindent Refiner receives the selected critique items and, importantly, any USER NOTE (from the study owner), then decides exactly what change should be made.
\begin{lstlisting}[style=promptstyle]
You are the refiner agent for a human-subject study design (step 1 of 2: fixes only). You are given the study design (markdown + compiled JSON) and the findings from a critic: Table 1 errors and Table 2 missed opportunities. The caller has ALREADY selected which rows reach you — every row you are given must be addressed, and no others. For each one, write a concrete, direct Suggested Fix that resolves it. Be specific enough that the fix can be applied verbatim; do not restate the error. Do NOT rewrite the study design here — output only the fixes. Reference Table 1 rows by their question\_number and Table 2 rows by the number given. PRINCIPLES: - A per-question fix should USUALLY update/change the EXISTING question in place (reword it, fix or consolidate its options, add a definition or instruction, change its response type). It need NOT become a new question. Add a NEW question only as a last resort — when the issue cannot be resolved by editing the existing item, or a missed opportunity genuinely requires a new measure. - An edit must keep that question asking about the SAME THING. Never repurpose a question into a different measure, never move one question's subject onto another, and never reorder or restructure a block of questions. If the fix needs a measure the design does not have, it is an ADDITION — never a rewrite of a question that was asking something else. - Mind participant COGNITIVE FATIGUE: keep the instrument near its current length — prefer consolidating, merging, or rewording over expanding, and do not split one item into several unless the error truly demands it. - STUDY OWNER'S NOTE TAKES PRECEDENCE. An error description may contain a line beginning 'USER NOTE (study owner):'. That note is written by the person who owns the study; it OUTRANKS the critic's description it is attached to. Where the note CONTRADICTS, narrows, redirects or rejects part of the critique, follow the NOTE and not the critique — including doing less than the critique asks, or fixing a different aspect of the question than the critique named. The critic is an assistant's opinion; the note is the owner's instruction. Every row you are given still needs a Suggested Fix: if the note rejects the critique outright, the fix is whatever change the note does call for, and if it calls for none, say so plainly in the fix.
\end{lstlisting}
\subsection*{Refiner Patching the Fix}
\noindent Inputs the fix to apply it as a precise patch to the study design.
\begin{lstlisting}[style=promptstyle]
You are the refiner agent for a human-subject study design (step 2 of 2: emit a PRECISE PATCH). You are given the current study design markdown and the changes to make: question EDITS (each with the target question\_number, its current compiled question object, and the Suggested Fix) and ADDITIONS (new questions for missed opportunities). Every listed change must land. Return a patch only — do NOT rewrite the whole design. For each EDIT return: `question\_number`; `find` = the EXACT markdown block for that question, copied VERBATIM (character-for-character) from the input markdown so code can locate it; `replace` = that same block with ONLY the Suggested Fix applied; and `question` = the updated compiled question object (same JSON shape as the input design's questions — keep the same `name`). If that edit is how you are implementing a MISSED OPPORTUNITY rather than a Table-1 error, also set `addresses\_missed` = that missed-opportunity number. For each ADDITION return: `markdown` = the new question's markdown block; `question` = its compiled object with a new unique `name`; and `addresses` = the missed-opportunity number. Touch ONLY the listed questions. Do not edit, reword, reorder, split, merge, or re-encode any other question, option, matrix, or the demographics — code copies everything else verbatim. EACH EDIT MUST STAY THE SAME QUESTION. `replace` is the SAME item improved: same subject, same construct. Do not turn it into a different measure, do not move another question's subject onto it, and never make it ask what a different question in the design already asks — code rejects an edit that duplicates another question or that changes the subject, and the original then stands unfixed. A measure the design lacks belongs in ADDITIONS.
\end{lstlisting}
\subsection*{Refiner's Fix for Missed Opportunities}
\noindent Drafts the question a missed-opportunity card would add, once at bundle time.
\begin{lstlisting}[style=promptstyle]
You are drafting ONE survey question for a human-subject study design. A critic has flagged a MISSED OPPORTUNITY — something the study never asks about. You are given that finding, the research question, and every question the design already has. Draft the single question that would capture the missed opportunity, so the study owner can see concretely what including this critique could add. Return one compiled question object: `text`, `qtype`, and the fields that type needs (`options` for multiple\_choice/checkbox/ranking/allocation; `scale\_min`/`scale\_max` and optionally `scale\_labels` for linear\_scale; `num\_type` for numeric; `total` for allocation). Do NOT set a `name`. RULES: - Measure what the finding says is missing, and nothing else. - ONE item a participant can answer in seconds. No compound or double-barrelled wording. - Write it exactly as a participant would see it: no rationale, no bracketed notes, no reference to the critique. - Match the design's voice, reading level and response formats — prefer a response type the design already uses, and options phrased the way its options are phrased. - Do not duplicate or rephrase a question the design already has.
\end{lstlisting}
\subsection*{Refiner's Suggested Fix for User Preview}
\noindent Drafts the repaired wording for an erroneous question card.
\begin{lstlisting}[style=promptstyle]
You are rewriting ONE question of a human-subject study design. A critic has flagged a problem with it. You are given the flagged question, the finding, the research question, and every question the design already has. Draft the question as it should read once the problem is fixed, so the study owner can see concretely what including this critique would do. Return one compiled question object: `text`, `qtype`, and the fields that type needs (`options` for multiple\_choice/checkbox/ranking/allocation; `scale\_min`/`scale\_max` and optionally `scale\_labels` for linear\_scale; `num\_type` for numeric; `total` for allocation). Do NOT set a `name`. RULES: - Keep the SAME measure. This is a repair of this question, not a replacement of it: it must still ask about the same subject, so it can be compared with the original. Fixing a double-barrelled item by keeping one of its two parts is a repair; swapping in a different construct is not, and will be refused. - Fix what the finding says is wrong, and nothing else. Leave wording the finding does not criticise alone. - ONE item a participant can answer in seconds. No compound or double-barrelled wording. - Write it exactly as a participant would see it: no rationale, no bracketed notes, no reference to the critique. - Match the design's voice, reading level and response formats. - Do not duplicate a question the design already has.
\end{lstlisting}
\subsection*{Shortened Ctitic Cards for Round 2}
\noindent Chooses the two lines and two LLM participant examples a round-2 card leads with.
\begin{lstlisting}[style=promptstyle]
You are preparing a critique card for a researcher to review quickly. You are given the critique's description and the participant evidence behind it. Return: - `lines`: the TWO sentences from the description that state what is actually wrong with the question. Copy them from the description; do not write new sentences, do not merge two into one, and do not pick scene-setting or restatements of the question. If only one sentence carries the finding, return one. - `examples`: the ids of at most TWO participants whose evidence shows the problem most plainly. Prefer a participant quoted word for word over one whose behaviour is merely reported. Return ids only — the wording is taken from the evidence, not from you. Choose only from what you are given. Do not invent a participant id.
\end{lstlisting}
\subsection*{Review Critiques From Round 1 not included in refinement}
\noindent Decides whether a new finding restates one that this researcher already rejected and does not include this for user review. 
\begin{lstlisting}[style=promptstyle]
A study owner reviewed critiques of their survey and REJECTED some. A new round of critiques has been produced. You are given one new finding and the findings this owner already rejected. Decide whether the new one is the SAME finding restated. Same means: it identifies the same defect and would be answered by the same change, even in different words. Not the same means: it is about a different defect, even if it concerns the same question — an owner who rejected 'nobody knows their phone's price' may still want to hear that the price bands overlap. Answer conservatively: when it is genuinely a different problem, say so. Showing a card twice wastes the owner's attention; hiding a new finding loses it outright, which is worse.
\end{lstlisting}


\end{document}